\documentclass[notitlepage,twocolumn,letterpaper,natbib,aps,prd,amsmath,amsfonts,nofootinbib,preprintnumbers,superscriptaddress,secnumarabic]{revtex4-1}
\pdfoutput=1
\usepackage{amssymb,amsmath,latexsym,mathrsfs}
\usepackage{url}
\usepackage{enumitem}
\usepackage{graphicx}
\usepackage{booktabs}
\usepackage[usenames,dvipsnames]{color}
\usepackage[breaklinks,colorlinks,urlcolor=blue,citecolor=blue,linkcolor=magenta]{hyperref}
\usepackage{multirow}
\usepackage{float}
\usepackage{cases}
\usepackage{blindtext}
\usepackage{pifont}
\usepackage{hhline}
\usepackage{mathtools}
\usepackage[section]{placeins}
\usepackage[normalem]{ulem}
\usepackage{amsmath}
\numberwithin{equation}{section}

\def\hc{\text{h.c.}}

\usepackage{xcolor}
\definecolor{linkcolor}{rgb}{0.0, 0.47, 0.75}
\definecolor{citecolor}{rgb}{1.0, 0.5, 0.0}
\hypersetup{
  linkcolor  = linkcolor,
  citecolor  = linkcolor,
  urlcolor   = linkcolor,
  colorlinks = true
}

\begin{document}

\title{A seesaw model for large neutrino masses in concordance with cosmology}
\preprint{CERN-TH-2022-180}

\author{Miguel Escudero}
\email{miguel.escudero@cern.ch}
\affiliation{Theoretical Physics Department, CERN, 1211 Geneva 23, Switzerland}

\author{Thomas Schwetz}
\email{schwetz@kit.edu}
\affiliation{Institut f\"ur Astroteilchenphysik, Karlsruher Institut f\"ur Technologie (KIT), Hermann-von-Helmholtz-Platz 1, 76344 Eggenstein-Leopoldshafen, Germany}

\author{Jorge Terol-Calvo}
\email{jorgetc@iac.es}
\affiliation{Instituto de Astrof\'isica de Canarias, C/ V\'ia L\'actea, s/n
E38205 - La Laguna, Tenerife, Spain}
\affiliation{Dpto. Astrof\'isica, Universidad de La Laguna, E38206 - La Laguna, Tenerife, Spain}

\date{\today}

\begin{abstract}
Cosmological constraints on the sum of the neutrino masses can be relaxed if the number density of active neutrinos is reduced compared to the standard scenario, while at the same time keeping the effective number of neutrino species $N_{\rm eff}\approx 3$ by introducing a new component of dark radiation. We discuss a UV complete model to realise this idea, which simultaneously provides neutrino masses via the seesaw mechanism. It is based on a $U(1)$ symmetry in the dark sector, which can be either gauged or global. In addition to heavy seesaw neutrinos, we need to introduce $\mathcal{O}(10)$ generations of massless sterile neutrinos providing the dark radiation. Then we can accommodate active neutrino masses with $\sum m_\nu \sim 1$~eV, in the sensitivity range of the KATRIN experiment. We discuss the phenomenology of the model and identify the allowed parameter space. We argue that the gauged version of the model is preferred, and in this case the typical energy scale of the model is in the 10~MeV to few~GeV range. 
\end{abstract}

\maketitle

\small
\tableofcontents
\normalsize

\section{Introduction}

Neutrinos play a crucial role during various stages of the cosmological evolution, see e.g., \cite{Dolgov:2002wy,Lesgourgues:2006nd} for reviews. In particular, non-zero neutrino masses will influence cosmological structure formation. Indeed, within the standard $\Lambda$CDM model, cosmology provides a tight bound on the sum of neutrino masses. From combined CMB and BAO observations, the Planck collaboration obtains~\cite{planck}:
\begin{equation}\label{eq:sum_bound}
  \sum m_\nu \equiv \sum_{i=1}^3m_i < 0.12\,{\rm eV} \,(95\%\,{\rm CL}) \ ,
\end{equation}
where $m_i$ are the masses of the three neutrino mass states.
Depending on the specific cosmological data used, even stronger limits can be obtained, see
e.g.~\cite{DiValentino:2021hoh}. With data from ongoing/upcoming large-scale structure surveys by
DESI~\cite{DESI:2016fyo} and Euclid~\cite{Amendola:2016saw},
sensitivities to $\sum m_\nu$ of $0.02\,\mathrm{eV}$ could be achieved in the very near future, see e.g.
\cite{Brinckmann:2018owf,Chudaykin:2019ock}. From neutrino oscillation data, a
minimal value of $\sum m_\nu \approx 0.06\,\mathrm{eV}$ for the normal neutrino mass ordering and $0.1\,\mathrm{eV}$ for inverted ordering is predicted for $m_{\rm lightest} = 0$. Hence, we can expect a positive detection of non-zero  neutrino mass from cosmology in $\Lambda$CDM in the next 5--7 years.

On the other hand, the search for the absolute neutrino mass scale is also one of the top priorities in terrestrial experimental physics. The best limit on the kinematical mass relevant for beta decay comes currently from the KATRIN experiment \cite{KATRIN:2019yun,KATRIN:2021uub} and reads
\begin{equation}\label{eq:KATRIN}
    m_\beta \equiv \left(\sum_{i=1}^3 |U_{ei}|^2 m_i^2\right)^{1/2} < 0.8 \,{\rm eV} \,(90\%\,{\rm CL}) \,,
\end{equation}
where $U_{ei}$ is the mixing matrix element of the mass state $i$ with the electron neutrino. The final sensitivity goal of KATRIN is 0.2~eV for $m_\beta$. For these mass ranges, neutrinos are quasi-degenerate and the quoted current limit and sensitivity correspond to $\sum m_\nu \approx 2.4$~eV and 0.6~eV, respectively, already excluded by the cosmological bound from eq.~\eqref{eq:sum_bound}.

If neutrinos are Majorana particles, they will induce neutrinoless double-beta decay, see Ref.~\cite{Agostini:2022zub} for a review. In the
absence of cancellations due to other exotic physics, the corresponding
decay rate is proportional to an effective Majorana mass:
\begin{equation}\label{eq:mbb}
  m_{\beta\beta} = \left|\sum_{i=1}^3 m_i U_{ei}^2\right| \ .
\end{equation}
The current constraint from the KamLAND-Zen experiment~\cite{KamLAND-Zen:2016pfg} reads
\begin{equation}\label{eq:mbb_bound}
  m_{\beta\beta} < 0.061 - 0.165 \, {\rm eV} 
 \quad (90\% \, {\rm CL}) \ ,
\end{equation}
where the indicated range corresponds to the uncertainty from nuclear
matrix elements. Comparable limits are obtained in \cite{GERDA:2020xhi,CUORE:2021gpk,EXO-200:2019rkq}, and
there is strong ongoing experimental effort to reach sensitivities in the range
$m_{\beta\beta} \approx 0.01 - 0.02$~eV~\cite{Giuliani:2019uno,Agostini:2022zub}. Depending the assumed nuclear matrix element and the unknown complex Majorana phases in $U_{ei}$, the bound in eq.~\eqref{eq:mbb_bound} corresponds to values of $\sum m_\nu \gtrsim 0.6$~eV, largely excluded by the cosmological bound eq.~\eqref{eq:sum_bound}. 

Hence, if the $\Lambda$CDM bound is taken at face value, terrestrial neutrino mass experiments are expected to obtain only upper limits for the neutrino mass, and detection prospects for neutrino-less double beta decay are challenging (depending on the neutrino mass ordering, see~\cite{Agostini:2022zub,Ettengruber:2022mtm}). In view of this situation, several non-standard scenarios have been considered in the literature, to relax cosmological neutrino mass bounds and make cosmology consistent with ``large'' (i.e., observable) neutrino masses. Such scenarios include neutrino decays~\cite{Chacko:2019nej,Escudero:2020ped,Escudero:2019gfk,Barenboim:2020vrr,Chacko:2020hmh,Abellan:2021rfq,Chen:2022idm}, neutrinos with a time-varying mass~\cite{Dvali:2016uhn,Dvali:2021uvk,Lorenz:2018fzb,Lorenz:2021alz,Esteban:2021ozz}, neutrinos with a temperature much lower than the thermal one supplemented with dark radiation~\cite{Farzan:2015pca,GAMBITCosmologyWorkgroup:2020htv}, and neutrinos with a momentum distribution function that deviates from the canonical Fermi-Dirac distribution~\cite{Cuoco:2005qr,Oldengott:2019lke,Alvey:2021sji}. For a discussion of phenomenological consequences of some of these scenarios, see \cite{Alvey:2021xmq}.

In this paper we focus on a mechanism similar to the one introduced by Farzan and Hannestad in \cite{Farzan:2015pca}, where the number density of massive (active) neutrinos is reduced while simultaneously populating a new component of dark radiation, see also \cite{Beacom:2004yd,GAMBITCosmologyWorkgroup:2020htv}. The goal of our work is to realise this mechanism within a UV complete model which simultaneously provides a model for neutrino masses. The model is based on a variant of the type-I seesaw mechanism and is sometimes called \textit{Minimal Extended Seesaw} \cite{Chun:1995js, Barry:2011wb, Zhang:2011vh, Heeck:2012bz, Ballett:2019cqp,Escudero:2020ped,Bringmann:2022aim, Ko:2014bka}.

The outline of our paper is as follows. In section~\ref{sec:FH_mechanism} we review the mechanism of Ref.~\cite{Farzan:2015pca} and highlight some differences in our realisation compared to the original paper. In section~\ref{sec:model} we introduce the model and the relevant parameters, where we discuss two versions, based either on a  global or a local $U(1)$ symmetry. The phenomenology is worked out in section~\ref{sec:pheno}, where we study various cosmological and astrophysical constraints and derive the available parameter space of the model. In section~\ref{sec:discussion} we provide further discussions of the viable parameter space and mention various phenomenological and cosmological consequences and predictions. We draw our conclusions in section~\ref{sec:conclusions}. The interested reader can find supplementary material in the appendices: in App.~\ref{sec:appendixPostrec} we comment on the possibility that the mechanism is active after recombination; in App.~\ref{sec:appendixNeff} we consider in detail the effective number of neutrino species within this mechanism; in App.~\ref{sec:appformuale} we collect a few useful formulae.

\section{Review of the mechanism}\label{sec:FH_mechanism}

Cosmological observations are not directly sensitive to the neutrino mass, but rather to the energy density in neutrinos and its evolution, $\Omega_\nu(z)$. As usual, $\Omega_\nu$ denotes the energy density relative to the critical density. Once a cosmological model is specified there is a direct connection between $\Omega_\nu(z)$ and $\sum m_\nu$ and a bound on the neutrino mass can be placed. The energy density in neutrinos is $\rho_\nu = \sum \left<E_{\nu} \right> n_\nu $, where $n_\nu$ denotes number density and the sum is over the three neutrino mass states. The sensitivity to the energy density in neutrinos in cosmology appears in two regimes, when they are relativistic and when they are non-relativistic. When neutrinos are ultrarelativistic $\left<E_\nu \right> \simeq \left<p_\nu \right>$, and the contribution of neutrinos to the energy density is parameterised by the number of effective ultrarelativistic neutrino species:
\begin{align}
    N_{\rm eff} \equiv \frac{8}{7}\left[\frac{11}{4}\right]^{4/3} \left[\frac{\rho_{\rm rad}-\rho_\gamma}{\rho_\gamma}\right]\,,
\end{align}
where $\rho_{\rm rad}$ is the energy density in relativistic species. We can clearly see that when only neutrinos and photons are present $N_{\rm eff} \propto \left<p_\nu \right> n_\nu$~\cite{Lesgourgues:2006nd}. This quantity, at the time of recombination is measured to be~\cite{planck}:
\begin{align}\label{eq:NeffCMB}
    N_{\rm eff} = 2.99 \pm 0.17\,,
\end{align}
which is in good agreement with the Standard Model prediction of $N_{\rm eff}^{\rm SM} =3.044(1)$~\cite{EscuderoAbenza:2020cmq,Akita:2020szl,Froustey:2020mcq,Bennett:2020zkv}.

When neutrinos become non-relativistic, which happens when $\left<p_{\nu} \right>\simeq m_\nu$, neutrinos start contributing to the expansion rate as dark matter and their energy density is given by $\rho_\nu = \sum m_\nu n_\nu$. It has been explicitly shown that current cosmological observations are insensitive to the exact distribution function of neutrinos~\cite{Alvey:2021sji}. This means that CMB observations can only place a bound on the energy density in non-relativistic neutrinos which Planck CMB data constrains to be~\cite{planck,Alvey:2021sji}:
\begin{align}
\Omega_\nu h^2  < 1.3\times 10^{-3} \quad [{\rm 95\% \,CL}]\,.
\end{align}
Since $\Omega_\nu h^2 \equiv {\sum m_\nu n_\nu^0} h^2/{\rho_{\rm crit}}$ this bound can be seen as a bound on the product of the neutrino mass and the neutrino number density today:
\begin{align}\label{eq:NeutrinoMassBound}
\sum m_\nu \times \left[\frac{n_\nu^0}{56\,{\rm cm^{-3}}}\right]  < 0.12\,{\rm eV} \quad [{\rm 95\% \,CL}] \,,
\end{align}
where $n_{\nu}^0$ refers to the background number density of neutrinos today per helicity state, which in the Standard Model is $n_\nu^0\simeq56\,{\rm cm^{-3}}$~\cite{Lesgourgues:2006nd}. 

Eq.~\eqref{eq:NeutrinoMassBound} highlights a way to relax the cosmological neutrino mass bound. Since what is constrained is a product of number density and mass, reducing the number density of neutrinos would relax the neutrino mass bound accordingly. Importantly, since $N_{\rm eff} \propto \left<p_\nu \right> n_\nu$, if one reduces the number density of neutrinos $N_{\rm eff}$ will decrease, but from eq.~\eqref{eq:NeffCMB} we see that $N_{\rm eff}$ measurements are compatible with the Standard Model prediction. This means that if one wants to reduce the neutrino number density before recombination one should also add new light or massless species beyond the Standard Model to compensate for the decrease of $N_{\rm eff}$ due to the decrease of $n_\nu$. This was precisely the idea of Farzan and Hannestad in~\cite{Farzan:2015pca}. For this mechanism to work, both the reduction of the neutrino number density and the addition of new massless dark radiation should happen before recombination\footnote{In Appendix~\ref{sec:appendixPostrec} we study the possibility of actually realizing the mechanism after recombination. We show that while it is in principle possible the regions of parameter space is significantly more restricted than if the mechanism operates before recombination.}. In addition, this should certainly happen after proton to neutron conversions have frozen out in the early Universe (around $T_\gamma \sim 0.7\,{\rm MeV}$), because otherwise the successful predictions of Big Bang Nucleosynthesis (BBN) will be spoiled. Nevertheless, since CMB observations are only sensitive to the Universe's evolution at $z\lesssim 2\times 10^5$, see e.g.~\cite{Allahverdi:2020bys}, or equivalently $T_\gamma \lesssim 10\,{\rm eV}$, there is plenty of time for this to happen.

\begin{figure}
\begin{center}
\includegraphics[width=0.45\textwidth]{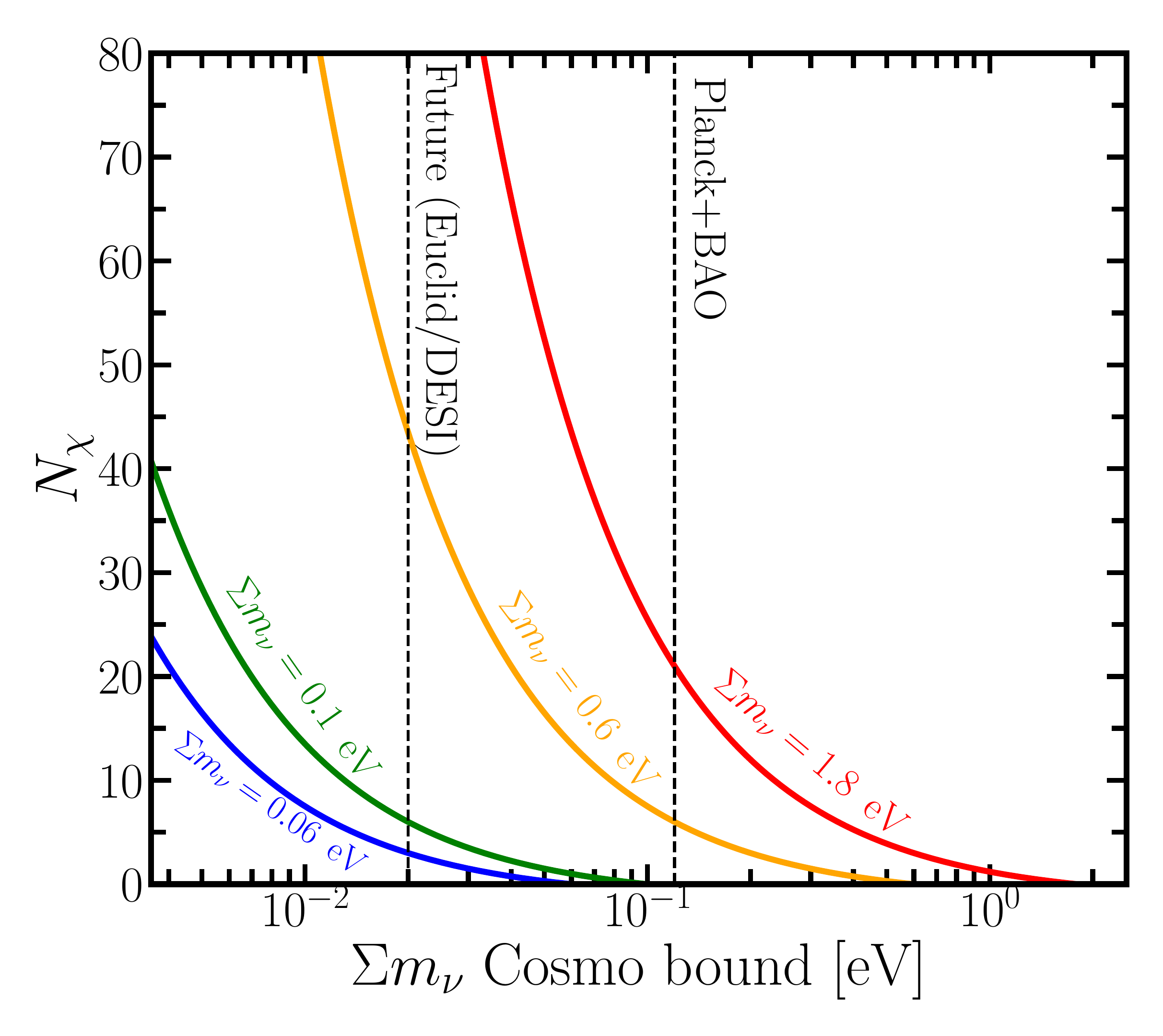}
\caption{Number of massless fermions $\chi$ with $g_\chi = 4$ degrees of freedom needed to make the standard cosmological bound (shown on the horizontal axis) consistent with different values of the sum of the neutrino masses $\sum m_\nu$. The vertical dashed lines indicate the current cosmological bound from Planck+BAO data, eq.~\eqref{eq:sum_bound}, and the prospect for future cosmological observations (0.02~eV).}
\label{fig:NMassless}
\end{center}
\end{figure}

Farzan and Hannestad~\cite{Farzan:2015pca} pointed out a way to achieve the two requirements outlined above: have a large number, $N_\chi$, of massless particles that thermalise with neutrinos after BBN but before recombination, at $10\,{\rm eV} \lesssim T_\gamma \lesssim 100\,{\rm keV}$. Since after neutrino decoupling at $T_\gamma \lesssim 2\,{\rm MeV}$ neutrinos do not interact with the Standard Model plasma, neutrinos cannot be produced anymore and therefore the production of new particles will be at the expense of neutrinos. In this case, the number of effective relativistic neutrino species in the early Universe is almost unchanged from its SM value $N_{\rm eff}\simeq 3$\footnote{In~\cite{Farzan:2015pca} it was mentioned that $N_{\rm eff}$ does not change in this mechanism. However, the production of particles out of equilibrium always leads to some entropy generation which does indeed make $N_{\rm eff}$ slightly larger than 3.044. The small difference is however negligible for practical purposes, see Appendix~\ref{sec:appendixNeff} for more details.}, but the number density  decreases and the current cosmological neutrino mass bound becomes:
\begin{align}\label{eq:NeutrinoMassBound_Mech}
\sum m_\nu < 0.12\,{\rm eV} \,(1+ g_\chi N_\chi/6) \quad [{\rm 95\% \,CL}] \,.
\end{align}
Here $g_\chi$ corresponds to the number of internal degrees of freedom of the massless BSM particle $\chi$ per species and $N_\chi$ is the number of species. Fig.~\ref{fig:NMassless} explicitly shows the number of new massless species needed to relax the cosmological neutrino mass bound as a function of the true neutrino mass for $g_\chi = 4$, as this is the case for the most relevant model of the two we will present later. We see that for example, for the case of $\sum m_\nu = 0.6\,{\rm eV}$ (which is the sensitivity limit of KATRIN), $N_\chi \sim 6$ would be needed to avoid the current Planck bound.

\begin{figure*}[t]
  \centering
  \includegraphics[width=0.95\textwidth]{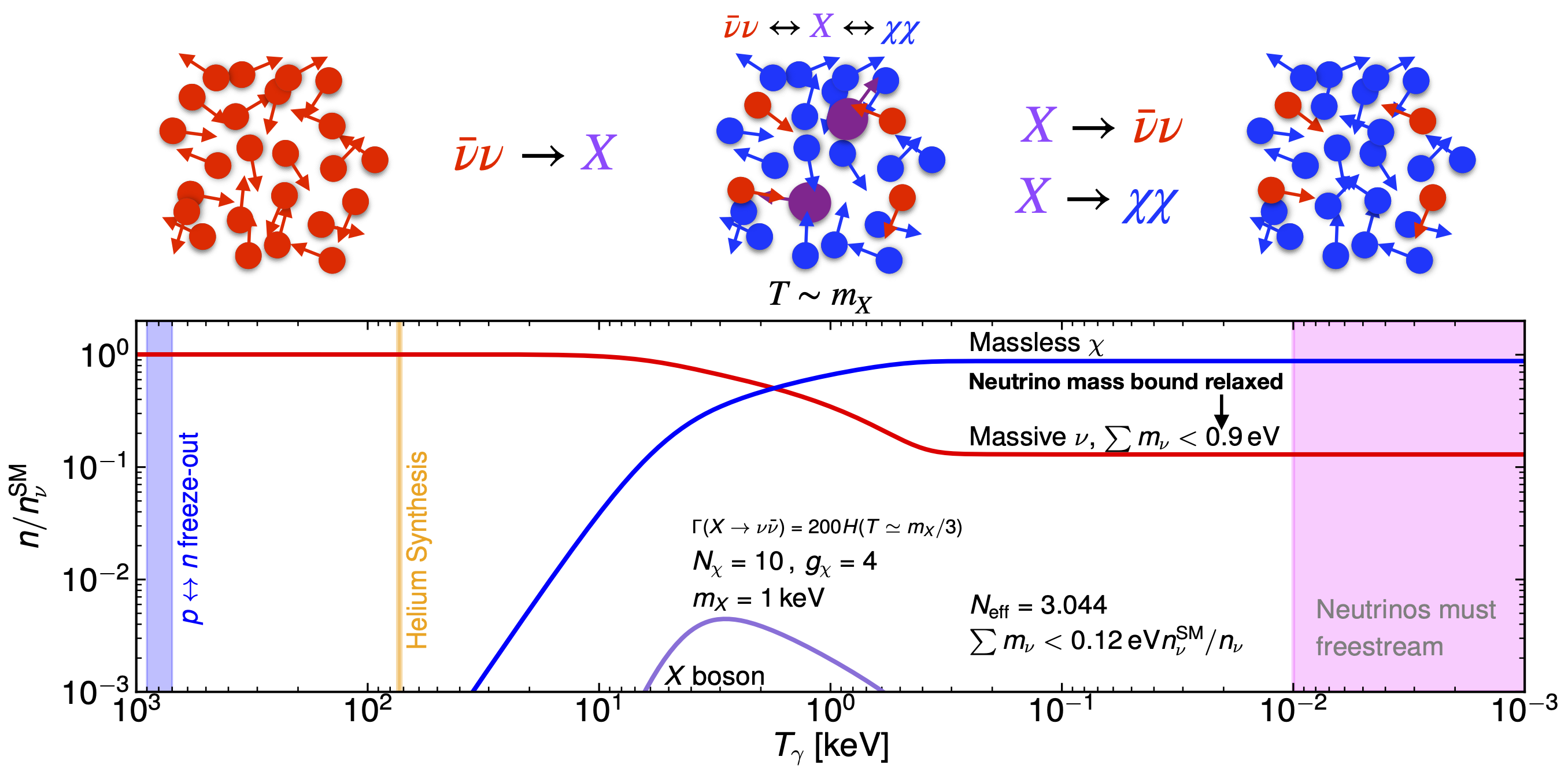}
  \vspace{-0.2cm}
  \caption{Illustration of the mechanism of Farzan and Hannestad~\cite{Farzan:2015pca} to reduce the neutrino number density between BBN and recombination. We show the relative number densities of active neutrinos (red), $N_\chi = 10$ generations of massless sterile fermions (blue), and the mediator boson $X$ with mass $m_X = 1$~keV (purple). 
  For reference we show relevant events taking place in the early Universe, see e.g.~\cite{Pospelov:2010hj}, as well as the region of temperatures at which neutrinos (or other relativistic species) should be freestreaming~\cite{Taule:2022jrz}.}
  \label{fig:mechanism}
\end{figure*}

An important question in this mechanism is how can neutrinos thermalise with a large number of new massless BSM species between BBN and recombination. Ref.~\cite{Farzan:2015pca} considers resonantly enhanced scattering between neutrinos and these new species, mediated by a new boson $X$ with a mass $10\,{\rm eV}\lesssim m_X\lesssim 100\,{\rm keV}$ via the process $\bar{\nu}\nu \to (X) \to \chi\chi$ where $\chi$ here represents one of the massless states. This requirement can actually be relaxed and what is really minimally required is that the new boson thermalises with neutrinos and that it interacts efficiently with a large number of massless species beyond the Standard Model (SM). Thus, the two requirements for this to work are:
\begin{align}\label{eq:Xthermalisation}
    \!\!\!\!\!\!\!\!\!\!\!\!\!\!\!\!\!\!\!\!\!\!\!\!\!\!\!\!\!\!
    {\rm 1)}\quad \quad \left<\Gamma(\bar{\nu}\nu \to X)\right> > H \,,
\end{align}
and 
\begin{align}\label{eq:Xdecay}
   {\rm 2)}\quad \quad  \left<\Gamma(X \to \chi_i +{\rm anything})\right> > H \,,
\end{align}
both for $10\,{\rm eV} \lesssim  T\lesssim 100\,{\rm keV}$. To illustrate the mechanism and its main ingredients we show in fig.~\ref{fig:mechanism} the evolution of neutrino and dark-sector particle densities as a function of photon temperature. For the parameters chosen in the plot, the bound on the sum of neutrino masses can be relaxed to 0.9~eV.

\section{A Seesaw Model for large neutrino masses and dark radiation}\label{sec:model}

In this section we discuss a specific model realisation of the mechanism described in the previous section, which in addition provides a framework to generate neutrino masses, following closely the discussion of Ref.~\cite{Escudero:2020ped}, section~4. 
The beyond-SM ingredients of the model are:
\begin{itemize}
    \item 
three fermion singlets $N_R$ (``right-handed neutrinos'') which play the usual role to generate active neutrino masses as in the type-I seesaw, 
\item
a new abelian symmetry $U(1)_X$ which can be either global or local,
\item
a scalar $\Phi$ with $U(1)_X$ charge $+1$, and 
\item
a set of $N_{\chi}$ fermions $\chi$ with $U(1)_X$ charge $-1$. 
\end{itemize}
With these assignments we can write the following BSM terms in the Lagrangian:
\begin{equation} \label{eq:MES}
-\mathcal{L} = 
\overline{N_R}\,Y_\nu\,\ell_L\,\widetilde{H}^\dagger
+ \frac{1}{2} \, \overline{N_R}\,M_R\,N_R^{c}
+ \overline{N_R} Y_\Phi\,\chi_L\,\Phi
+ \, \hc \, .
\end{equation}
Here $H$ and $\ell_L$ are the SM Higgs and lepton doublets, respectively, and $\widetilde{H}=i\,\tau_2\,H^*$, $M_R$ is the $3\times 3$ Majorana mass matrix for $N_R$, and 
$Y_\nu$ and $Y_\Phi$ are $3\times 3$ and $3\times N_\chi$ Yukawa matrices, respectively.
As we are interested in ``large'' neutrino masses, possibly in the quasi-degenerate regime, we need 3 right-handed neutrinos $N_R$\footnote{We note that the mixing pattern of very degenerate neutrinos is particularly sensitive to radiative corrections~\cite{Georgi:1998bf,Ellis:1999my,Casas:1999tp}. In specific flavor models this poses constraints on the scale of the origin of neutrino masses, see e.g. for some constructions~\cite{deMedeirosVarzielas:2008uzc,Boudjemaa:2008jf}. }.
Here and in the following we keep $SU(2)_L$ and flavour indices contractions implicit. 
The scalar potential is 
\begin{align}
    V &=\mu^2_H H^\dagger H+\lambda_H \left(H^\dagger H\right)^2 \nonumber\\
    &+\mu_\Phi^2|\Phi|^2+\lambda_\Phi|\Phi|^4 +\lambda_{H\Phi}|\Phi|^2 H^\dagger H\,,
\end{align}
with $\mu^2$ and $\mu_\Phi^2$ parameters with dimensions of [mass]$^2$ and $\lambda_H,\lambda_\Phi, \lambda_{H\Phi}$ dimensionless. We assume $\lambda_{H\Phi} = 0$, i.e., no mixing between the two scalar fields. With this assumption we avoid that $\Phi$ gets thermalised in the early Universe due to its interactions with the SM Higgs. Electroweak symmetry breaking takes place in the usual way, with
\begin{equation} \label{eq:EWvev}
\langle H \rangle = \frac{1}{\sqrt{2}} \, \left( \begin{array}{c}
0 \\
v_{\rm EW}
\end{array} \right) \, ,
\end{equation}
with $v_{\rm EW} \simeq 246$~GeV denoting the SM Higgs vacuum expectation value (VEV). The breaking of the $U(1)_X$ takes place when $\Phi$ develops a VEV
\begin{equation} \label{eq:U(1)vev}
\langle \Phi \rangle = \dfrac{v_\Phi}{\sqrt{2}} \,,
\end{equation}
with $v_\Phi^2 = -\mu_\Phi^2/\lambda_\Phi$.

\subsection{Neutrino mixing}
\label{sec:mixing}

After symmetry breaking, several terms in the Yukawa Lagrangian in eq.~\eqref{eq:MES} induce mixing in the neutral lepton sector. In the basis $n = \left(\nu_L^c,
N_R, \chi_L^c\right)$, the fermion mass terms can be written as
\begin{equation} \label{eq:Basisn}
-\mathcal{L}_m =
\frac{1}{2} \overline{n^c} \, \mathcal M_n \, n +\, \hc \, ,
\end{equation}
with the $(6 + N_\chi) \times (6 + N_\chi)$ mass matrix given by 
\begin{equation}
\mathcal M_n=\begin{pmatrix}
0 & m_D & 0\\
m_D^T & M_R & \Lambda\\
0 & \Lambda^T &0
\end{pmatrix},
\label{eq:MassMatrix}
\end{equation}
where $m_D = \frac{v_{\rm EW}}{\sqrt{2}} \, Y_\nu$ and $\Lambda = \frac{v_\Phi}{\sqrt{2}}  \, Y_\Phi$. We assume the following  hierarchy between the entries of the mass matrix: 
\begin{equation}\label{eq:hierarchy}
\Lambda \ll m_D \ll M_R \,,
\end{equation}
where these relations are understood for the typical scales relevant for the matrices.

The block-diagonalisation of the mass matrix leads to the masses of the 3 active neutrinos, the 3 heavy neutrinos and the $N_\chi$ massless sterile neutrinos
\begin{equation}
\mathcal M_n^D=\begin{pmatrix}
m_{\rm active} & 0 & 0\\
0 & m_{\rm heavy} & 0\\
0 & 0 & m_{\rm sterile}
\end{pmatrix},\label{eq:DiagonalMassMatrix}
\end{equation}
with 
\begin{align}
    m_{\rm active}&\approx m_D\, M_R^{-1}\, m_D^T +  \Lambda\, \Lambda^T\, M_R^{-1} \approx m_D\, M_R^{-1}\, m_D^T,\nonumber\\
     m_{\rm heavy} &\approx M_R + m_D\, M_R^{-1}\, m_D^T + \Lambda \, \Lambda^T\, M_R^{-1}  \approx M_R, \nonumber\\
    m_{\rm sterile} &= 0,
\end{align}
where $m_{\rm active} = U_\nu^* \, \widehat m_\nu \, U_\nu^\dagger$. Adopting the diagonal mass basis for charged lepton, $U_\nu$ is the PMNS mixing matrix, given in terms of $3$ mixing angles and $3$ CP-violating phases (including Majorana phases), while $\widehat m_\nu$ is a diagonal matrix containing the physical neutrino mass eigenvalues $m_i$.
There are $N_\chi$ states which are exactly massless at tree level, due to the rank of the matrix \eqref{eq:MassMatrix}. Loop contributions to $m_{\rm sterile}$ are small enough to consider the $N_\chi$ states effectively massless \cite{Escudero:2020ped}.

The mass basis is obtained by rotating the fields with the unitary matrix $W$ which induces a mixing between the different states:
\begin{equation}
\begin{pmatrix}
\tilde{\nu}\\
\tilde{N}\\
\tilde{\chi}
\end{pmatrix} 
= W^\dagger \begin{pmatrix}
\nu_{L}^c\\
N_R\\
\chi_L^c
\end{pmatrix} \,,
\label{eq:Mixing}
\end{equation}
where we have introduced the notation $\tilde{\nu},\, \tilde{N},\, \tilde{\chi}$ to denote the active neutrinos, heavy neutrinos, massless sterile neutrino in the mass basis, respectively.  Following e.g., \cite{Grimus:2000vj} one can find the
mixing matrix at leading order, taking into account the hierarchy in eq.~\eqref{eq:hierarchy}:
\begin{equation}
\small
 W=\begin{pmatrix}
1 & m_D^*\,  (M_R^{-1})^\dagger & -(m_D^{-1})^T\, \Lambda\\
-M_R^{-1}\, m_D^T & 1 & 0\\
\Lambda^\dagger\, (m_D^{-1})^* & 0 & 1
\end{pmatrix}\begin{pmatrix}
U_{\nu} & 0 & 0\\
0 & 1 & 0\\
0 & 0 & 1
\end{pmatrix}.
\label{eq:MixingMatrix}
\end{equation}
Without loss of generality, we have adopted a basis where the right-handed neutrino mass matrix $M_R$ is diagonal. 

In order to simplify the discussion, we will adopt below the one-flavour approximation for the active and heavy neutrinos and introduce mixing angles
\begin{align}
    \theta_{\nu N} = \frac{m_D}{M_R} \,,\quad \theta_{\nu\chi} = \frac{\Lambda}{m_D} \,,
    \label{eq:Mixings}
\end{align}
describing the mixing between active neutrinos and the heavy and massless states, respectively.  With our assumption eq.~\eqref{eq:hierarchy}, both angles are small. We need to keep $N_\chi$ flavors of massless sterile states and $\theta_{\nu\chi}$ represents the mixing between each of them and the active neutrinos. Finally, using the seesaw relation $m_\nu = m_D^2/M_R = \theta_{\nu N}^2 M_R$ we  
will eliminate $m_D$ (or $\theta_{\nu N}$) and $\Lambda$ and consider $m_\nu$, $M_R$ and $\theta_{\nu\chi}$ as independent parameters.

In the following we discuss the relevant interaction terms and distinguish the particularities of the global and gauged versions of the model.

\subsection{Global $U(1)_X$}
\label{sec:global}

Let us decompose the complex scalar $\Phi$ into two real fields as $\Phi = \dfrac{1}{\sqrt{2}}\left(v_\Phi + \rho + i\phi\right)$, where we take $v_\Phi$ real without loss of generality. The real part $\rho$ has a mass $m_\rho$ of order $|\mu_\Phi|$, while $\phi$ corresponds to the Goldstone boson. We assume that in addition to the spontaneously breaking of the $U(1)_X$ global symmetry also explicit breaking terms are present, e.g.\ arising from higher-dimensional terms of the scalar potential, inducing a mass term for the imaginary part $\phi$. Hence, the pseudo-Goldstone mass $m_\phi$ is an additional independent parameter in the global version of the model. 

The relevant processes for our mechanism are $X\leftrightarrow \nu\, \nu$ and 
$X\leftrightarrow \nu\, \chi$, 
where for the global case $X$ can be the scalar $\rho$ or the pseudoscalar $\phi$. These interactions arise from the third term in eq.~\eqref{eq:MES}
through the mixing of the neutral particles $\nu_L,\, N_R$ and $\chi_L$. In the mass basis and after Spontaneous Symmetry Breaking (SSB) we have for the interaction of the scalars with two active neutrinos
\begin{align}\label{eq:OperatorGlobal}
    \overline{N_R} Y_\Phi\,\chi_L\,\Phi + \hc \supset 
    - \overline{\tilde{\nu}} \, \lambda_{\rho/\phi}^{\nu\nu} \dfrac{1}{\sqrt{2}}(\rho - i \gamma_5 \phi)\, \tilde{\nu}^c + \hc \,
\end{align}
with the coupling 
\begin{align}
   \lambda_{\rho/\phi}^{\nu\nu} &= \dfrac{1}{\sqrt{2}} U_\nu^\dagger \,m_D^*\, \left(M_R^{-1}\right)^\dagger\, Y_\Phi\, \Lambda^T m_D^{-1}\, U_\nu^* \nonumber\\
    & =  \frac{1}{v_\Phi}  \, \widehat m_\nu \, U_\nu^T\, \left(m_D^{-1}\right)^\dagger\, \Lambda\, \Lambda^T m_D^{-1} \, U_\nu^* \nonumber\\
    & \to \frac{m_\nu}{v_\Phi} \, \theta_{\nu\chi}^2 \,.
\end{align}
We have used that $\widehat m_\nu = U_\nu^T\, m_D\, M_R^{-1}\, m_D^T\, U_\nu$ and $\Lambda = Y_\Phi\, v_\Phi/\sqrt{2}$. The last line holds in the one-flavour approximation. An analogous calculation for the coupling $\rho/\phi \leftrightarrow \nu\, \chi$ leads to (again in the one-flavour approximation)
$\lambda_{\rho/\phi}^{\nu\chi} = \theta_{\nu \chi} m_\nu/ v_\Phi$,
whereas there is no direct vertex for $\rho/\phi \leftrightarrow \chi\, \chi$, as at leading order there is no mixing between $N_R$ and $\chi_L$. 
Summarizing, in the global realisation of the model we have:
\begin{subequations}\label{eq:globalcouplings}
\begin{align}
\lambda_{\rho/\phi}^{\nu\nu} &= \frac{m_\nu}{v_\Phi} \, \theta_{\nu\chi}^2 \,,
\label{eq:lambdaGlobal_aa}\\
\lambda_{\rho/\phi}^{\nu\chi} &= \frac{m_\nu}{v_\Phi} \, \theta_{\nu\chi} \,,
\label{eq:lambdaGlobal_as}\\
\lambda_{\rho/\phi}^{\chi\chi} &= 0  \,.
\label{eq:lambdaGlobal_ss}
\end{align}
\end{subequations}
We observe that the couplings $\lambda_{\rho/\phi}^{\nu\chi}$ and $\lambda_{\rho/\phi}^{\nu\nu}$ are suppressed by the ratio $m_\nu/v_\Phi$ as well as one or two powers of the active-sterile mixing $\theta_{\nu\chi}$, respectively. 

\subsection{Gauged $U(1)_X$}
\label{sec:gauged}

Let us now consider the case of $U(1)_X$ being a gauge symmetry. In this case, besides the particle content of the global case above we have to add another set of $N_\chi$ fermions charged under the new symmetry with the same charge but opposite chirality, $\chi_R$, in order to cancel the gauge anomaly introduced by the $\chi_L$. 
With the introduction of $\chi_R$, two new terms arise in the Lagrangian, an interaction with the singlet $N_R$ and the scalar $\Phi$ analogous to the last term of eq.~\eqref{eq:MES}, and a vector-like mass term for the $\chi$ field. These terms would change the picture of the neutrino mass generation. These terms can be forbidden by postulating a discrete $\mathbb{Z}_2$ symmetry under which all particles are even except $\chi_R$. Hence, in the following we will assume that these terms are absent.

The breaking of the $U(1)_X$ will give mass to the associated gauge boson, $Z^\prime$, and the would-be Goldstone boson $\phi$ becomes the longitudinal polarisation of the $Z'$.
We assume no tree-level mixing between the SM $U(1)_{\rm em}$ and the  $U(1)_X$ gauge fields; there will still be loop contributions \cite{Gherghetta:2019coi,Bauer:2022nwt} though negligible for phenomenology in our case as there is no particle content charged simultaneously under $U(1)_Y$ or $SU(2)_L$ and $U(1)_X$.

The interactions of fermions with the new gauge boson are described by 
\begin{equation}
\mathcal{L} = \sum_{f=\chi_L,\chi_R}\,  Q_f\,  g_X\,  Z^\prime_\mu\,  \Bar{f}\, \gamma^\mu\,  f,
\label{eq:ZprimeCoupling}
\end{equation}
where $g_X$ is the $U(1)_X$ gauge coupling and $f$ are the $N_\chi$ fermions charged under the $U(1)_X$ with charges $Q_{\chi_L} = Q_{\chi_R} = -1$.
For the interaction of active neutrinos with the $Z^\prime$ we have 
\begin{align}\label{eq:OperatorGauge}
    &g_X\,  Z^\prime_\mu\,  \overline{\chi_L}\, \gamma^\mu\,  \,\chi_L + \hc \supset  \\   
    &  \dfrac{m_{Z^\prime}}{v_\Phi}\, Z^\prime_\mu\, \overline{\tilde{\nu}^c}\, U_\nu^T\, \left(m_D^{-1}\right)^\dagger\, \Lambda^* \gamma^\mu\, \Lambda^T \left(m_D^{-1}\right)\, U_\nu^*\, \tilde{\nu}^c + \hc \nonumber ,
\end{align}
where we have used $m_{Z^\prime} = g_X\, v_\Phi$. In analogy to the global case we introduce couplings for the $Z' \leftrightarrow \nu\, \nu / \nu\, \chi / \chi\,\chi$ interactions
in the one-flavour approximation:
\begin{subequations}\label{eq:gaugecouplings}
\begin{align}
  \lambda_{Z'}^{\nu\nu} &=  \frac{m_{Z'}}{v_\Phi} \, \theta_{\nu\chi}^2 \,, 
  \label{eq:lambdaGauge_aa}\\
  \lambda_{Z'}^{\nu\chi} &=  \frac{m_{Z'}}{v_\Phi} \, \theta_{\nu\chi} \,,\label{eq:CouplingZas}\\
  \lambda_{Z'}^{\chi\chi} &=  \frac{m_{Z'}}{v_\Phi}   \,. \label{eq:gauge_lambda_ss}
\end{align}
\end{subequations}
In comparison to eq.~\eqref{eq:globalcouplings}, here also a $Z'\leftrightarrow\chi\, \chi$ vertex is present.
Note that these interactions cannot be induced by the $\chi_R$ gauge interaction, since there is no mixing between these particles and the active neutrinos under our assumption of a $\mathbb{Z}_2$ symmetry. However, $\chi_R$ will contribute to the mechanism described in sec.~\ref{sec:FH_mechanism} as relativistic species in thermal equilibrium with the neutrinos through the process $\nu\, \nu \leftrightarrow Z^\prime \leftrightarrow \overline{\chi_R}\, \chi_R$. That means that in this case $g_\chi = 4$ in eq.~\eqref{eq:NeutrinoMassBound_Mech}, and therefore requiring a smaller $N_\chi$ compared to the global case. Note that while the interaction $\phi/\rho \to \nu \, \chi$ is crucial for the mechanism in the global case to create the new light species, in the gauge case the $Z^\prime \to \chi \, \chi$ interaction will be more efficient to do so than the analogous with $Z^\prime$ due to the suppression with $\theta_{\nu\chi}$, see eqs.~\eqref{eq:gaugecouplings}. 

The real scalar $\rho$ is also present in the gauged version and will induce scalar-mediated interactions with $\lambda_\rho^{\nu\nu}$, $\lambda_\rho^{\nu\chi}$ according to eqs.~\eqref{eq:lambdaGlobal_aa}, \eqref{eq:lambdaGlobal_as}. In fact, it thermalises, however, for values in the parameter space that are ruled out by other constraints on the interaction $\nu\, \nu\leftrightarrow Z^\prime$ as we will discuss later. 

\subsection{Parameter summary}

Let us now summarise the discussion above and list the independent parameters of the model, adopting the one-flavour approximation. As discussed in sec.~\ref{sec:mixing}, the neutrino sector can be parameterised by the three parameters
\begin{align}
    &m_\nu, \, M_R, \, \theta_{\nu\chi} \,,
\end{align}
being the active neutrino mass, the heavy right-handed neutrino mass, and the mixing between active and massless sterile neutrinos, respectively.
For numerical estimates we fix $m_\nu = 0.2$~eV, at the final KATRIN sensitivity.
As we will see below the value of $M_R$ is irrelevant for the phenomenology of the mechanism itself. However, the value of $M_R$ can be constrained by perturbativity requirements and we will be able to make predictions for the required range of $M_R$, which can in turn lead to cosmological consequences in the very early Universe. Regarding $\theta_{\nu\chi}$, we work within the one-flavor approximation and thus consider that the mixing between all the $\chi_L$ states and $\nu_a$ is comparable in size. As we will see, this parameter is constrained mainly by BBN observations and we will place constraints that take into account the number of $\chi_L$ states, $N_\chi$. 

In the scalar and interaction sector we take as independent parameters
\begin{align}
    v_\Phi,\, m_\rho , \, m_X = 
    \left\{\begin{array}{l}
        m_\phi \quad\text{(global)} \\
        m_{Z'} \quad\!\text{(gauge)} 
    \end{array}\right. \,.
\end{align}
We use the VEV and the real-scalar mass $m_\rho$ as independent parameters of the scalar potential.
In the global $U(1)_X$ case, the pseudo-Goldstone boson mass $m_\phi$ is another independent parameter, whereas in the gauge version, we trade the gauge coupling constant $g_X$ by taking the VEV and the $Z'$ mass as independent. 

In summary, the most relevant parameters to be determined by phenomenology  are $\theta_{\nu\chi}$, $v_\Phi$ and $m_X$, where $X = \phi/\rho$ ($Z'$) in the global (gauge) symmetry case. As outlined in sections~\ref{sec:global} and \ref{sec:gauged}, all the relevant interaction rates can be expressed in terms of these few parameters. In what follows we study the available parameter space in terms of these parameters.

\section{Viable parameter space of the model}\label{sec:pheno}

\begin{figure*}[t]
  \centering
  \includegraphics[width= \columnwidth]{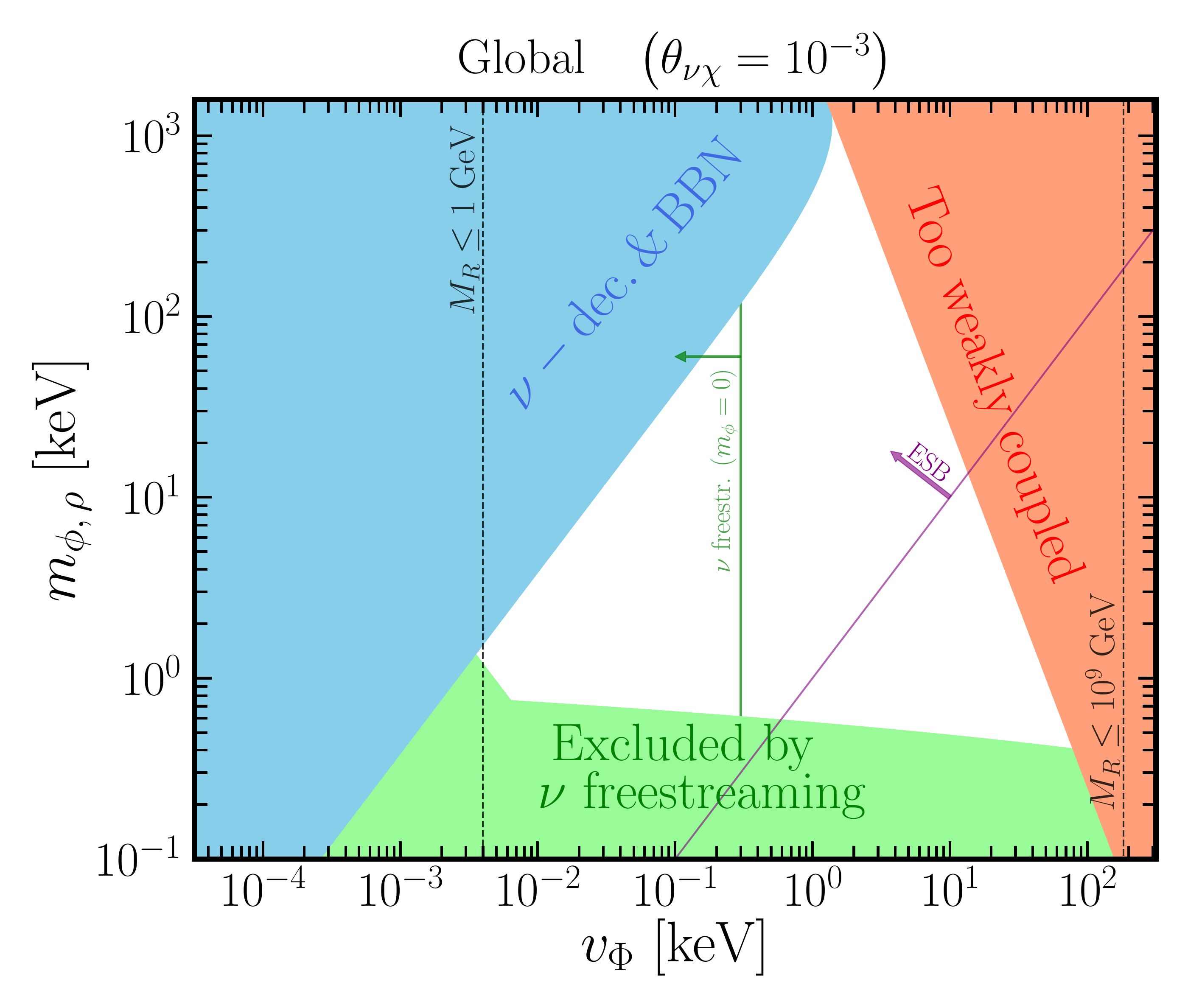}
  \includegraphics[width=\columnwidth]{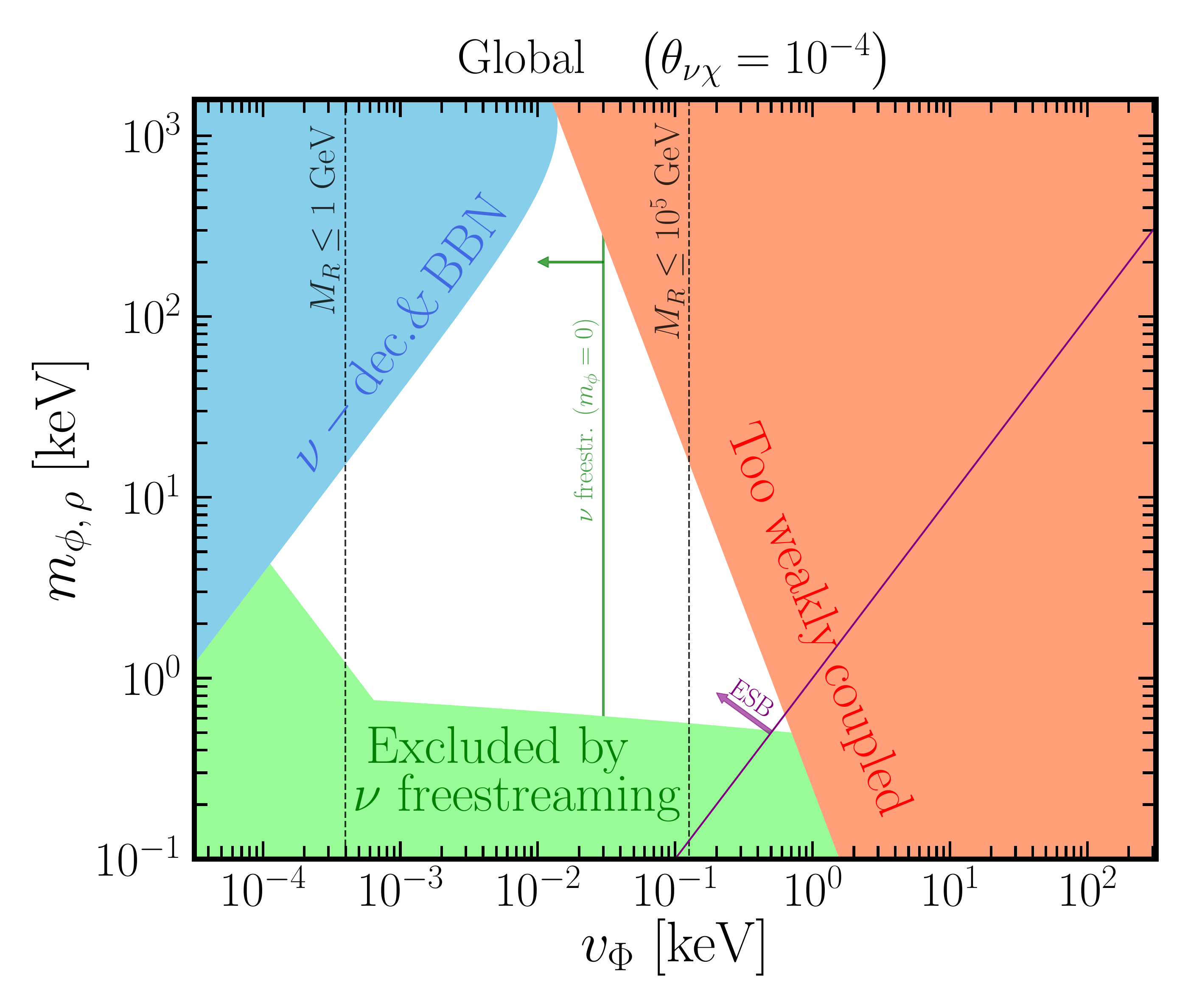}
  \caption{Regions of the parameter space of the global $U(1)_X$ model excluded by several cosmological bounds for a value of the mixing between active and massless sterile neutrinos, $\theta_{\nu\chi} = 10^{-3}$ (left) and $10^{-4}$ (right). The white region is allowed. Vertical dashed black lines correspond to the maximum $M_R$ value in GeV given by the requirement of perturbativity for $Y_\Phi$, see eq.~\eqref{eq:MNmax} or by the requirement of $\lambda_{H\Phi}\leq 10^{-6}$ when stronger. The purple line indicates the region where $m_\phi > v_\Phi$, where the explicit breaking (ESB) of the $U(1)_X$ symmetry by the scalar mass would dominate over the spontaneous breaking. The vertical green line highlights parameter space excluded from neutrino freestreaming in the specific case $m_\phi = 0$.}
  \label{fig:GlobalRegion}
\end{figure*}

\begin{figure*}[t]
  \centering
  \includegraphics[width= \columnwidth]{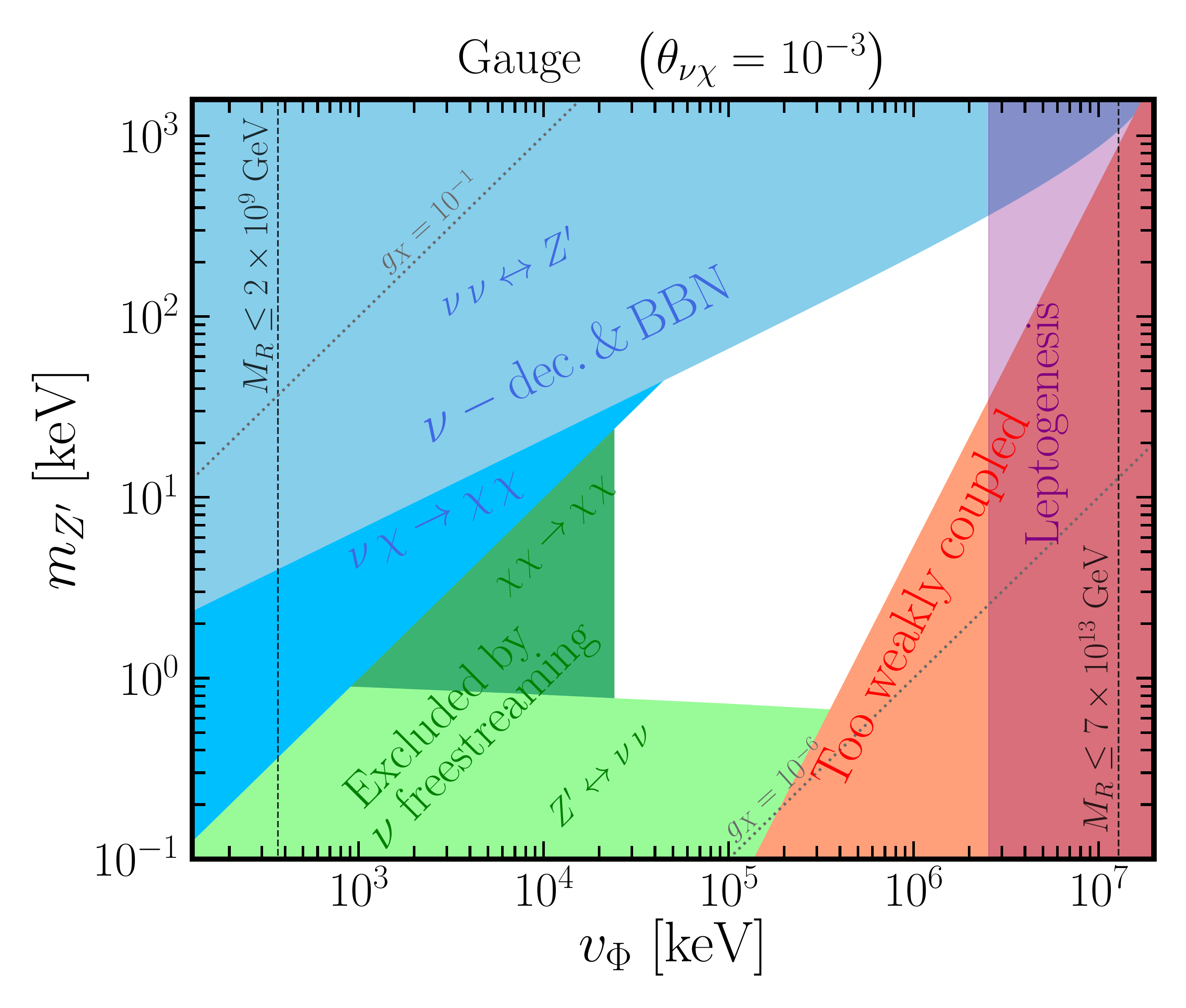}
  \includegraphics[width=\columnwidth]{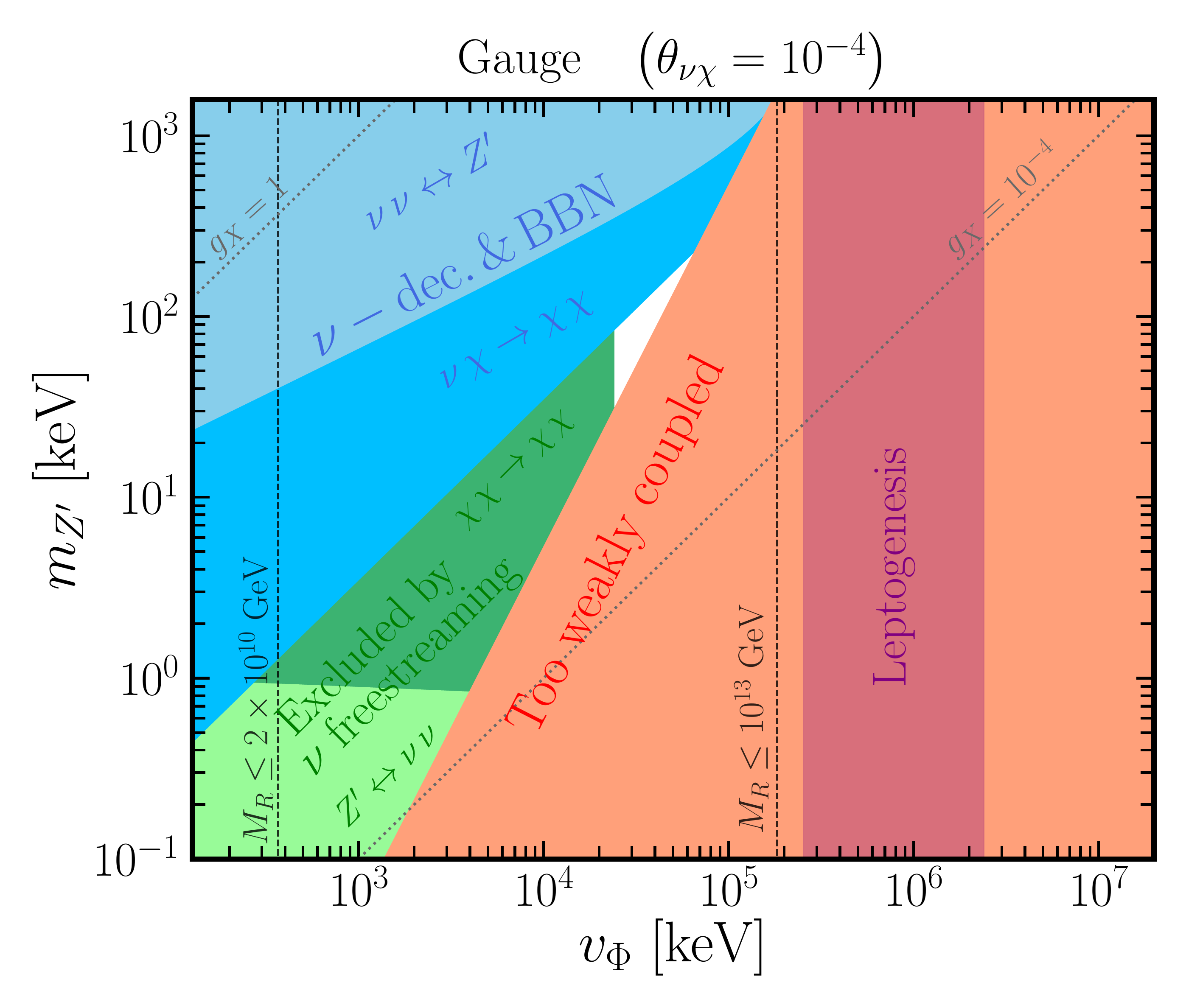}
  \caption{Regions of the parameter space of the gauge $U(1)_X$ model excluded by several cosmological bounds for a value of the mixing between active and massless sterile neutrinos, $\theta_{\nu\chi} = 10^{-3}$ (left) and $10^{-4}$ (right). The white region is allowed. Dotted black lines correspond to the maximum $M_R$ value in GeV given by the requirement of perturbativity for $Y_\Phi$, see eq.~\eqref{eq:MNmax}, or by the requirement of $\lambda_{H\Phi}\leq 10^{-6}$ when stronger. The grey dotted lines indicate regions of constant value of the gauge coupling constant $g_X = m_{Z'}/v_\Phi$. We also indicate the region where standard thermal leptogenesis can work (purple shading).
  }
  \label{fig:GaugeRegion}
\end{figure*}

In this section we consider all relevant phenomenological and cosmological constraints in order to narrow down the parameter space to a region in which the mechanism can be fully realised and the bounds on the neutrino mass significantly relaxed. Figs.~\ref{fig:GlobalRegion} and~\ref{fig:GaugeRegion} show the results for the global and gauge cases, respectively, in terms of the mass of the new boson, $m_X$, and the vacuum expectation value of the $\Phi$ field, $v_\Phi$, for two representative values of the mixing between active and sterile states $\theta_{\nu\chi} = 10^{-3}$ and $10^{-4}$. 
Below we elaborate upon all of these constraints and considerations.

\definecolor{myred}{rgb}{1.0, 0.627, 0.478}
\bigskip

{\Large {\color{myred} $\bullet$$\bullet$$\bullet$}} \textit{Thermalisation:} As discussed in sec.~\ref{sec:FH_mechanism} the mechanism requires the $X$ boson to thermalise with neutrinos and $\chi$ states after proton-neutron freeze-out and before recombination at $10\,{\rm eV}\lesssim T_\gamma \lesssim 0.7 \,{\rm MeV}$. Calculating the thermally averaged decay rate and comparing it to the expansion rate of the Universe allows us to narrow down the range of couplings for which this happens (see appendix~\ref{sec:appformuale} for formulae). In our models, the rate $\Gamma(X\to \nu \chi)$ is always larger than $\Gamma(X\to \nu {\nu})$ due to the mixing angle suppression, see eqs.~\eqref{eq:globalcouplings} and \eqref{eq:gaugecouplings}. However, we assume that initially no $\chi$s are present in the plasma, and therefore the thermalisation requirement applies to the process
$X\leftrightarrow\nu{\nu}$, controlled by the coupling
$\lambda_{X}^{\nu {\nu}}$. Since the thermally average decay rate peaks at $T\sim m_X/3$ we demand $\left<\Gamma(X\leftrightarrow\nu\nu)\right> \gtrsim H (T = m_X/3)$, which approximately implies
\begin{align}\label{eq:thermal}
    \lambda_{X}^{\nu {\nu}} \gtrsim 4 \times 10^{-12} \,\sqrt{\frac{m_X}{{\rm keV}}} \,. 
\end{align}
In figs.~\ref{fig:GlobalRegion} and~\ref{fig:GaugeRegion} we highlight in  red the region of parameter space for $m_X$ and $v_\Phi$ for which the $X$ boson will not reach thermal equilibrium with active neutrinos in the early Universe. The thermalisation requirement leads to an upper bound on the scalar VEV, where the dependence on $m_X$ follows from combining eq.~\eqref{eq:thermal} with either \eqref{eq:lambdaGlobal_aa} or \eqref{eq:lambdaGauge_aa}.

\definecolor{myblue1}{rgb}{0.529, 0.808, 0.921}
\bigskip

{\Large {\color{myblue1} $\bullet$$\bullet$$\bullet$}} \textit{BBN Constraints on $X$--$\nu$ interactions:} The new boson $X$ cannot be in thermal equilibrium at temperatures above $T_\gamma > 0.7\,{\rm MeV}$ for two reasons: 1) It would reduce the number density of neutrinos which in turn participate in $p\leftrightarrow n$ conversions, and 2) it would contribute by itself and with the new $\chi$ states to the expansion rate at the time of BBN since neutrino decoupling happened at a very similar temperature $T \sim 2 \,{\rm MeV}$. In order to ensure that the $X$ boson does not spoil the success of BBN we can place a bound again on the interaction rate between the $X$ boson and active neutrinos:
\begin{align}
    \left<\Gamma(\nu{\nu} \to X)\right> \lesssim H (T =0.7\,{\rm MeV}) \,,
\end{align} 
which in terms of the relevant coupling approximately reads:
\begin{align}
    \lambda_{X}^{\nu {\nu}} \lesssim  10^{-7} \,\frac{{\rm keV}}{m_X} \,.
\end{align}
This constraint is shown in blue in figs.~\ref{fig:GlobalRegion} and~\ref{fig:GaugeRegion}. In passing, we note that the coupling $\lambda_{X}^{\nu {\nu}}$ is also bounded by possible $\nu\nu \to X X $ processes occurring before neutrino decoupling and leading to a large $\Delta N_{\rm eff}$. For this not to happen, then $\lambda_{X}^{\nu {\nu}}\lesssim 10^{-5}$~\cite{Escudero:2019gfk}. However, we never probe such large couplings in the parameter space of interest in our study.

\definecolor{mygreen}{rgb}{0.596, 0.984, 0.596}
\bigskip

{\Large {\color{mygreen} $\bullet$$\bullet$$\bullet$}}  \textit{CMB Constraints on $X$--$\nu$ interactions:} The interaction between $X$ particles and neutrinos and sterile massless states can leave an imprint on CMB observations if it occurs sufficiently close to recombination as this would alter neutrino freestreaming and distort the CMB power spectra. A recent model-independent analysis of Planck legacy data has shown that provided that the $\nu$--$X$ interactions are not efficient at $z < 10^5$ there are no CMB constraints~\cite{Taule:2022jrz}. We will use this as a constraint on the parameter space, requiring that 
\begin{align}
    \left<\Gamma({\nu}\nu \to X)\right>  < H \,\,\,  \text{at}\,\,\,z < 10^{5}\,.
\end{align}
In addition, since in our scenario a large fraction of the energy density in relativistic particles is in the form of massless sterile states $\chi$ we will also require 
\begin{align}
    \left<\Gamma(X \leftrightarrow \chi + \chi/{\nu}  )\right> < H \,\,\, \text{at}\,\,\,z <  10^{5}\,.
\end{align}
The combination of these bounds is shown in light green in figs.~\ref{fig:GlobalRegion} and~\ref{fig:GaugeRegion} and essentially implies $m_X \gtrsim 1$~keV. 

Furthermore, there are constraints coming from neutrino freestreaming from possible $2\to 2$ processes. In particular, one needs to make sure that $\chi \chi \leftrightarrow \chi \chi $ interactions as mediated by a $Z'$ or $\phi$ are not efficient at $z < 10^5$. For the scalar case this is automatically fulfilled because $\lambda_{\phi}^{\chi \chi} =0$. However, for the gauge case it is not since $\lambda_{Z'}^{\chi\chi} = m_{Z'}/v_\Phi = g_X$. By enforcing:
\begin{align}
     \left<\Gamma( \chi  \chi \leftrightarrow  \chi  \chi )\right> < H \,\,\, \text{at}\,\,\,z <  10^{5}\,,
\end{align}
we find a constraint on $v_\Phi$:
\begin{align}\label{eq:freestr-vPhi}
     v_\Phi > 3\times 10^4\,{\rm keV}\,,
\end{align}
for the gauge case. This bound is shown with a darker green colour in fig.~\ref{fig:GaugeRegion} and restricts or even close the allowed parameter space for smaller values of $\theta_{\nu\chi}$.

For the scalar case we still should consider the process $\chi \nu  \leftrightarrow \chi \nu$. By demanding the same requirement on this process we find a small region of parameter space to be excluded which is highlighted in green in the left corner of the available parameter space and that follows a diagonal shape in fig.~\ref{fig:GlobalRegion}.

\bigskip

{\Large {\color{gray} $\bullet$$\bullet$$\bullet$}}  \textit{Astrophysical considerations:} Since our $X$ boson interacts with neutrinos it can be subject to constraints from astrophysical considerations, in particular from core collapse supernova. Supernova cores have temperatures $T\sim 30\,{\rm MeV}$ and release almost all their binding energy in the form of neutrinos on a timescale of $t\sim \mathcal{O}(10)\,{\rm s}$. The observed neutrino spectrum of SN1987A is in broad agreement with that expected from standard core collapse supernova simulations, see e.g. for a review~\cite{Janka:2012wk}. In this context, there are two bounds one can place. Firstly, the $X$ particle should not be copiously produced and escape on a timescale shorter than $t\sim \mathcal{O}(10)\,{\rm s}$, otherwise the supernova will cool much faster than what has been observed in SN1987A~\cite{Raffelt:1996wa}. Requiring that the luminosity in $X$ states is smaller than the one from neutrinos in the standard scenario rules out couplings in the range~\cite{Fiorillo:2022cdq}: 
\begin{align}
   4\times 10^{-6} \,  \frac{\rm keV}{m_X} \lesssim \, \lambda_X^{\nu\nu} \, \lesssim 10^{-4} \, \frac{\rm keV}{m_X} \,,
\end{align}
for $ {\rm keV}$ scale bosons. Secondly, it has been very recently pointed out~\cite{Fiorillo:2022cdq} that even if the luminosity of $X$ particles emitted by the supernova is substantially smaller than that of active neutrinos there could still be constraints due to the lack of high energy events in the time window where the SN1987A signal was observed. For $m_X < {\rm MeV}$, the reported exclusion range corresponds to 
\begin{align}
   3\times 10^{-7} \, \frac{\rm keV}{m_X} \lesssim \, \lambda_X^{\nu\nu} \, \lesssim  10^{-4}\,  \frac{\rm keV}{m_X} \,.
\end{align}
As such, we find that these constraints are weaker than the one we impose from BBN consistency in our scenario. However, a future galactic supernova could push further these limits. A galactic supernova at a distance of 10~kpc detected by Hyper-Kamiokande detector would test values of the coupling down to $\lambda_X^{\nu\nu} \sim 3 \times 10^{-9}\,  \frac{\rm keV}{m_X}$~\cite{Akita:2022etk}.

\bigskip

{\Large {\color{gray} {$\bullet$$\bullet$$\bullet$}}}  \textit{BBN Constraints on the mixing between active and sterile neutrinos:} The massless sterile neutrinos we consider are subject to BBN constraints on their own because they mix with active neutrinos and therefore can be produced via  collisions and oscillations in the early Universe. In addition, since these states are lighter than active neutrinos they feature an enhanced resonant production~\cite{Barbieri:1990vx}. The production rate for these sterile neutrinos peaks at $T \simeq 10\,{\rm MeV}({|\Delta m^2|/0.1{\rm eV^2}})^{1/6}$~\cite{Dolgov:2002wy}, which is well above neutrino decoupling and BBN. The main effect of these additional states will then be to contribute to the energy density in the Universe both at the time of BBN and recombination. Here we take the production rate of sterile neutrinos from~\cite{Abazajian:2005gj} and integrate it up to the time of neutrino decoupling, $T_{\nu}^{\rm dec} \simeq 2 \,{\rm MeV}$, in order to obtain this contribution to the number of ultrarelativistic neutrino species in the early Universe:
\begin{align}
    \left.\Delta N_{\rm eff}\right|_{\chi} \simeq 0.014 \sum_{\chi=1}^{N_{\chi}} \frac{|\theta_{e\chi}|^2 + 0.8(|\theta_{\mu \chi}|^2 + |\theta_{\tau \chi}|^2)}{10^{-6}} \left(\frac{m_\nu}{0.1\,{\rm eV}}\right) \,.
\end{align}
Note that for simplicity to obtain this expression we have neglected destruction of sterile neutrinos in the collision rates. By solving the relevant Boltzmann equations we have explicitly checked that this is a good approximation provided that $\Delta N_{\rm eff}\lesssim 0.3$ for a given new species. 

Assuming that the mixing is similar for each species and applying a bound of $\Delta N_{\rm eff}<0.3$ which is representative of both Planck data~\cite{planck} and  global BBN analyses~\cite{Pisanti:2020efz}, we can find a bound on $\theta_{\nu \chi}$ and $N_\chi$ which reads:
\begin{align}\label{eq:BBNtheta}
    |\theta_{\nu \chi}| \lesssim 10^{-3} \sqrt{\frac{10}{N_\chi}}\, \sqrt{\frac{0.2\, {\rm eV}}{m_\nu}}\,,
\end{align}
where here $m_\nu$ refers to the mass of an individual and almost degenerate active neutrino. 
In terms of the parameters of interest in our study this means that:
\begin{align}\label{eq:BBNrho}
\theta_{\nu \chi} = \frac{\Lambda}{m_D} \lesssim 10^{-4}-10^{-3}\,,
\end{align}
where the ranges are taken by varying $0.1\,{\rm eV}<m_\nu <1\,{\rm eV}$ and $N_\chi\lesssim 50$ as relevant for a range of scenarios as seen in fig.~\ref{fig:NMassless}. This explains why in Figures~\ref{fig:GlobalRegion} and~\ref{fig:GaugeRegion} we take as benchmarks $|\theta_{\nu\chi}| = 10^{-3}$ and $|\theta_{\nu\chi}| = 10^{-4}$. Choosing smaller values of $\theta_{\nu\chi}$ would move the allowed regions to smaller values of $v_\Phi$, as a seen from eqs.~\eqref{eq:globalcouplings}, \eqref{eq:gaugecouplings}. This would lead to non-perturbative gauge couplings or similar inconsistencies in the global case (see discussion below). For the gauged version, for mixing angles significantly below $10^{-4}$ the allowed region above the free streaming bound on $v_\Phi$, eq.~\eqref{eq:freestr-vPhi}, would disappear. Therefore, the preferred parameter region is close to the upper bounds for $\theta_{\nu\chi}$ discussed above.

With active neutrinos close to the eV scale and massless sterile neutrinos, we obtain a mass-squared difference $\Delta m^2\sim 1$~eV$^2$, potentially relevant for short-baseline oscillation experiments \cite{Boser:2019rta,Dentler:2018sju}. However, mixing angles in the range indicated in eq.~\eqref{eq:BBNrho} are too small to be tested in oscillation experiments\footnote{In appendix \ref{sec:ThetaBBN} we discuss the possibility to evade this constraint on $\theta_{\nu\chi}$ for the gauge case thanks to the $\chi$ self-interactions induced potential.}.

\section{Discussion}\label{sec:discussion}

Let us summarise the main results from the various constraints discussed in the previous section, referring to figs.~\ref{fig:GlobalRegion} and \ref{fig:GaugeRegion}. We find a closed region of parameter space for the mediator mass $m_X$ and scalar VEV $v_\Phi$, where the mechanism can work. The mediator mass is restricted roughly to the range 1~keV~$\lesssim m_X\lesssim 1$~MeV, both in the global and gauged version of the model. For the global case, the scalar VEV is roughly in the range 1~eV~$\lesssim v_\Phi\lesssim 10$~keV for $\theta_{\nu\chi} = 10^{-3}$, and roughly one order of magnitude smaller for 
$\theta_{\nu\chi} = 10^{-4}$. In the gauged version, we obtain larger scales for $v_\Phi$, ranging roughly from 10~MeV to few~GeV for the case $\theta_{\nu\chi} = 10^{-3}$, but only viable for 20--200~MeV for $\theta_{\nu\chi} = 10^{-4}$ as a result of the $v_\Phi$ dependence  of the neutrino freestreaming constraint on $\chi$ self-interactions. 

The different scales for the VEV in the global and gauged versions, as well as the dependence on the mixing angle follow from the parametric dependence of the coupling constants shown in eqs.~\eqref{eq:globalcouplings} and \eqref{eq:gaugecouplings}, being proportional to $m_\nu/v_\Phi$ ($m_{Z'}/v_\Phi$) in the global (gauged) version. For the global case, a shift in the neutrino mass would have a similar effect as changing $\theta_{\nu\chi}$, with lower masses moving the excluded regions to smaller $v_\Phi$. This follows directly from eqs.~\eqref{eq:globalcouplings}.

Let us now discuss in some more detail the parameter region for the global $U(1)_X$ version. In the scalar sector, the parameters of the model are the scalar VEV $v_\Phi$, the mass of the real scalar, $m_\rho$, and the mass of the pseudo-Goldstone boson $m_\phi$, which we consider as an independent parameter responsible for the explicit breaking of the $U(1)_X$ symmetry. The diagonal line in fig.~\ref{fig:GlobalRegion} indicates the condition $m_X = v_\Phi$. Considering the case $X=\phi$, we see that in large regions of the parameter space (for $\theta_{\nu\chi} = 10^{-4}$ even all of the viable parameter space), we have $m_\phi > v_\Phi$. This means actually that the explicit symmetry breaking (ESB) happens at scales higher than the spontaneous breaking, which in some sense contradicts the notion of the global symmetry. We consider this configuration as theoretically inconsistent, and we remain only with the small triangle to the right of the diagonal line labelled ESB in fig.~\ref{fig:GlobalRegion} (left), which however, still implies a mass for the pseudo-Goldstone not too far from the VEV, and also requires ``large'' mixing angles $\theta_{\nu\chi}$, saturating the bound from BBN. Furthermore, we note that there is another cosmological element that makes the region where $m_{\rho/\phi} > v_\Phi$ theoretically unappealing. The reason is that in such regions of parameter space the $U(1)_X$ symmetry would only be spontaneously broken in the early Universe at $T\sim v_\Phi$. That in turn means that actually the scalars may not have gotten their masses until $T\ll m_{\rho/\phi} $ which would prevent them from actually thermalising with the neutrino sector of the plasma because of the strong dependence of the interaction rate with $m_{\rho/\phi}$.

Alternatively, we can consider the situation $X=\rho$, i.e., the real part of the complex scalar plays the role of the mediator particle. In this case we can avoid explicit symmetry breaking at all and keep $\phi$ to be a massless Goldstone. Note that $\rho$ and $\phi$ have the same couplings to the fermions, eqs.~\eqref{eq:globalcouplings}, and therefore, the phenomenological discussion from sec.~\ref{sec:pheno} applies equally to $\rho$ and $\phi$. For the mass of $\rho$ we have $m_\rho = \sqrt{2\lambda_\Phi} v_\Phi$. Therefore, the perturbativity requirement for the quartic coupling, $\lambda_\Phi < \sqrt{4\pi}$, implies that $m_\rho$ cannot be much larger than the VEV: $m_\rho \lesssim 2.7 v_\Phi$.  Hence, we see that again we are restricted to the small region to the right of the diagonal line  in fig.~\ref{fig:GlobalRegion}. In addition, in this case there is an additional constraint related to the presence of the massless pseudo-scalar $\phi$, which now can lead to 2-to-2 processes such as $\chi \chi \to \phi \phi$ and $\nu \nu \to \phi \phi$, which can suppress neutrino freestreaming. For this not to happen, these processes cannot be in thermal equilibrium at $z< 10^5$~\cite{Taule:2022jrz}. Explicit calculations show that the coupling mediating this process then should be $\lambda \lesssim 7\times 10^{-7} $~\cite{Forastieri:2019cuf}. In particular, this bound will apply to $\lambda_{\phi}^{\nu \chi}$, see eq.~\eqref{eq:lambdaGlobal_as}, and can be interpreted as a lower bound on the scalar VEV that reads: 
\begin{align}
    v_\Phi > 0.3\,{\rm keV} \, \frac{|\theta_{\nu\chi}|}{10^{-3}} \left(\frac{m_\nu}{0.2\,{\rm eV}}\right)\,.
\end{align}
This constraint is shown as a vertical green line in fig.~\ref{fig:GlobalRegion} and we see that this bound, together with the perturbativity requirement for the quartic coupling, puts also severe restrictions to the parameter space in the case of global $U(1)_X$ with no explicit symmetry breaking ($m_\phi = 0$).

We conclude that the global symmetry version of the model is severely constrained by perturbativity and theoretical consistency arguments. This is not the case for the gauged version: we indicate in fig.~\ref{fig:GaugeRegion} by diagonal lines the values of the gauge coupling
$g_X = m_{Z'}/v_\Phi$, and we can see that the allowed region does not have any perturbativity problem since the couplings are sufficiently small (unless $\theta_{\nu\chi}$ is not chosen much smaller than the values adopted in the figure, see also the discussion in sec.~\ref{sec:pheno}). In addition, since $g_X \ll 1$, we note that the $U(1)_X$ symmetry will be spontaneously broken in the early Universe at $T \sim v_\Phi \gg m_X$. Since the interactions between $Z'$ and fermions are most efficient at $T\sim m_{Z'}/3$ we conclude that the $U(1)_X$ symmetry is broken and $Z'$ is massive by the time the mechanism we describe to reduced the neutrino number density is active.

In the following subsections we discuss further cosmological consequences of the model.

\subsection{Heavy right-handed neutrinos, $\Delta N_{\rm eff}$, and leptogenesis}

The mass of the heavy neutrinos $M_R$ is essentially unconstrained, as for given $m_\nu, v_\Phi$ and $\theta_{\nu\chi}$ we still can choose the Yukawa couplings $Y_\Phi$. However, we can derive an upper bound on $M_R$ from two arguments. First, by requiring perturbativity for $Y_\Phi$, i.e.\ that $Y_\Phi$ has to be smaller than $\sqrt{4\pi}$. In the one flavour approximation this reads
\begin{equation}
    Y_\Phi = \dfrac{\sqrt{m_\nu M_R}}{v_\Phi}\theta_{\nu \chi} \leq \sqrt{4\pi} \,,
\end{equation}
which gives an upper bound on $M_R$:
\begin{equation}
      M_R \leq 4\pi\dfrac{v_\Phi^2}{m_\nu\, \theta_{\nu \chi}^2} \,. \label{eq:MNmax}
\end{equation}
Second, although we set $\lambda_{H\Phi}=0 $ at the electroweak scale, it can get a sizeable value via radiative corrections. In particular, a box diagram with $\nu,\ \chi$ and $N_R$ in the loop can generate this mixing between the two scalar fields. Thus, we require that this contribution has to be smaller than $10^{-6}$ to avoid $\Phi$ thermalisation before BBN~\cite{Escudero:2021rfi}. Using SARAH~\cite{Staub:2008uz,Staub:2015kfa} we find that the one loop contribution to the $\lambda_{H\Phi}$ coupling reads as follows: 
\begin{equation}
    \lambda_{H\Phi}^{\rm loop} \simeq \dfrac{Y_\nu^2\, Y_\Phi^2}{4\pi^2} = \dfrac{\theta_{\nu\chi}^2\, m_\nu^2\, M_R^2}{4 \pi^2\, v_\Phi^2\, v_{EW}^2}.
\end{equation}
The requirement $ \lambda_{H\Phi}^{\rm loop}  < 10^{-6}$ gives an upper bound on $M_R$:
\begin{equation}
      M_R \leq 2\pi\dfrac{v_\Phi\, v_{EW}}{m_\nu\, \theta_{\nu \chi}}10^{-6} \,. \label{eq:MNmaxScalar}
\end{equation}
These bounds are shown in  figs.~\ref{fig:GlobalRegion} and \ref{fig:GaugeRegion} as vertical dashed lines. For values of $v_\Phi \lesssim 10^2$ keV, the bound coming from perturbativity dominates while for $v_\Phi \gtrsim 10^2$ keV the scalar mixing sets an stronger bound.
We see that in the global case the heavy neutrinos are relatively light, with masses below the TeV scale in a large fraction of the parameter space. However, the mixing between active and heavy neutrinos given by eq.~\eqref{eq:Mixings} is too small to be tested in collider searches, see e.g.~\cite{Abada:2022wvh}. In the gauge case we have upper bounds on $M_R$ in the range between $10^9\,{\rm GeV}$ and $10^{14}\,{\rm GeV}$. In general the heavy neutrinos are not going to interfere with the mechanism altering the neutrino number density before recombination, but nevertheless we can wonder about other cosmological implications they can lead to. 

Right handed neutrinos that give masses to the active neutrinos are expected to thermalise in the early Universe with the SM plasma via inverse decays of the form $H \, L \to N$, see e.g.~\cite{Buchmuller:2004nz}. That means that if the reheating temperature of the Universe was $T_{\rm RH} > M_R$ then we expect these sterile neutrinos to have cosmological implications. Perhaps the most widely studied consequence of them is baryogenesis via leptogenesis, see e.g.~\cite{Davidson:2008bu}. In addition, in our scenario these states interact with a large number of very light species, which can lead to a primordial $\Delta N_{\rm eff}$ that is effective both at the time of BBN and recombination and would therefore exclude the model. The key parameter controlling the contribution to $\Delta N_{\rm eff}$ and also the possible effect on the generation of a lepton asymmetry is the branching ratio into the dark sector:
\begin{align}\label{eq:BrN}
        {\rm BR}(N\to \phi\,\chi) = \dfrac{2\, N_\chi\, \theta_{\nu\chi}^2\, v_{\rm EW}^2 }{2\, N_\chi\, \theta_{\nu\chi}^2\, v_{\rm EW}^2 +  v_\Phi^2}\,.
\end{align}

The interaction rate of heavy sterile neutrinos with the SM plasma is directly determined by the Yukawa coupling $Y_\nu$ which in the type-I seesaw mechanism is intimately related to the neutrino mass, $m_\nu \simeq Y_\nu^2 v_{\rm EW}^2/2 M_R$. We thus expect the sterile neutrinos to reach thermal equilibrium in the early Universe for $K_{\rm SM} \gtrsim 1$, where
\begin{equation}\label{eq:K_SM}
K_{\rm SM} \equiv \frac{\Gamma(N\to H L )}{H(T\simeq M_R)}  \simeq 200\, \left(\frac{m_\nu}{0.2\,{\rm eV}}\right) \,.    
\end{equation}
Clearly, if these states have efficient interactions with $\chi_L$ and $\phi$ then they will generate a large number density of massless states that will contribute to $\Delta N_{\rm eff}$. In addition, these new interactions, if efficient, can make the distribution function of heavy sterile neutrinos very close to thermal and therefore suppress the generation of a primordial lepton asymmetry.

$\Delta N_{\rm eff}$ can be written as
\begin{align}
    \Delta N_{\rm eff} &\equiv \frac{8}{7}\left(\frac{11}{4}\right)^{4/3}\left[\frac{\rho_{\rm DS}}{\rho_\gamma}\right] \, ,
\end{align}
where $\rho_{\rm DS}$ refers to the energy density in the dark sector, i.e. $\chi_L$ and $\phi$. This contribution is bounded from above because the maximum temperature the $\phi-\chi$ population could have obtained is the SM one, in which case we can use entropy conservation to find:
\begin{align}
     \Delta N_{\rm eff} \leq \frac{8}{7}\left(\frac{11}{4}\right)^{4/3} \left[\frac{g_\chi\frac{7}{8} N_\chi + g_{\phi}}{2}\right] \left[\frac{g_{\star S}^{\rm SM}(T_{\rm today})}{g_{\star S}^{\rm SM}(T \simeq M_R)}\right]^{4/3}\nonumber\,.
\end{align} 
Using $g_{\star S}^{\rm SM}(T_{\rm today}) \simeq 3.9$ and $g_{\star S}^{\rm SM}(T \simeq M_R) \simeq 105$, this yields a maximum contribution to $\Delta N_{\rm eff}$ of 
\begin{align}\label{eq:maxDNeff}
    \Delta N_{\rm eff}|_{\rm max} \simeq 0.027 \left[1 + \frac{7}{8} g_\chi N_\chi\right] \,.
\end{align}
For instance, for typical values we have $\Delta N_{\rm eff}|_{\rm max}(N_\chi = 10, g_\chi =2)\simeq 0.5$, which is already excluded. 

By solving explicitly the Boltzmann equations for the number density of heavy sterile neutrinos in the early Universe allowing them to decay into these new very light species we find that in order to satisfy the CMB and BBN bound $\Delta N_{\rm eff} \lesssim 0.3$ we require:
\begin{align}
   K_{\rm SM} \times {\rm BR}(N\to \phi \,\chi) \lesssim 1\,,
\end{align}
where this expression is rather insensitive to $N_\chi$ provided that $N_\chi \gtrsim 5$.\footnote{This simply means that the rate of interactions of sterile neutrinos with the dark sector should be slower than the expansion rate at $T\simeq M_R$:
$\left<{\Gamma(N\to \phi \,\chi)}\right> \lesssim H(T\simeq M_R)$.} From eq.~\eqref{eq:K_SM} we obtain $10 < K_{\rm SM} < 800$, where the limits come from taking $\sqrt{|\Delta m_{\rm sol}^2|}<  m_\nu <0.8\,{\rm eV}$. Putting this together, we see that the branching ratio of sterile neutrinos is bounded to be 
\begin{align}\label{eq:BRLepto}
 {\rm BR}(N\to \phi \,\chi)\lesssim (10^{-3}-0.1)\,.
\end{align}

Using eq.~\eqref{eq:BrN}, we find that for the global version of the model the full viable parameter space is incompatible with this condition. Therefore in the global case we need to require $T_{\rm RH} < M_R$, such that the heavy neutrinos cannot thermalise. Since in the global case the right handed neutrinos could have masses $M_R < 100\,{\rm GeV}$ this means that one would need to invoke a mechanism that explains the baryon asymmetry of the Universe without the help of sphalerons processes. This argument again makes the global version of the model less attractive compared to the gauged case, in addition to the discussion above.

For the gauge case there is some parameter space fulfilling eq.~\eqref{eq:BRLepto} which is displayed as a vertical band in purple labelled ``Leptogenesis'' in fig.~\ref{fig:GaugeRegion}. For these values of $v_\Phi$ the contribution to $\Delta N_{\rm eff}$ remains small and we can be confident that the usual thermal leptogenesis  mechanism will not be distorted because the interactions with the new dark sector will always be slower than the expansion rate of the Universe and therefore will not impact the potential generation of a primordial lepton asymmetry. Of course, for the standard thermal leptogenesis mechanism to be successful the sterile neutrino mass cannot be arbitrarily low and should be $M_R\gtrsim 10^{8}\,{\rm GeV}$~\cite{Davidson:2002qv}. For the rest of the parameter space (to the left of the purple band), in order to be in agreement with BBN and CMB bounds on $\Delta N_{\rm eff}$, we need to require $T_{\rm RH} < M_R$ and the usual thermal leptogenesis will not be operative. However, this parameter space is compatible with having $T_{\rm RH}$ large enough to allow for sphaleron processes.

\subsection{Active neutrino decay} 

The existence of interacting fermions lighter than active neutrinos allow neutrinos to decay. Furthermore, if active neutrinos were to decay on a cosmological timescale this will diminishing their density today, see e.g.~\cite{Escudero:2020ped}. Two body decays $\nu \to \chi \,\, \phi/\rho/Z'$ are forbidden  in our scenario because the $\phi/\rho/Z'$ bosons are heavier than active neutrinos. As such, the only possible decay channels at tree-level are $\nu \to 3 \chi$ and $\nu_{i} \to  \nu_{j}  \chi \chi$. The first cannot happen in the global case because $\lambda^{\chi\chi}_{\phi \rho} = 0$, see eq.~\eqref{eq:globalcouplings}. The rate for the second is very small specially for the region of degenerate active neutrino masses we are interested in. 

However, the  $\nu \to 3 \chi$  decay is possible for the gauge mediated case through the gauge interactions $\lambda_{Z'}^{\nu\chi} \overline{\nu} \gamma^\mu \chi$ and $g_X\,  Z^\prime_\mu\,  \overline{\chi}\, \gamma^\mu\,  \,\chi$. Assuming the mixing between active and sterile neutrinos is similar for all the $N_\chi$ sterile species, we get $2\, N_\chi^2$ possible decay channels and therefore the total active neutrino decay rate will be $\Gamma_{\rm tot}^{\nu} = 2\, N_\chi^2\, \Gamma_{\nu \to 3 \chi}$. Taking the decay rate calculated in \cite{Escudero:2020ped} we can compute  the lifetime of the active neutrinos with the parameters of our model in terms of the age of the universe, $t_U\simeq 13.8$ Gyr, and the number of new sterile species, which reads:
\begin{align}
    \tau_{\nu} = 4\times 10^3\, t_U \, \frac{\left(\dfrac{v_\Phi}{1\, \rm MeV} \right)^4}{\left(\dfrac{N_\chi}{6} \right)^2\, \left(\dfrac{m_\nu}{0.2 \, \rm eV} \right)^5 \left(\dfrac{|\theta_{\nu\chi}|}{10^{-3}} \right)^2 } \,.
\end{align}
Given this equation it can be seen from fig.~\ref{fig:GaugeRegion} that neutrinos do not decay on cosmological timescales in the regions of parameter space in which the mechanism is viable.

\section{Conclusions}\label{sec:conclusions}

We have discussed a UV complete model to realise a mechanism to relax cosmological neutrino mass bounds to the level that the neutrino mass becomes observable in terrestrial experiments such as KATRIN and/or neutrinoless double-beta decay. The mechanism is based on an idea put forward by Farzan and Hannestad in Ref.~\cite{Farzan:2015pca}. The main ingredient is the depletion of the cosmological abundance of active neutrinos by populating instead a sufficient number of massless species, see eq.~\eqref{eq:NeutrinoMassBound_Mech} and figs.~\ref{fig:NMassless} and~\ref{fig:mechanism}.

Our model realisation is based on a version of the seesaw mechanism sometimes called ``minimal extended seesaw'' with the following main features: we have 3 heavy right-handed neutrinos, responsible for generating active neutrino masses via the type-I seesaw. In addition we introduce a large number of additional sterile neutrino species, which remain massless and provide the dark radiation. In order to achieve the required relaxation of the neutrino mass bound to $\simeq 1$~eV for the sum of neutrino masses, we need $\mathcal{O}(10)$ species of massless sterile neutrinos. To realise the conversion of active to sterile neutrinos we introduce a $U(1)_X$ symmetry, which can be either global or gauged and is spontaneously broken by a complex scalar field. The (pseudo-)scalar or vector bosons act as mediators between active neutrinos and the dark sector.

The depletion of active neutrinos should happen between BBN and recombination. This restricts the mass of the mediator particle to the range between 1~keV and 1~MeV. We provide a detailed discussion of the relevant phenomenology and cosmological constraints in sec.~\ref{sec:pheno} and find that the mechanism can work only in a closed region of parameter space. The version with the global symmetry is severely constrained by theoretical consistency arguments and perturbativity requirements. Furthermore, it is incompatible with thermal leptogenesis and requires a reheating temperature below the electro-weak scale. In contrast, the gauged version of the model is safely in the perturbative regime, is partially compatible with thermal leptogenesis and allows for reheating temperatures above the electro-weak scale. In the gauged version of the model, the VEV breaking the $U(1)_X$ symmetry can be between 10~MeV and a few GeV when the mixing $|\theta_{\nu\chi}|= 10^{-3}$ while for $|\theta_{\nu\chi}|= 10^{-4}$ only a narrow window of VEVs is allowed $20\,{\rm MeV} \lesssim v_\Phi \lesssim 100\,{\rm MeV}$.

In conclusion, we have presented a relatively simple extension of the Standard Model, explaining active neutrino masses close to the eV scale, featuring a non-standard neutrino cosmology. The main signature of the model is the possibility of observing a non-zero neutrino mass in KATRIN and/or a signal in neutrinoless double-beta decay, which would be excluded within the standard $\Lambda$CDM cosmology. The model requires the presence of $\mathcal{O}(10)$ species of massless sterile neutrinos, whose theoretical motivation remains to be identified, see however~\cite{Arkani-Hamed:2016rle,Buchmuller:2007zd,Ellis:2007wz} for extensions of the Standard Model with large number of sterile neutrinos or BSM species. 

Looking forward, in this scenario the number density of active neutrinos forming the cosmic neutrino background (CNB) is smaller than in $\Lambda$CDM. However, these neutrinos have a smaller temperature and therefore can cluster much more efficiently in the solar system. This in turn enhances the direct detection prospects of the CNB within this cosmology~\cite{Alvey:2021xmq}. In addition, while in this work we have focused on relaxing the current cosmological neutrino mass bound, the essence of the model remains relatively unchanged and will be relevant for the interpretation of upcoming cosmological neutrino mass measurements from DESI/Euclid, see Fig~\ref{fig:NMassless}. 

\bigskip

\textbf{Acknowledgements.}
This project has received support from the European Union's Horizon 2020 research and innovation programme under the Marie Sklodowska-Curie grant agreement No 860881-HIDDeN. The work of JTC is supported by the Ministerio de Ciencia e Innovaci\'on under FPI contract PRE2019-089992 of the SEV-2015-0548 grant. JTC is very grateful for the hospitality of the Theoretical Astroparticle Physics group of KIT during the months spent there when most of this project was carried out and to CERN-TH where this project was finalised. JTC would also like to thank Andrés Castillo and Jorge Martin Camalich for useful discussion. We are very grateful to Toshi Ota, Enrique Fernandez Martinez and Pilar Coloma for pointing out the relevance of $\chi$ self-interactions in the gauge $U(1)$ case. We are also very grateful to Xiaoyong Chu for pointing out the potential relevance of exponential production of $\chi$ particles prior to BBN.
ME and TS thank the Instituto de Fisica Teorica (IFT UAM-CSIC) in Madrid for support via the Centro de Excelencia Severo Ochoa Program under Grant CEX2020-001007-S, during the Extended Workshop “Neutrino Theories”, where part of this work was developed.
\appendix

\section{The Mechanism after recombination}\label{sec:appendixPostrec}

In the main text we have considered the possibility of decreasing the number density of neutrinos before recombination in order to relax the cosmological neutrino mass bound. However, a priori it is not really needed that the depletion of the neutrino number density happens before recombination, it can also happen after. In this appendix we comment on this possibility.

First, let us note that the mean momentum of a neutrino at the time of recombination is $\left<p_\nu\right> \simeq 3 T_\nu \simeq 0.6\,{\rm eV}$. As such, the primary CMB spectra can only be affected by  neutrinos that are $m_\nu\gtrsim 0.6\,{\rm eV}$. Indeed, the WMAP bound on the neutrino mass within $\Lambda$CDM was $\sum m_\nu < 1.3\,{\rm eV}$ at 95\% CL~\cite{WMAP:2012nax}, very close to the estimate given above. Nevertheless, the CMB is indirectly sensitive to less massive neutrinos via their effect on the growth of the large scale structures in the Universe. This sensitivity steams from the fact that such large scale structures can lens the CMB photons emitted at the surface of last scattering and detected today, see e.g.~\cite{Lattanzi:2017ubx}. The energy density in non-relativistic neutrinos suppresses this growth and CMB observations are therefore sensitive to significantly lighter neutrinos, e.g. $\sum m_\nu < 0.24\,{\rm eV}$ at 95\% CL is the bound from Planck CMB observations alone~\cite{planck}. Importantly, the main contribution to the lensing of CMB photons happens only at moderately low redshifts, $z\lesssim 6$, and this means that provided that $m_\nu< 0.5\,{\rm eV}$ Planck CMB observations are really sensitive to $\Omega_\nu h^2$ at $z\lesssim 6$. Similar considerations apply for data from Baryon Acoustic Oscillations. Put together, this means that in principle the mechanism described in the main text could work equally well provided that the neutrino number density is reduced at $z\gtrsim 10$. 

To concretely see how this could work we first note that neutrinos become non-relativistic at $z_{\rm nr}\simeq 200 \,m_\nu/0.1\,{\rm eV}$. In the scenario we considered in the main text we need a particle $X$ that can decay to neutrinos and therefore should have a mass $m_X > 2\,m_\nu$. That means that the mass of this particle would need to be in a rather narrow range $0.1\,{\rm eV}\lesssim m_X \lesssim 1\,{\rm eV}$. In addition, this particle should interact efficiently with neutrinos at $z \lesssim 10^3$ but not at $z \gtrsim 10^3$. This is simply because otherwise this new state will suppress neutrino freestreaming and distort the CMB~\cite{Taule:2022jrz}. Thus, we can appreciate the three requirements are that the interactions are not effective until $z\lesssim 10^{3}$, that the mediator is in the right mass window $0.1\,{\rm eV}\lesssim m_X \lesssim 1\,{\rm eV}$, and that the interactions occur rapidly as compared with the timescale for expansion. We believe that these three conditions can be met although we think that this will only occur in a very limited region of parameter space for $m_\nu$ and $m_X$ and for the interactions between $X$ and neutrinos. To confidently study the viable parameter space in this case a solution of the relevant Boltzmann equations would be needed which is beyond the scope of this study.

Finally, one can consider the possibility of depleting the neutrino number density by neutrino annihilations into massless states, $\nu\bar{\nu}\to \chi \bar{\chi}$. This in fact was considered a long time ago~\cite{Beacom:2004yd}, but was shown indeed not to be possible because of the suppression of neutrino freestreaming from these interactions~\cite{Hannestad:2004qu}.

\section{Entropy production and the value of~$N_{\rm eff}$}\label{sec:appendixNeff}

The thermodynamic processes discussed in the main text that depletes the number density of neutrinos in the early Universe leads to entropy production. This in particular means that we do not expect $N_{\rm eff}$ to be exactly 3.044. However, for large $N_\chi$ we expect the difference not to be significant. The reason that $N_{\rm eff}\neq 3.044$ is because the production and decay of a massive particle, $X$, always leads to some extra energy density production stored in its decay products. Of course, the more $\chi$ states there are the less significant will be this release of energy because the fractional number density of $X$ states will be small. Since we are working in a regime where the relevant processes are very efficient with respect to the characteristic expansion timescale $t_U \sim 1/H$, we can use equilibrium thermodynamics to elucidate this matter.

We should split the discussion in two events that we consider happen instantaneously. First, we consider that at $T\gg m_X$ the $X$ bosons and the $\chi$ fermions thermalise instantaneously generating a coupled system with temperature $T_{\rm eq}$ and chemical potential $\mu_i^{\rm eq}$. Since all the particles are massless energy conservation implies:
\begin{align}\label{eq:energycons}
\!\!\!\!\!\rho_\nu(T_\nu,0) = \rho_\nu(T_{\rm eq},\mu_\nu^{\rm eq}) + \rho_\chi(T_{\rm eq},\mu_\chi^{\rm eq}) + \rho_X(T_{\rm eq},\mu_X^{\rm eq})\,.
\end{align}
In addition, since we are considering processes $X\leftrightarrow \nu {\nu}$ and $X\leftrightarrow \chi {\chi}$ we know the chemical potentials of these species are related to be $2\mu_\nu^{\rm eq} = 2\mu_\chi^{\rm eq} = \mu_X^{\rm eq}$\footnote{Note that in the global case considered in the main text the relevant processes are $X\leftrightarrow \nu {\nu} $ and $X\leftrightarrow \nu {\chi}$. Although they are different they also lead to the condition $2\mu_\nu^{\rm eq} = 2\mu_\chi^{\rm eq} = \mu_X^{\rm eq}$. Similarly, while the relevant processes in the gauge case are $X\leftrightarrow \chi {\chi} $, $X\leftrightarrow \nu {\chi}$, and $X\leftrightarrow \chi {\chi}$, we end up having the same result.}.

In addition to energy conservation, we know that the processes we consider respect a conservation law for the number density of the various species:
\begin{align}\label{eq:numbercons1}
   n_\nu(T_\nu,0) = n_\nu(T_{\rm eq},\mu_\nu^{\rm eq}) +  n_\chi(T_{\rm eq},\mu_\chi^{\rm eq}) + 2 n_X(T_{\rm eq},\mu_X^{\rm eq})\,,
\end{align}
where here the 2 comes from the fact that a pair of neutrinos or $\chi$ particles can be interchanged by an $X$ state. Solving these two equations simultaneously provides $T_{\rm eq}$ and $\mu_\nu^{\rm eq}$ from a given $T_\nu$.

After thermalisation happens the Universe expands and eventually the $X$ boson decays away at $T\ll m_X$. Since we are in thermal equilibrium, in this system entropy density is conserved and number density is also conserved as before. This explicitly means:
\begin{align}
\!\!\!\! a^3(s_\nu(T,\mu)  + s_\chi(T,\mu)  + s_X(T,2\mu)) = \text{constant}\label{eq:entropycons2}\,,\\
\!\!\!\! a^3(n_\nu(T,\mu)  + n_\chi(T,\mu) +  2n_X(T,2\mu)) = \text{constant}\label{eq:numbercons2}\,.
\end{align}
Thus, solving eqs.~\eqref{eq:entropycons2} and~\eqref{eq:numbercons2} simultaneously using as initial condition $T_{\rm eq}$ and $\mu_{\rm eq}$ allows one to find $T_f$ and $\mu_f$. Namely, the final temperature and chemical potential of both neutrinos and massless $\chi$ species as relevant for CMB observations once the $X$ particle has disappeared from the plasma.

It is important to highlight that in order to obtain the decrease in the neutrino number density after this process one does not actually need to solve these equations explicitly. The number density conservation in eq.~\eqref{eq:numbercons1} and in eq.~\eqref{eq:numbercons2} imply
\begin{align}
   n_\nu(T_\nu,0) = n_\nu(T_f,\mu_{f}) +  n_\chi(T_{f},\mu_{f}) \,.
\end{align}
Since the $\chi$ particles are fermions we can easily see that\footnote{While in the main text we consider ultralight fermions the same consideration can be done for bosons. In such a case a compact formula cannot be obtained. A numerical fit to the result is: $n_\nu(T_f,\mu_{f}) = n_\nu^{\rm SM} /\left({1+g_\chi N_\chi/6 + 0.07}\right)$ which is accurate to better than 0.5\% for any $N_\chi$.}:
\begin{align}
   n_\nu(T_f,\mu_{f}) = n_\nu^{\rm SM} \frac{1}{1+g_\chi N_\chi/6}\,,
\end{align}
as highlighted by eq.~\eqref{eq:NeutrinoMassBound_Mech}. 

\begin{figure}[t]
\begin{center}
\includegraphics[width=0.46\textwidth]{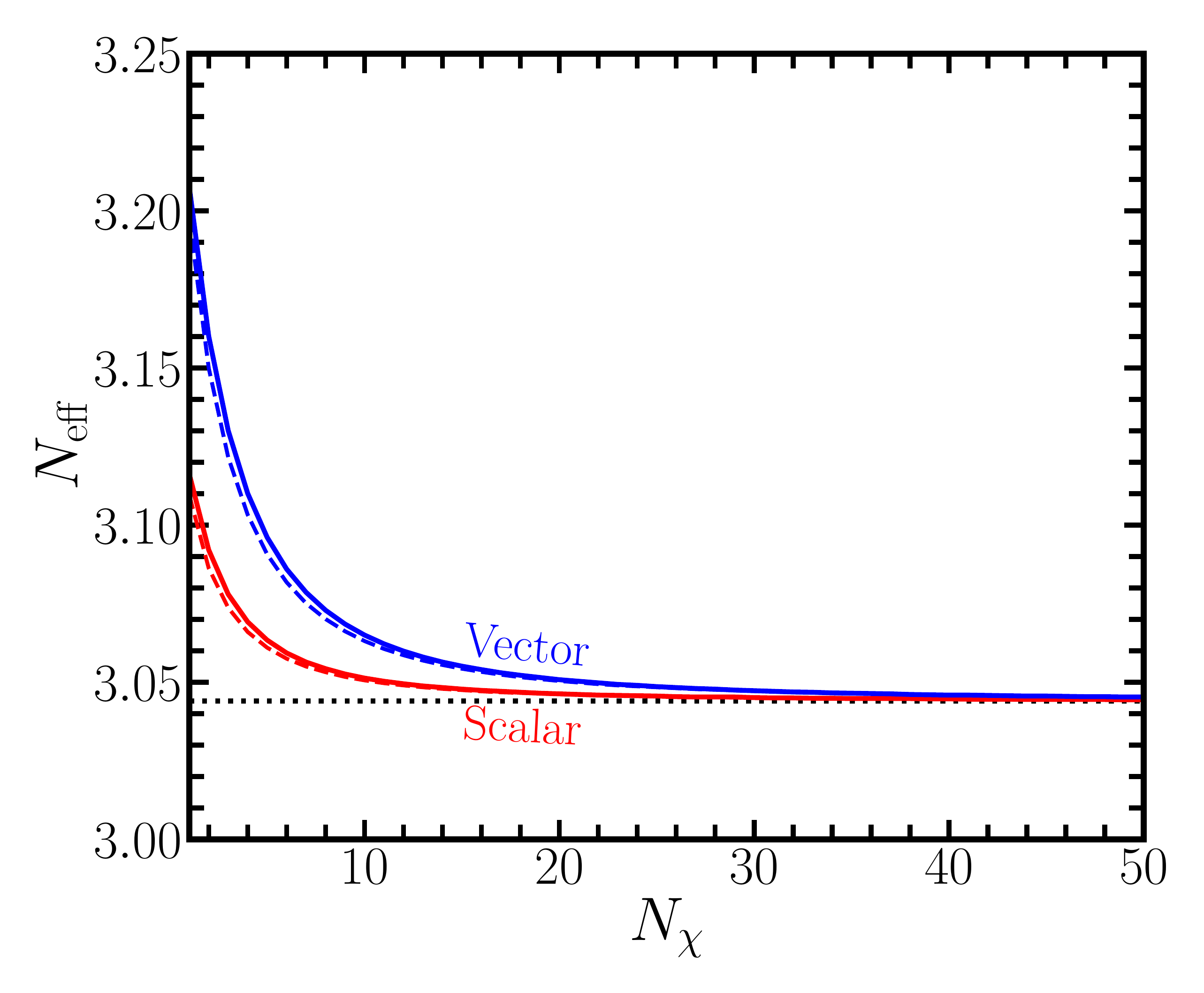}
\vspace{-0.2cm}
\caption{$N_{\rm eff}$ as relevant for CMB observations as a function of the number of massless species that eventually thermalise with neutrinos $N_\chi$. In blue when the mediator is a vector boson with $g_X = 3$ and in red  when we have a scalar mediator with $g_X = 1$. Dashed lines correspond to the case of $\chi$ being bosons and solid lines to the case of $\chi$ being bosons. The black dotted line shows the SM prediction of $N_{\rm eff} =3.044$. }
\label{fig:Neff}
\end{center}
\end{figure}

Obtaining $N_{\rm eff}$, however, does require solving explicitly for these four equations simultaneously. In fig.~\ref{fig:Neff} we show its value as a function of $N_\chi$. We can appreciate that unless $N_\chi$ is small, $N_\chi\lesssim 3$ the contribution to $N_{\rm eff}$ at the time of recombination should be undetectable even with an ultrasensitive Stage-IV CMB experiment.

Finally, we note that actually the processes we consider lead to a significant production of entropy because a large number of new relativistic species are produced. Importantly this does not change the number density of photons and therefore have not relevant impact on what the number density of baryons or dark matter particles should be. For completeness, however we report an approximate fitting function of $g_{\star S}$ which reads:
\begin{align}
    g_{\star S} \simeq  3.5 + \frac{6}{10}  \log(N_\chi^{7/10})\,,
\end{align}
and that we see depends only logarithmically upon $N_\chi$.

\section{Decay Rates}\label{sec:appformuale}

Here we outline relevant formulas used to compute the various of the constraints in the main text. The thermally averaged rate of the process $a + b \leftrightarrow X$ with $X$ being either a scalar, pseudoscalar or a vector boson and $a,\, b$ being two light fermions is given by \cite{EscuderoAbenza:2020cmq} :
\begin{equation}\label{eq:DecRateAverage}
    \langle\Gamma\rangle = \Gamma_X\, \dfrac{g_X}{4} \left( \dfrac{m_X}{T_a} \right)^2\, K_1\left( \dfrac{m_X}{T_a} \right),
\end{equation}
where $\Gamma_X$ is the decay rate of $X$ in vacuum, $g_X$ is the number of degrees of freedom of $X$ (1 for the scalar/pseudoscalar and 3 for the $Z^\prime$), $m_X$ is the mass of $X$, $T_a$ is the temperature of the fermions (in case of neutrinos, after decoupling we will approximate $T_\gamma/T_\nu\simeq 1.4$) and $K_1$ is a modified Bessel function of the first kind. The decay rate in vacuum for the scalar and pseudoscalar is
\begin{equation}\label{eq:DecRate}
    \Gamma_X =  \dfrac{\lambda^2}{16 \pi} m_X,
\end{equation}
and for the vector boson
\begin{equation}
    \Gamma_X =  \dfrac{\lambda^2}{8 \pi} m_X,
\end{equation}
with $\lambda$ the corresponding coupling, see eqs.~\eqref{eq:globalcouplings} and~\eqref{eq:gaugecouplings}. For the cross sections of 2 to 2 processes with interaction coupling $g$ and mediator $X$ we have used 
\begin{equation}
    \sigma \sim \dfrac{g^4}{16\pi^2}\dfrac{1}{m_X^4}T^2,
\end{equation}
when they are mediated by a massive ($m_X\gtrsim T$) mediator and 
\begin{equation}
    \sigma \sim \dfrac{g^4}{16\pi^2}\dfrac{1}{T^2},
\end{equation}
when the mediator is light ($m_X\ll T$).

\onecolumngrid

\bigskip
\bigskip

\noindent \textbf{Note added:} \textit{Appendices \ref{sec:appendixAdenddum_1} and \ref{sec:addendum2} correspond to the Addendum published as \cite{Addendum}.} 

\section{Adenddum: Impact of self-interactions on BBN and CMB constraints on the mixing between active and sterile neutrinos}\label{sec:ThetaBBN}


As discussed in Section~\ref{sec:pheno}, the mixing between $\chi$ and $\nu$ states is bounded by $N_{\rm eff}$ constraints resulting from the oscillations between them prior to neutrino decoupling in the early Universe at $T\sim \,{\rm MeV}$. In the main text we reported constraints on the mixing $\theta_{\nu \chi}$ assuming that $\chi$ particles were not interacting with the plasma in any appreciable way and did not feel any potential energy in the thermal bath at $T> 2\,{\rm MeV}$. Despite the small couplings and number densities this is actually not strictly true and very tiny interactions can still modify and relax the constraints on $\theta_{\nu \chi}$ for the gauge case. The reason is that in the gauge version of the mechanism, the $\chi$ states feel a potential energy arising from the forward scattering between them in resemblance to the MSW effect in the sun. In particular, the potential for $\chi$ species can be written at the 1-loop level as~\cite{Weldon:1982bn,Dasgupta:2013zpn,Chu:2015ipa}\footnote{Note that compared to~\cite{Dasgupta:2013zpn,Chu:2015ipa}, we choose to parametrize the dark sector distribution functions as $f_{\chi/Z'} = \xi_{\chi/Z'} [e^{p/T_\nu} \pm 1]^{-1} $, meaning the dark sector particles have $\left<E\right> \simeq 3T_\nu$, but their number density is allowed to be smaller than the equilibrium value, with $\xi_{\chi/Z'} = n_{\chi/Z'}/n_{\chi/Z'}^{\rm eq}$. On the other hand, Refs.~\cite{Dasgupta:2013zpn,Chu:2015ipa} parametrize the dark sector with just a dark sector temperature, $T_s$. Although the functional form of the potentials only match with $\xi_{\chi/Z'} = 1$ and $T_s = T_\nu$, in the relevant regions of parameter space the results for the energy density of $\chi$ states one finds at the end are very similar.}: 
\begin{align}\label{eq:Potential_chichi}
    V_{\chi \chi} = \frac{g_X^2 T_\nu^2}{8 E} \left[\frac{2}{3}\frac{n_{Z'}}{n_{Z'}^{\rm eq}}+ \frac{1}{3}\frac{n_\chi}{n_\chi^{\rm eq}}\right] \,,
\end{align}
where $E$ is the energy of a given $\chi$ state, $T_\nu$ is the neutrino temperature, $g_X$ is the $U(1)_X$ gauge coupling, $n$ are number densities, and $n^{\rm eq}$ refers to the equilibrium density of a given species. The factors of $1/3$ and $2/3$ arise from the contribution from fermions and bosons in the loop respectively, see~\cite{Weldon:1982bn}. Note that in addition, the oscillations between $\nu-\chi$ are also subject to effects coming from the active neutrino thermal potential in the Standard Model ($V_{\nu\nu}$) which is relevant in the early Universe~\cite{Notzold:1987ik}.

There are two implications from these thermal potentials. Firstly, if $|V_{\nu \nu}-V_{\chi \chi}| \gg m_\nu^2/(2E_\nu) $ then the in-medium mixing angle will be strongly suppressed in the early Universe, which can happen already for rather small couplings, $g_X \gtrsim 10^{-7}$. Secondly, in our particular case of interest, the $\chi$ species is \emph{lighter} than active neutrinos which means that it is resonantly produced in the early Universe. Since the thermal potentials are in turn dependent upon the $\chi$ number density this could potentially augment the resonant production.

In order to take into account the impact of these potentials on the production rate of $\chi$ states in the early Universe we proceed as in~\cite{Escudero:2020ped} and write down evolution equations for the neutrino temperature and the number density of $\chi$ states following~\cite{Escudero:2018mvt,EscuderoAbenza:2020cmq} and we solve for them using a modified version of the online~\href{https://github.com/MiguelEA/nudec_BSM}{\texttt{NUDEC\_BSM}} code. We take into account the $\nu-\chi$ oscillations and collisions by considering the collision term from~\cite{Abazajian:2005gj} including the potential energy in Eq.~\eqref{eq:Potential_chichi} in addition to the Standard Model one for neutrinos~\cite{Notzold:1987ik}. For simplicity, in Eq.~\eqref{eq:Potential_chichi} we assume the same relative number density of $Z'$ and $\chi$ states, i.e. ${n_\chi}/{n_\chi^{\rm eq}} = {n_{Z'}}/{n_{Z'}^{\rm eq}}$. This is motivated because of the strong self-interactions between these states.

\begin{figure}
\begin{center}
\includegraphics[width=0.55\textwidth]{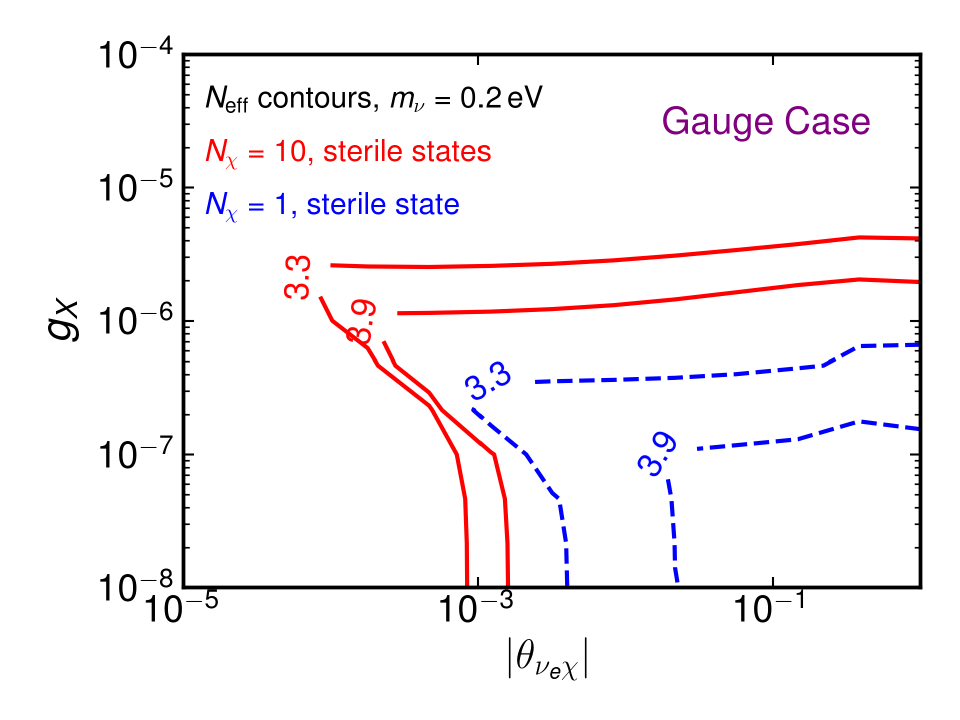} 
\caption{$N_{\rm eff}$ isocontours for $N_\chi =1$ and $N_{\chi} = 10$ taking into account self-interactions between the $\chi$ states in the early Universe as a function of the $\nu-\chi$ mixing and the $U(1)_X$ gauge coupling.}
\label{fig:Neffcontours}
\end{center}
\end{figure}

In Figure~\ref{fig:Neffcontours}, we show isocontours of $N_{\rm eff}$ as relevant for BBN and CMB observations for two scenarios with $N_\chi = 1$ and $N_\chi= 10$. For the case with $N_\chi = 10$ we assume that all massless sterile neutrinos have the same mixing angle.  For illustration we consider the case $m_\nu = 0.2\,{\rm eV}$ as the bounds do not change substantially for the parameter space of interest, $m_\nu < 0.8\,{\rm eV}$. 

Current BBN and CMB constraints require $N_{\rm eff} \lesssim 3.3$ and we therefore notice that the cosmological bound on $\theta_{\nu \chi}$ can be evaded for rather small values of $g_X$. In particular, for the most relevant case of $N_\chi = 10$ we find that the cosmological bound on the mixing angle can be evaded provided that $g_X > 10^{-6}$. We do notice, however, an interesting region for $10^{-6} \gtrsim g_X \gtrsim 10^{-7}$ where actually the bound on $\theta_{\nu \chi}$ would be stronger than the one expected for a vanishing $g_X$. This can be explained as follows: the thermal potential depends on $n_\chi/n_\chi^{\rm eq}$ and there is resonant $\chi$ production when $\Delta m^2/(2E) \simeq V_{\chi \chi}(E) $. In the expanding Universe since $E_\nu$ decreases $\Delta m^2/(2E)$ increases, but $V_{\chi \chi}$ can also increase via the production of $\chi$ particles and therefore the $n_\chi/n_\chi^{\rm eq}$ ratio. That can allow $\chi$ particles to be in resonance for longer and in consequence it enhances the production. We have explicitly checked that indeed for this region of parameter space across a significant fraction of the thermal evolution $V_{\chi \chi} \simeq \Delta m^2/(2E_\nu)$ is fulfilled. 

\begin{figure*}[t]
\centering
\includegraphics[width= 0.45\columnwidth]{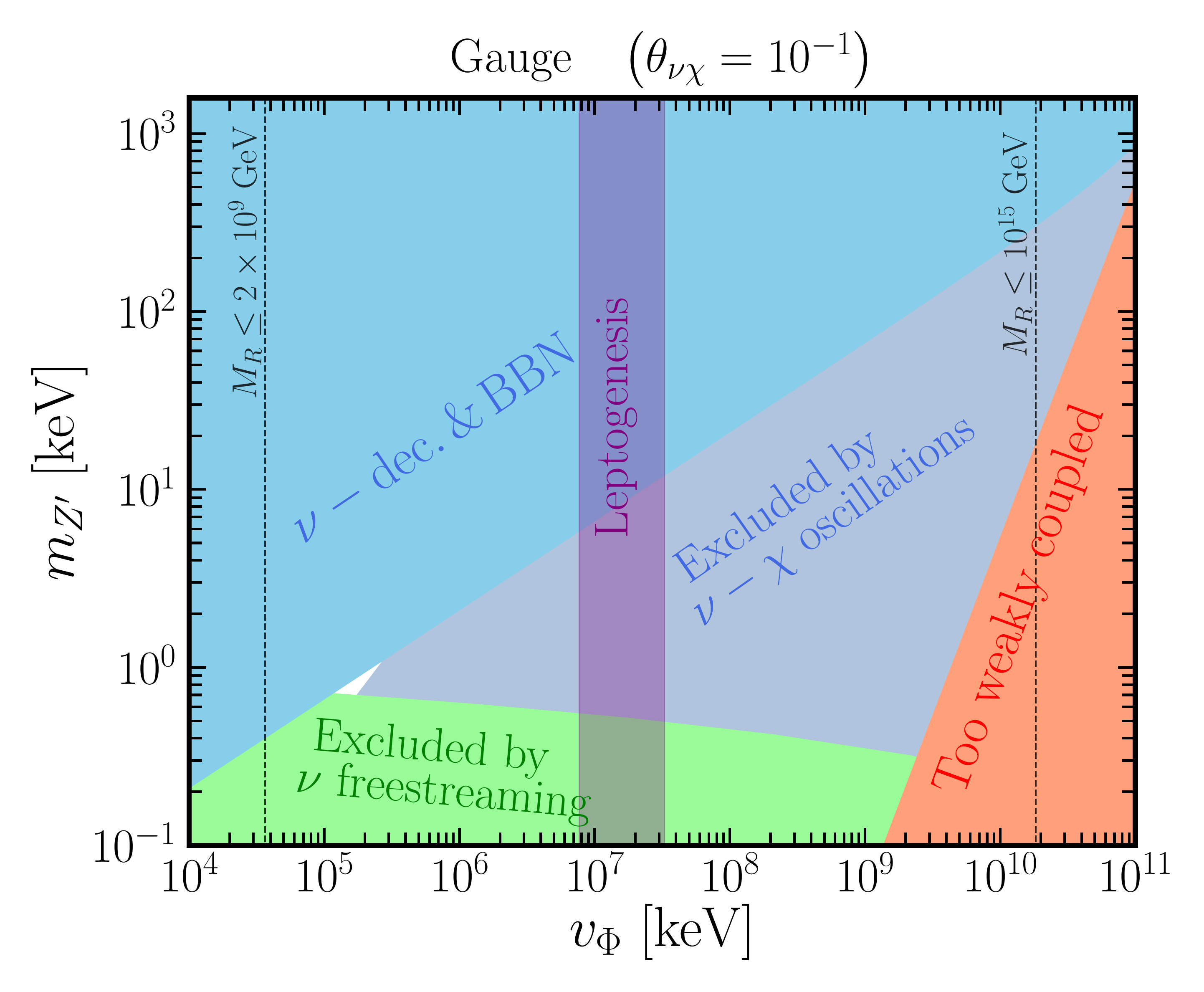}
\includegraphics[width=0.45\columnwidth]{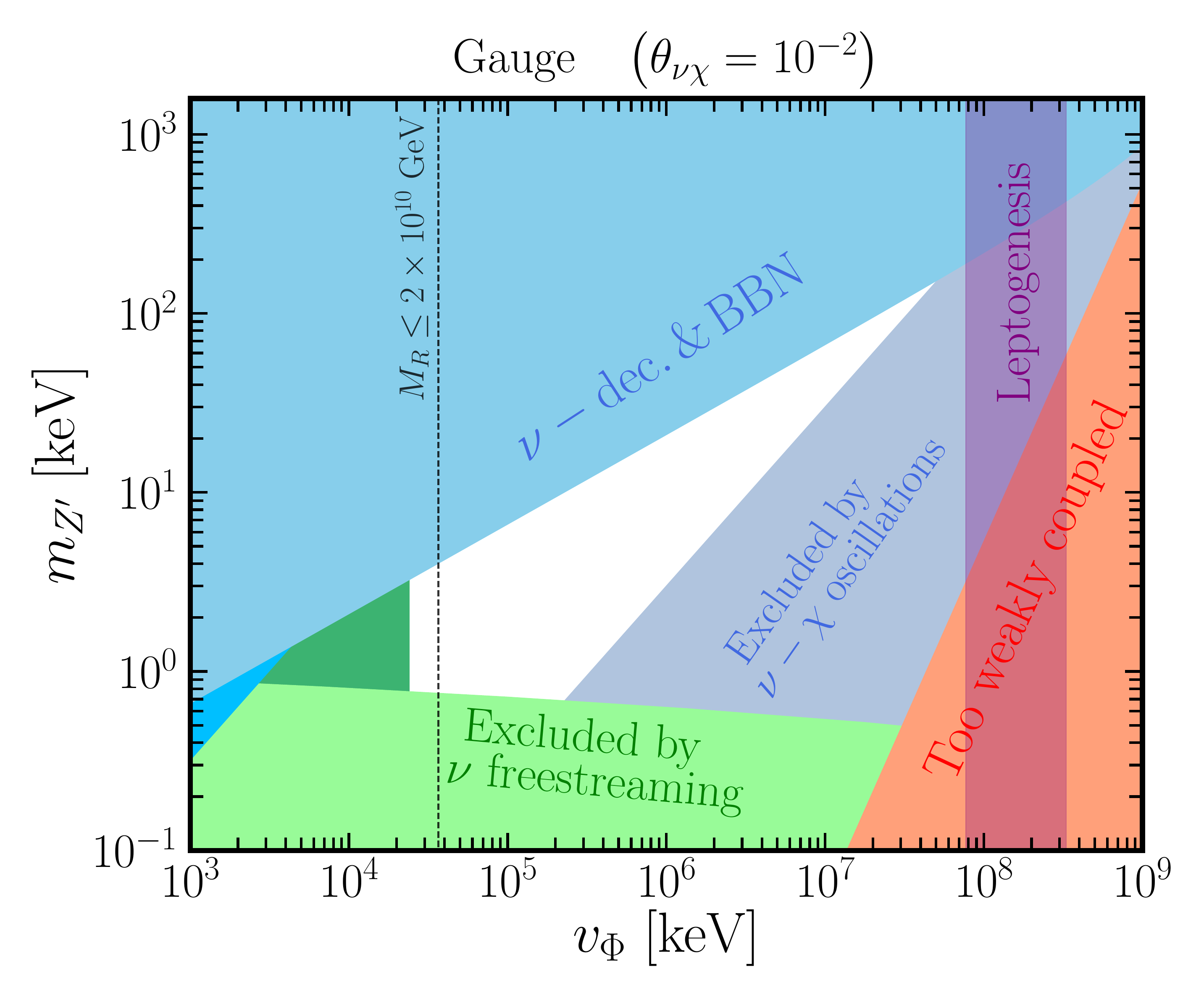}
\caption{As Figure~\ref{fig:GaugeRegion} but for larger $\nu-\chi$ mixing angles, $\theta_{\nu\chi} = 10^{-1}$ and $\theta_{\nu \chi} = 10^{-2}$. 
The grey region is excluded by the BBN \& CMB constraint $N_{\rm eff} < 3.3$ according to Figure~\ref{fig:Neffcontours} for the case of $N_\chi = 10$.}
\label{fig:ParameterSpace}
\end{figure*}

In Figure~\ref{fig:ParameterSpace}, we highlight the parameter space in the $m_X$-$v_\Phi$ plane that opens up as a result of taking into account the $\chi$ self-interactions for the gauge case. In grey we highlight regions of parameter space where $g_X$ is not large enough to evade the cosmological bound on $\theta_{\nu \chi}$. We clearly see that the parameter space for $\theta_{\nu \chi} = 0.1$ is very small while for $\theta_{\nu \chi} = 0.01$ substantial parameter space is allowed. We note that this could open up the possibility of detecting $\chi$ states in neutrino oscillation experiments as they can be sensitive to $\theta_{\nu \chi} \sim 0.1$, see e.g.~\cite{Boser:2019rta} for a review. For $\theta_{\nu \chi} \lesssim 10^{-3}$ no new constraint appears and the allowed parameter space shown in Fig.~\ref{fig:GaugeRegion} remains unchanged.

In the global $U(1)$ version of the model, there is a potential mediated by a $N-\phi$ loop. We will consider $M_N\gg 100\,{\rm MeV}$ and in this regime of parameter space direct calculation shows that\footnote{Technically speaking a potential mediated by the same interaction could contribute to the gauge version of the model too. However, it will be tiny because the masses of sterile neutrinos in that case are much larger and also it will become exponentially small for $T < m_\rho < M_N$ and therefore it is totally negligible.}
\begin{align}\label{eq:newpotentialscalar}
    V_{\chi \chi} = Y_\Phi^2 \frac{2\pi}{45} \frac{T_\nu^4}{M_N^4} \frac{n_\phi}{n_\phi^{\rm eq}}  E_\nu\,,
\end{align}
where here $Y_\Phi$ is the scalar coupling in Eq.~\eqref{eq:MES} of the main text and we work in the single flavor approximation. We clearly see that this potential is strongly suppressed by the sterile neutrino mass $M_N$, and that the phenomenology will be determined by $Y_\Phi$ and $M_N$ which are related to $v_\Phi$ and $\theta_{\nu \chi}$. It is easy to understand in which regime would this potential suppress the oscillations between $\nu_L$ and $\chi_L$ in the early Universe and this corresponds to $V_{\chi \chi} > \Gamma_{\rm osc}$ at $T\sim {\rm MeV}$. The reason is that since neutrinos decouple at $2-3\,{\rm MeV}$ even if the oscillation rate between them and $\chi$ states is large, no $\chi$ species will be generated as there are no more collisions in the plasma. By using the relations between $M_N$, $m_\nu$, and $v_\Phi$ of Section~\ref{sec:model}, we can turn this into a condition for $v_\phi$ which reads:
\begin{align}
        v_\phi < {\rm keV} \,\left[\frac{\theta_{\nu \chi}}{0.1} \right] \, \left[\frac{2}{Y_\Phi}\right]^{3/4} \,\quad [{\rm no} \, N_{\rm eff}\,{\rm bound}]\,.
\end{align}
That is, for $v_\phi$ smaller than this number, the $\chi$ production in the early Universe will be significantly damped and the $N_{\rm eff}$ bound will not apply. By directly solving the appropriate Boltzmann equations we have explicitly checked that this condition depends only very mildly on $m_\nu$ or $N_{\chi}$. The question to be addressed is whether this can open up relevant parameter space in the mechanism for mixings $\theta_{\nu \chi} > 10^{-3}$. The answer depends upon the value one chooses for $Y_{\Phi}$, but this is in turn related to the sterile neutrino mass, $Y_{\Phi} = \theta_{\nu \chi} \sqrt{m_\nu M_N}/v_\Phi$. In this version of the model, the sterile neutrinos decay efficiently in the early Universe into the dark sector, and therefore in order to comply with $N_{\rm eff}$ bounds $T_{\rm RH}> M_N$. Taking the most plausible scenario with $T_{\rm RH} > 200\,{\rm GeV}$ leads to the following constraint 
\begin{align}
        v_\phi > {\rm keV} \left[\frac{\theta_{\nu \chi}}{0.1} \right] \left[\frac{2}{Y_\Phi}\right]\,\quad [M_N > 200\,{\rm GeV}]\,. 
\end{align}
We therefore see that these two bounds are conflicting, or in other words, that in the regime of parameter space where the potential in Eq.~\eqref{eq:newpotentialscalar} could be phenomenologically relevant, the plasma would be actually filled with a thermal bath of $\phi$ and $\chi$ states already and this is excluded by both BBN and the CMB.

\section{Addendum: Impact of exponential $\chi$ production on BBN constraints on $N_{\rm eff}$} \label{sec:addendum2}


In Section \ref{sec:discussion} we discussed the constraint on neutrino interactions with the dark mediator as its decay would produce an initial abundance of $\chi$ before BBN. However in the published version it was not mentioned that, in the gauge case, the $\chi$ states can also be produced and reach equilibrium with neutrinos before BBN via the process $\nu\chi \to \chi\chi$, provided an initial primordial population of $\chi$ is present.\footnote{This effect is absent in the global case, as there is no direct coupling of the scalars to two $\chi$ particles, see eq.~\eqref{eq:globalcouplings}}
Though the initial abundance of $\chi$ is very very small, it experiences an exponential growth~\cite{Bringmann:2021tjr,Bringmann:2022aim}. For this not to happen before BBN, we require
\begin{equation}
    \left<\Gamma(\nu\chi \to \chi\chi)\right> \lesssim H (T =0.7\,{\rm MeV}) \,,
\end{equation}
setting an upper bound on the gauge coupling,
\begin{equation}
   g_X \lesssim 1.1 \times 10^{-3}\, \left(\dfrac{10^{-3}}{\theta_{\nu\chi}}\right)^{\frac{1}{2}}.
\end{equation}
This bound does not significantly affect the allowed parameter space for $\theta_{\nu\chi} \gtrsim 10^{-3}$, but removes part of the (already small) allowed region for $\theta_{\nu\chi} = 10^{-4}$ as shown in fig.~\ref{fig:GaugeRegion}.

\bibliography{biblio}

\begin{thebibliography}{88}%
\makeatletter
\providecommand \@ifxundefined [1]{%
 \@ifx{#1\undefined}
}%
\providecommand \@ifnum [1]{%
 \ifnum #1\expandafter \@firstoftwo
 \else \expandafter \@secondoftwo
 \fi
}%
\providecommand \@ifx [1]{%
 \ifx #1\expandafter \@firstoftwo
 \else \expandafter \@secondoftwo
 \fi
}%
\providecommand \natexlab [1]{#1}%
\providecommand \enquote  [1]{``#1''}%
\providecommand \bibnamefont  [1]{#1}%
\providecommand \bibfnamefont [1]{#1}%
\providecommand \citenamefont [1]{#1}%
\providecommand \href@noop [0]{\@secondoftwo}%
\providecommand \href [0]{\begingroup \@sanitize@url \@href}%
\providecommand \@href[1]{\@@startlink{#1}\@@href}%
\providecommand \@@href[1]{\endgroup#1\@@endlink}%
\providecommand \@sanitize@url [0]{\catcode `\\12\catcode `\$12\catcode
  `\&12\catcode `\#12\catcode `\^12\catcode `\_12\catcode `\%12\relax}%
\providecommand \@@startlink[1]{}%
\providecommand \@@endlink[0]{}%
\providecommand \url  [0]{\begingroup\@sanitize@url \@url }%
\providecommand \@url [1]{\endgroup\@href {#1}{\urlprefix }}%
\providecommand \urlprefix  [0]{URL }%
\providecommand \Eprint [0]{\href }%
\providecommand \doibase [0]{http://dx.doi.org/}%
\providecommand \selectlanguage [0]{\@gobble}%
\providecommand \bibinfo  [0]{\@secondoftwo}%
\providecommand \bibfield  [0]{\@secondoftwo}%
\providecommand \translation [1]{[#1]}%
\providecommand \BibitemOpen [0]{}%
\providecommand \bibitemStop [0]{}%
\providecommand \bibitemNoStop [0]{.\EOS\space}%
\providecommand \EOS [0]{\spacefactor3000\relax}%
\providecommand \BibitemShut  [1]{\csname bibitem#1\endcsname}%
\let\auto@bib@innerbib\@empty
\bibitem [{\citenamefont {Dolgov}(2002)}]{Dolgov:2002wy}%
  \BibitemOpen
  \bibfield  {author} {\bibinfo {author} {\bibfnamefont {A.~D.}\ \bibnamefont
  {Dolgov}},\ }\href {\doibase 10.1016/S0370-1573(02)00139-4} {\bibfield
  {journal} {\bibinfo  {journal} {Phys. Rept.}\ }\textbf {\bibinfo {volume}
  {370}},\ \bibinfo {pages} {333} (\bibinfo {year} {2002})},\ \Eprint
  {http://arxiv.org/abs/hep-ph/0202122} {arXiv:hep-ph/0202122} \BibitemShut
  {NoStop}%
\bibitem [{\citenamefont {Lesgourgues}\ and\ \citenamefont
  {Pastor}(2006)}]{Lesgourgues:2006nd}%
  \BibitemOpen
  \bibfield  {author} {\bibinfo {author} {\bibfnamefont {J.}~\bibnamefont
  {Lesgourgues}}\ and\ \bibinfo {author} {\bibfnamefont {S.}~\bibnamefont
  {Pastor}},\ }\href {\doibase 10.1016/j.physrep.2006.04.001} {\bibfield
  {journal} {\bibinfo  {journal} {Phys. Rept.}\ }\textbf {\bibinfo {volume}
  {429}},\ \bibinfo {pages} {307} (\bibinfo {year} {2006})},\ \Eprint
  {http://arxiv.org/abs/astro-ph/0603494} {arXiv:astro-ph/0603494} \BibitemShut
  {NoStop}%
\bibitem [{\citenamefont {Aghanim}\ \emph {et~al.}(2020)\citenamefont {Aghanim}
  \emph {et~al.}}]{planck}%
  \BibitemOpen
  \bibfield  {author} {\bibinfo {author} {\bibfnamefont {N.}~\bibnamefont
  {Aghanim}} \emph {et~al.} (\bibinfo {collaboration} {Planck}),\ }\href
  {\doibase 10.1051/0004-6361/201833910} {\bibfield  {journal} {\bibinfo
  {journal} {Astron. Astrophys.}\ }\textbf {\bibinfo {volume} {641}},\ \bibinfo
  {pages} {A6} (\bibinfo {year} {2020})},\ \Eprint
  {http://arxiv.org/abs/1807.06209} {arXiv:1807.06209 [astro-ph.CO]}
  \BibitemShut {NoStop}%
\bibitem [{\citenamefont {Di~Valentino}\ \emph {et~al.}(2021)\citenamefont
  {Di~Valentino}, \citenamefont {Gariazzo},\ and\ \citenamefont
  {Mena}}]{DiValentino:2021hoh}%
  \BibitemOpen
  \bibfield  {author} {\bibinfo {author} {\bibfnamefont {E.}~\bibnamefont
  {Di~Valentino}}, \bibinfo {author} {\bibfnamefont {S.}~\bibnamefont
  {Gariazzo}}, \ and\ \bibinfo {author} {\bibfnamefont {O.}~\bibnamefont
  {Mena}},\ }\href {\doibase 10.1103/PhysRevD.104.083504} {\bibfield  {journal}
  {\bibinfo  {journal} {Phys. Rev. D}\ }\textbf {\bibinfo {volume} {104}},\
  \bibinfo {pages} {083504} (\bibinfo {year} {2021})},\ \Eprint
  {http://arxiv.org/abs/2106.15267} {arXiv:2106.15267 [astro-ph.CO]}
  \BibitemShut {NoStop}%
\bibitem [{\citenamefont {Aghamousa}\ \emph {et~al.}(2016)\citenamefont
  {Aghamousa} \emph {et~al.}}]{DESI:2016fyo}%
  \BibitemOpen
  \bibfield  {author} {\bibinfo {author} {\bibfnamefont {A.}~\bibnamefont
  {Aghamousa}} \emph {et~al.} (\bibinfo {collaboration} {DESI}),\ }\href@noop
  {} {\  (\bibinfo {year} {2016})},\ \Eprint {http://arxiv.org/abs/1611.00036}
  {arXiv:1611.00036 [astro-ph.IM]} \BibitemShut {NoStop}%
\bibitem [{\citenamefont {Amendola}\ \emph {et~al.}(2018)\citenamefont
  {Amendola} \emph {et~al.}}]{Amendola:2016saw}%
  \BibitemOpen
  \bibfield  {author} {\bibinfo {author} {\bibfnamefont {L.}~\bibnamefont
  {Amendola}} \emph {et~al.},\ }\href {\doibase 10.1007/s41114-017-0010-3}
  {\bibfield  {journal} {\bibinfo  {journal} {Living Rev. Rel.}\ }\textbf
  {\bibinfo {volume} {21}},\ \bibinfo {pages} {2} (\bibinfo {year} {2018})},\
  \Eprint {http://arxiv.org/abs/1606.00180} {arXiv:1606.00180 [astro-ph.CO]}
  \BibitemShut {NoStop}%
\bibitem [{\citenamefont {Brinckmann}\ \emph {et~al.}(2019)\citenamefont
  {Brinckmann}, \citenamefont {Hooper}, \citenamefont {Archidiacono},
  \citenamefont {Lesgourgues},\ and\ \citenamefont
  {Sprenger}}]{Brinckmann:2018owf}%
  \BibitemOpen
  \bibfield  {author} {\bibinfo {author} {\bibfnamefont {T.}~\bibnamefont
  {Brinckmann}}, \bibinfo {author} {\bibfnamefont {D.~C.}\ \bibnamefont
  {Hooper}}, \bibinfo {author} {\bibfnamefont {M.}~\bibnamefont
  {Archidiacono}}, \bibinfo {author} {\bibfnamefont {J.}~\bibnamefont
  {Lesgourgues}}, \ and\ \bibinfo {author} {\bibfnamefont {T.}~\bibnamefont
  {Sprenger}},\ }\href {\doibase 10.1088/1475-7516/2019/01/059} {\bibfield
  {journal} {\bibinfo  {journal} {JCAP}\ }\textbf {\bibinfo {volume} {01}},\
  \bibinfo {pages} {059} (\bibinfo {year} {2019})},\ \Eprint
  {http://arxiv.org/abs/1808.05955} {arXiv:1808.05955 [astro-ph.CO]}
  \BibitemShut {NoStop}%
\bibitem [{\citenamefont {Chudaykin}\ and\ \citenamefont
  {Ivanov}(2019)}]{Chudaykin:2019ock}%
  \BibitemOpen
  \bibfield  {author} {\bibinfo {author} {\bibfnamefont {A.}~\bibnamefont
  {Chudaykin}}\ and\ \bibinfo {author} {\bibfnamefont {M.~M.}\ \bibnamefont
  {Ivanov}},\ }\href {\doibase 10.1088/1475-7516/2019/11/034} {\bibfield
  {journal} {\bibinfo  {journal} {JCAP}\ }\textbf {\bibinfo {volume} {11}},\
  \bibinfo {pages} {034} (\bibinfo {year} {2019})},\ \Eprint
  {http://arxiv.org/abs/1907.06666} {arXiv:1907.06666 [astro-ph.CO]}
  \BibitemShut {NoStop}%
\bibitem [{\citenamefont {Aker}\ \emph {et~al.}(2019)\citenamefont {Aker} \emph
  {et~al.}}]{KATRIN:2019yun}%
  \BibitemOpen
  \bibfield  {author} {\bibinfo {author} {\bibfnamefont {M.}~\bibnamefont
  {Aker}} \emph {et~al.} (\bibinfo {collaboration} {KATRIN}),\ }\href {\doibase
  10.1103/PhysRevLett.123.221802} {\bibfield  {journal} {\bibinfo  {journal}
  {Phys. Rev. Lett.}\ }\textbf {\bibinfo {volume} {123}},\ \bibinfo {pages}
  {221802} (\bibinfo {year} {2019})},\ \Eprint
  {http://arxiv.org/abs/1909.06048} {arXiv:1909.06048 [hep-ex]} \BibitemShut
  {NoStop}%
\bibitem [{\citenamefont {Aker}\ \emph {et~al.}(2022)\citenamefont {Aker} \emph
  {et~al.}}]{KATRIN:2021uub}%
  \BibitemOpen
  \bibfield  {author} {\bibinfo {author} {\bibfnamefont {M.}~\bibnamefont
  {Aker}} \emph {et~al.} (\bibinfo {collaboration} {KATRIN}),\ }\href {\doibase
  10.1038/s41567-021-01463-1} {\bibfield  {journal} {\bibinfo  {journal}
  {Nature Phys.}\ }\textbf {\bibinfo {volume} {18}},\ \bibinfo {pages} {160}
  (\bibinfo {year} {2022})},\ \Eprint {http://arxiv.org/abs/2105.08533}
  {arXiv:2105.08533 [hep-ex]} \BibitemShut {NoStop}%
\bibitem [{\citenamefont {Agostini}\ \emph {et~al.}(2022)\citenamefont
  {Agostini}, \citenamefont {Benato}, \citenamefont {Detwiler}, \citenamefont
  {Men\'endez},\ and\ \citenamefont {Vissani}}]{Agostini:2022zub}%
  \BibitemOpen
  \bibfield  {author} {\bibinfo {author} {\bibfnamefont {M.}~\bibnamefont
  {Agostini}}, \bibinfo {author} {\bibfnamefont {G.}~\bibnamefont {Benato}},
  \bibinfo {author} {\bibfnamefont {J.~A.}\ \bibnamefont {Detwiler}}, \bibinfo
  {author} {\bibfnamefont {J.}~\bibnamefont {Men\'endez}}, \ and\ \bibinfo
  {author} {\bibfnamefont {F.}~\bibnamefont {Vissani}},\ }\href@noop {} {\
  (\bibinfo {year} {2022})},\ \Eprint {http://arxiv.org/abs/2202.01787}
  {arXiv:2202.01787 [hep-ex]} \BibitemShut {NoStop}%
\bibitem [{\citenamefont {Gando}\ \emph {et~al.}(2016)\citenamefont {Gando}
  \emph {et~al.}}]{KamLAND-Zen:2016pfg}%
  \BibitemOpen
  \bibfield  {author} {\bibinfo {author} {\bibfnamefont {A.}~\bibnamefont
  {Gando}} \emph {et~al.} (\bibinfo {collaboration} {KamLAND-Zen}),\ }\href
  {\doibase 10.1103/PhysRevLett.117.109903, 10.1103/PhysRevLett.117.082503}
  {\bibfield  {journal} {\bibinfo  {journal} {Phys. Rev. Lett.}\ }\textbf
  {\bibinfo {volume} {117}},\ \bibinfo {pages} {082503} (\bibinfo {year}
  {2016})},\ \bibinfo {note} {[Addendum: Phys. Rev.
  Lett.117,no.10,109903(2016)]},\ \Eprint {http://arxiv.org/abs/1605.02889}
  {arXiv:1605.02889 [hep-ex]} \BibitemShut {NoStop}%
\bibitem [{\citenamefont {Agostini}\ \emph {et~al.}(2020)\citenamefont
  {Agostini} \emph {et~al.}}]{GERDA:2020xhi}%
  \BibitemOpen
  \bibfield  {author} {\bibinfo {author} {\bibfnamefont {M.}~\bibnamefont
  {Agostini}} \emph {et~al.} (\bibinfo {collaboration} {GERDA}),\ }\href
  {\doibase 10.1103/PhysRevLett.125.252502} {\bibfield  {journal} {\bibinfo
  {journal} {Phys. Rev. Lett.}\ }\textbf {\bibinfo {volume} {125}},\ \bibinfo
  {pages} {252502} (\bibinfo {year} {2020})},\ \Eprint
  {http://arxiv.org/abs/2009.06079} {arXiv:2009.06079 [nucl-ex]} \BibitemShut
  {NoStop}%
\bibitem [{\citenamefont {Adams}\ \emph {et~al.}(2021)\citenamefont {Adams}
  \emph {et~al.}}]{CUORE:2021gpk}%
  \BibitemOpen
  \bibfield  {author} {\bibinfo {author} {\bibfnamefont {D.~Q.}\ \bibnamefont
  {Adams}} \emph {et~al.} (\bibinfo {collaboration} {CUORE}),\ }\href@noop {}
  {\  (\bibinfo {year} {2021})},\ \Eprint {http://arxiv.org/abs/2104.06906}
  {arXiv:2104.06906 [nucl-ex]} \BibitemShut {NoStop}%
\bibitem [{\citenamefont {Anton}\ \emph {et~al.}(2019)\citenamefont {Anton}
  \emph {et~al.}}]{EXO-200:2019rkq}%
  \BibitemOpen
  \bibfield  {author} {\bibinfo {author} {\bibfnamefont {G.}~\bibnamefont
  {Anton}} \emph {et~al.} (\bibinfo {collaboration} {EXO-200}),\ }\href
  {\doibase 10.1103/PhysRevLett.123.161802} {\bibfield  {journal} {\bibinfo
  {journal} {Phys. Rev. Lett.}\ }\textbf {\bibinfo {volume} {123}},\ \bibinfo
  {pages} {161802} (\bibinfo {year} {2019})},\ \Eprint
  {http://arxiv.org/abs/1906.02723} {arXiv:1906.02723 [hep-ex]} \BibitemShut
  {NoStop}%
\bibitem [{\citenamefont {Giuliani}\ \emph {et~al.}(2019)\citenamefont
  {Giuliani}, \citenamefont {Gomez~Cadenas}, \citenamefont {Pascoli},
  \citenamefont {Previtali}, \citenamefont {Saakyan}, \citenamefont
  {Sch\"affner},\ and\ \citenamefont {Sch\"onert}}]{Giuliani:2019uno}%
  \BibitemOpen
  \bibfield  {author} {\bibinfo {author} {\bibfnamefont {A.}~\bibnamefont
  {Giuliani}}, \bibinfo {author} {\bibfnamefont {J.~J.}\ \bibnamefont
  {Gomez~Cadenas}}, \bibinfo {author} {\bibfnamefont {S.}~\bibnamefont
  {Pascoli}}, \bibinfo {author} {\bibfnamefont {E.}~\bibnamefont {Previtali}},
  \bibinfo {author} {\bibfnamefont {R.}~\bibnamefont {Saakyan}}, \bibinfo
  {author} {\bibfnamefont {K.}~\bibnamefont {Sch\"affner}}, \ and\ \bibinfo
  {author} {\bibfnamefont {S.}~\bibnamefont {Sch\"onert}} (\bibinfo
  {collaboration} {APPEC Committee}),\ }\href@noop {} {\  (\bibinfo {year}
  {2019})},\ \Eprint {http://arxiv.org/abs/1910.04688} {arXiv:1910.04688
  [hep-ex]} \BibitemShut {NoStop}%
\bibitem [{\citenamefont {Ettengruber}\ \emph {et~al.}(2022)\citenamefont
  {Ettengruber}, \citenamefont {Agostini}, \citenamefont {Caldwell},
  \citenamefont {Eller},\ and\ \citenamefont {Schulz}}]{Ettengruber:2022mtm}%
  \BibitemOpen
  \bibfield  {author} {\bibinfo {author} {\bibfnamefont {M.}~\bibnamefont
  {Ettengruber}}, \bibinfo {author} {\bibfnamefont {M.}~\bibnamefont
  {Agostini}}, \bibinfo {author} {\bibfnamefont {A.}~\bibnamefont {Caldwell}},
  \bibinfo {author} {\bibfnamefont {P.}~\bibnamefont {Eller}}, \ and\ \bibinfo
  {author} {\bibfnamefont {O.}~\bibnamefont {Schulz}},\ }\href {\doibase
  10.1103/PhysRevD.106.073004} {\bibfield  {journal} {\bibinfo  {journal}
  {Phys. Rev. D}\ }\textbf {\bibinfo {volume} {106}},\ \bibinfo {pages}
  {073004} (\bibinfo {year} {2022})},\ \Eprint
  {http://arxiv.org/abs/2208.09954} {arXiv:2208.09954 [hep-ph]} \BibitemShut
  {NoStop}%
\bibitem [{\citenamefont {Chacko}\ \emph {et~al.}(2020)\citenamefont {Chacko},
  \citenamefont {Dev}, \citenamefont {Du}, \citenamefont {Poulin},\ and\
  \citenamefont {Tsai}}]{Chacko:2019nej}%
  \BibitemOpen
  \bibfield  {author} {\bibinfo {author} {\bibfnamefont {Z.}~\bibnamefont
  {Chacko}}, \bibinfo {author} {\bibfnamefont {A.}~\bibnamefont {Dev}},
  \bibinfo {author} {\bibfnamefont {P.}~\bibnamefont {Du}}, \bibinfo {author}
  {\bibfnamefont {V.}~\bibnamefont {Poulin}}, \ and\ \bibinfo {author}
  {\bibfnamefont {Y.}~\bibnamefont {Tsai}},\ }\href {\doibase
  10.1007/JHEP04(2020)020} {\bibfield  {journal} {\bibinfo  {journal} {JHEP}\
  }\textbf {\bibinfo {volume} {04}},\ \bibinfo {pages} {020} (\bibinfo {year}
  {2020})},\ \Eprint {http://arxiv.org/abs/1909.05275} {arXiv:1909.05275
  [hep-ph]} \BibitemShut {NoStop}%
\bibitem [{\citenamefont {Escudero}\ \emph {et~al.}(2020)\citenamefont
  {Escudero}, \citenamefont {Lopez-Pavon}, \citenamefont {Rius},\ and\
  \citenamefont {Sandner}}]{Escudero:2020ped}%
  \BibitemOpen
  \bibfield  {author} {\bibinfo {author} {\bibfnamefont {M.}~\bibnamefont
  {Escudero}}, \bibinfo {author} {\bibfnamefont {J.}~\bibnamefont
  {Lopez-Pavon}}, \bibinfo {author} {\bibfnamefont {N.}~\bibnamefont {Rius}}, \
  and\ \bibinfo {author} {\bibfnamefont {S.}~\bibnamefont {Sandner}},\ }\href
  {\doibase 10.1007/JHEP12(2020)119} {\bibfield  {journal} {\bibinfo  {journal}
  {JHEP}\ }\textbf {\bibinfo {volume} {12}},\ \bibinfo {pages} {119} (\bibinfo
  {year} {2020})},\ \Eprint {http://arxiv.org/abs/2007.04994} {arXiv:2007.04994
  [hep-ph]} \BibitemShut {NoStop}%
\bibitem [{\citenamefont {Escudero}\ and\ \citenamefont
  {Fairbairn}(2019)}]{Escudero:2019gfk}%
  \BibitemOpen
  \bibfield  {author} {\bibinfo {author} {\bibfnamefont {M.}~\bibnamefont
  {Escudero}}\ and\ \bibinfo {author} {\bibfnamefont {M.}~\bibnamefont
  {Fairbairn}},\ }\href {\doibase 10.1103/PhysRevD.100.103531} {\bibfield
  {journal} {\bibinfo  {journal} {Phys. Rev. D}\ }\textbf {\bibinfo {volume}
  {100}},\ \bibinfo {pages} {103531} (\bibinfo {year} {2019})},\ \Eprint
  {http://arxiv.org/abs/1907.05425} {arXiv:1907.05425 [hep-ph]} \BibitemShut
  {NoStop}%
\bibitem [{\citenamefont {Barenboim}\ \emph {et~al.}(2021)\citenamefont
  {Barenboim}, \citenamefont {Chen}, \citenamefont {Hannestad}, \citenamefont
  {Oldengott}, \citenamefont {Tram},\ and\ \citenamefont
  {Wong}}]{Barenboim:2020vrr}%
  \BibitemOpen
  \bibfield  {author} {\bibinfo {author} {\bibfnamefont {G.}~\bibnamefont
  {Barenboim}}, \bibinfo {author} {\bibfnamefont {J.~Z.}\ \bibnamefont {Chen}},
  \bibinfo {author} {\bibfnamefont {S.}~\bibnamefont {Hannestad}}, \bibinfo
  {author} {\bibfnamefont {I.~M.}\ \bibnamefont {Oldengott}}, \bibinfo {author}
  {\bibfnamefont {T.}~\bibnamefont {Tram}}, \ and\ \bibinfo {author}
  {\bibfnamefont {Y.~Y.~Y.}\ \bibnamefont {Wong}},\ }\href {\doibase
  10.1088/1475-7516/2021/03/087} {\bibfield  {journal} {\bibinfo  {journal}
  {JCAP}\ }\textbf {\bibinfo {volume} {03}},\ \bibinfo {pages} {087} (\bibinfo
  {year} {2021})},\ \Eprint {http://arxiv.org/abs/2011.01502} {arXiv:2011.01502
  [astro-ph.CO]} \BibitemShut {NoStop}%
\bibitem [{\citenamefont {Chacko}\ \emph {et~al.}(2021)\citenamefont {Chacko},
  \citenamefont {Dev}, \citenamefont {Du}, \citenamefont {Poulin},\ and\
  \citenamefont {Tsai}}]{Chacko:2020hmh}%
  \BibitemOpen
  \bibfield  {author} {\bibinfo {author} {\bibfnamefont {Z.}~\bibnamefont
  {Chacko}}, \bibinfo {author} {\bibfnamefont {A.}~\bibnamefont {Dev}},
  \bibinfo {author} {\bibfnamefont {P.}~\bibnamefont {Du}}, \bibinfo {author}
  {\bibfnamefont {V.}~\bibnamefont {Poulin}}, \ and\ \bibinfo {author}
  {\bibfnamefont {Y.}~\bibnamefont {Tsai}},\ }\href {\doibase
  10.1103/PhysRevD.103.043519} {\bibfield  {journal} {\bibinfo  {journal}
  {Phys. Rev. D}\ }\textbf {\bibinfo {volume} {103}},\ \bibinfo {pages}
  {043519} (\bibinfo {year} {2021})},\ \Eprint
  {http://arxiv.org/abs/2002.08401} {arXiv:2002.08401 [astro-ph.CO]}
  \BibitemShut {NoStop}%
\bibitem [{\citenamefont {Abell\'an}\ \emph {et~al.}(2021)\citenamefont
  {Abell\'an}, \citenamefont {Chacko}, \citenamefont {Dev}, \citenamefont {Du},
  \citenamefont {Poulin},\ and\ \citenamefont {Tsai}}]{Abellan:2021rfq}%
  \BibitemOpen
  \bibfield  {author} {\bibinfo {author} {\bibfnamefont {G.~F.}\ \bibnamefont
  {Abell\'an}}, \bibinfo {author} {\bibfnamefont {Z.}~\bibnamefont {Chacko}},
  \bibinfo {author} {\bibfnamefont {A.}~\bibnamefont {Dev}}, \bibinfo {author}
  {\bibfnamefont {P.}~\bibnamefont {Du}}, \bibinfo {author} {\bibfnamefont
  {V.}~\bibnamefont {Poulin}}, \ and\ \bibinfo {author} {\bibfnamefont
  {Y.}~\bibnamefont {Tsai}},\ }\href@noop {} {\  (\bibinfo {year} {2021})},\
  \Eprint {http://arxiv.org/abs/2112.13862} {arXiv:2112.13862 [hep-ph]}
  \BibitemShut {NoStop}%
\bibitem [{\citenamefont {Chen}\ \emph {et~al.}(2022)\citenamefont {Chen},
  \citenamefont {Oldengott}, \citenamefont {Pierobon},\ and\ \citenamefont
  {Wong}}]{Chen:2022idm}%
  \BibitemOpen
  \bibfield  {author} {\bibinfo {author} {\bibfnamefont {J.~Z.}\ \bibnamefont
  {Chen}}, \bibinfo {author} {\bibfnamefont {I.~M.}\ \bibnamefont {Oldengott}},
  \bibinfo {author} {\bibfnamefont {G.}~\bibnamefont {Pierobon}}, \ and\
  \bibinfo {author} {\bibfnamefont {Y.~Y.~Y.}\ \bibnamefont {Wong}},\ }\href
  {\doibase 10.1140/epjc/s10052-022-10518-3} {\bibfield  {journal} {\bibinfo
  {journal} {Eur. Phys. J. C}\ }\textbf {\bibinfo {volume} {82}},\ \bibinfo
  {pages} {640} (\bibinfo {year} {2022})},\ \Eprint
  {http://arxiv.org/abs/2203.09075} {arXiv:2203.09075 [hep-ph]} \BibitemShut
  {NoStop}%
\bibitem [{\citenamefont {Dvali}\ and\ \citenamefont
  {Funcke}(2016)}]{Dvali:2016uhn}%
  \BibitemOpen
  \bibfield  {author} {\bibinfo {author} {\bibfnamefont {G.}~\bibnamefont
  {Dvali}}\ and\ \bibinfo {author} {\bibfnamefont {L.}~\bibnamefont {Funcke}},\
  }\href {\doibase 10.1103/PhysRevD.93.113002} {\bibfield  {journal} {\bibinfo
  {journal} {Phys. Rev. D}\ }\textbf {\bibinfo {volume} {93}},\ \bibinfo
  {pages} {113002} (\bibinfo {year} {2016})},\ \Eprint
  {http://arxiv.org/abs/1602.03191} {arXiv:1602.03191 [hep-ph]} \BibitemShut
  {NoStop}%
\bibitem [{\citenamefont {Dvali}\ \emph {et~al.}(2021)\citenamefont {Dvali},
  \citenamefont {Funcke},\ and\ \citenamefont {Vachaspati}}]{Dvali:2021uvk}%
  \BibitemOpen
  \bibfield  {author} {\bibinfo {author} {\bibfnamefont {G.}~\bibnamefont
  {Dvali}}, \bibinfo {author} {\bibfnamefont {L.}~\bibnamefont {Funcke}}, \
  and\ \bibinfo {author} {\bibfnamefont {T.}~\bibnamefont {Vachaspati}},\
  }\href@noop {} {\  (\bibinfo {year} {2021})},\ \Eprint
  {http://arxiv.org/abs/2112.02107} {arXiv:2112.02107 [hep-ph]} \BibitemShut
  {NoStop}%
\bibitem [{\citenamefont {Lorenz}\ \emph {et~al.}(2019)\citenamefont {Lorenz},
  \citenamefont {Funcke}, \citenamefont {Calabrese},\ and\ \citenamefont
  {Hannestad}}]{Lorenz:2018fzb}%
  \BibitemOpen
  \bibfield  {author} {\bibinfo {author} {\bibfnamefont {C.~S.}\ \bibnamefont
  {Lorenz}}, \bibinfo {author} {\bibfnamefont {L.}~\bibnamefont {Funcke}},
  \bibinfo {author} {\bibfnamefont {E.}~\bibnamefont {Calabrese}}, \ and\
  \bibinfo {author} {\bibfnamefont {S.}~\bibnamefont {Hannestad}},\ }\href
  {\doibase 10.1103/PhysRevD.99.023501} {\bibfield  {journal} {\bibinfo
  {journal} {Phys. Rev. D}\ }\textbf {\bibinfo {volume} {99}},\ \bibinfo
  {pages} {023501} (\bibinfo {year} {2019})},\ \Eprint
  {http://arxiv.org/abs/1811.01991} {arXiv:1811.01991 [astro-ph.CO]}
  \BibitemShut {NoStop}%
\bibitem [{\citenamefont {Lorenz}\ \emph {et~al.}(2021)\citenamefont {Lorenz},
  \citenamefont {Funcke}, \citenamefont {L\"offler},\ and\ \citenamefont
  {Calabrese}}]{Lorenz:2021alz}%
  \BibitemOpen
  \bibfield  {author} {\bibinfo {author} {\bibfnamefont {C.~S.}\ \bibnamefont
  {Lorenz}}, \bibinfo {author} {\bibfnamefont {L.}~\bibnamefont {Funcke}},
  \bibinfo {author} {\bibfnamefont {M.}~\bibnamefont {L\"offler}}, \ and\
  \bibinfo {author} {\bibfnamefont {E.}~\bibnamefont {Calabrese}},\ }\href
  {\doibase 10.1103/PhysRevD.104.123518} {\bibfield  {journal} {\bibinfo
  {journal} {Phys. Rev. D}\ }\textbf {\bibinfo {volume} {104}},\ \bibinfo
  {pages} {123518} (\bibinfo {year} {2021})},\ \Eprint
  {http://arxiv.org/abs/2102.13618} {arXiv:2102.13618 [astro-ph.CO]}
  \BibitemShut {NoStop}%
\bibitem [{\citenamefont {Esteban}\ and\ \citenamefont
  {Salvado}(2021)}]{Esteban:2021ozz}%
  \BibitemOpen
  \bibfield  {author} {\bibinfo {author} {\bibfnamefont {I.}~\bibnamefont
  {Esteban}}\ and\ \bibinfo {author} {\bibfnamefont {J.}~\bibnamefont
  {Salvado}},\ }\href {\doibase 10.1088/1475-7516/2021/05/036} {\bibfield
  {journal} {\bibinfo  {journal} {JCAP}\ }\textbf {\bibinfo {volume} {05}},\
  \bibinfo {pages} {036} (\bibinfo {year} {2021})},\ \Eprint
  {http://arxiv.org/abs/2101.05804} {arXiv:2101.05804 [hep-ph]} \BibitemShut
  {NoStop}%
\bibitem [{\citenamefont {Farzan}\ and\ \citenamefont
  {Hannestad}(2016)}]{Farzan:2015pca}%
  \BibitemOpen
  \bibfield  {author} {\bibinfo {author} {\bibfnamefont {Y.}~\bibnamefont
  {Farzan}}\ and\ \bibinfo {author} {\bibfnamefont {S.}~\bibnamefont
  {Hannestad}},\ }\href {\doibase 10.1088/1475-7516/2016/02/058} {\bibfield
  {journal} {\bibinfo  {journal} {JCAP}\ }\textbf {\bibinfo {volume} {02}},\
  \bibinfo {pages} {058} (\bibinfo {year} {2016})},\ \Eprint
  {http://arxiv.org/abs/1510.02201} {arXiv:1510.02201 [hep-ph]} \BibitemShut
  {NoStop}%
\bibitem [{\citenamefont {Renk}\ \emph {et~al.}(2021)\citenamefont {Renk} \emph
  {et~al.}}]{GAMBITCosmologyWorkgroup:2020htv}%
  \BibitemOpen
  \bibfield  {author} {\bibinfo {author} {\bibfnamefont {J.~J.}\ \bibnamefont
  {Renk}} \emph {et~al.} (\bibinfo {collaboration} {GAMBIT Cosmology
  Workgroup}),\ }\href {\doibase 10.1088/1475-7516/2021/02/022} {\bibfield
  {journal} {\bibinfo  {journal} {JCAP}\ }\textbf {\bibinfo {volume} {02}},\
  \bibinfo {pages} {022} (\bibinfo {year} {2021})},\ \Eprint
  {http://arxiv.org/abs/2009.03286} {arXiv:2009.03286 [astro-ph.CO]}
  \BibitemShut {NoStop}%
\bibitem [{\citenamefont {Cuoco}\ \emph {et~al.}(2005)\citenamefont {Cuoco},
  \citenamefont {Lesgourgues}, \citenamefont {Mangano},\ and\ \citenamefont
  {Pastor}}]{Cuoco:2005qr}%
  \BibitemOpen
  \bibfield  {author} {\bibinfo {author} {\bibfnamefont {A.}~\bibnamefont
  {Cuoco}}, \bibinfo {author} {\bibfnamefont {J.}~\bibnamefont {Lesgourgues}},
  \bibinfo {author} {\bibfnamefont {G.}~\bibnamefont {Mangano}}, \ and\
  \bibinfo {author} {\bibfnamefont {S.}~\bibnamefont {Pastor}},\ }\href
  {\doibase 10.1103/PhysRevD.71.123501} {\bibfield  {journal} {\bibinfo
  {journal} {Phys. Rev. D}\ }\textbf {\bibinfo {volume} {71}},\ \bibinfo
  {pages} {123501} (\bibinfo {year} {2005})},\ \Eprint
  {http://arxiv.org/abs/astro-ph/0502465} {arXiv:astro-ph/0502465} \BibitemShut
  {NoStop}%
\bibitem [{\citenamefont {Oldengott}\ \emph {et~al.}(2019)\citenamefont
  {Oldengott}, \citenamefont {Barenboim}, \citenamefont {Kahlen}, \citenamefont
  {Salvado},\ and\ \citenamefont {Schwarz}}]{Oldengott:2019lke}%
  \BibitemOpen
  \bibfield  {author} {\bibinfo {author} {\bibfnamefont {I.~M.}\ \bibnamefont
  {Oldengott}}, \bibinfo {author} {\bibfnamefont {G.}~\bibnamefont
  {Barenboim}}, \bibinfo {author} {\bibfnamefont {S.}~\bibnamefont {Kahlen}},
  \bibinfo {author} {\bibfnamefont {J.}~\bibnamefont {Salvado}}, \ and\
  \bibinfo {author} {\bibfnamefont {D.~J.}\ \bibnamefont {Schwarz}},\ }\href
  {\doibase 10.1088/1475-7516/2019/04/049} {\bibfield  {journal} {\bibinfo
  {journal} {JCAP}\ }\textbf {\bibinfo {volume} {04}},\ \bibinfo {pages} {049}
  (\bibinfo {year} {2019})},\ \Eprint {http://arxiv.org/abs/1901.04352}
  {arXiv:1901.04352 [astro-ph.CO]} \BibitemShut {NoStop}%
\bibitem [{\citenamefont {Alvey}\ \emph
  {et~al.}(2022{\natexlab{a}})\citenamefont {Alvey}, \citenamefont {Escudero},\
  and\ \citenamefont {Sabti}}]{Alvey:2021sji}%
  \BibitemOpen
  \bibfield  {author} {\bibinfo {author} {\bibfnamefont {J.}~\bibnamefont
  {Alvey}}, \bibinfo {author} {\bibfnamefont {M.}~\bibnamefont {Escudero}}, \
  and\ \bibinfo {author} {\bibfnamefont {N.}~\bibnamefont {Sabti}},\ }\href
  {\doibase 10.1088/1475-7516/2022/02/037} {\bibfield  {journal} {\bibinfo
  {journal} {JCAP}\ }\textbf {\bibinfo {volume} {02}},\ \bibinfo {pages} {037}
  (\bibinfo {year} {2022}{\natexlab{a}})},\ \Eprint
  {http://arxiv.org/abs/2111.12726} {arXiv:2111.12726 [astro-ph.CO]}
  \BibitemShut {NoStop}%
\bibitem [{\citenamefont {Alvey}\ \emph
  {et~al.}(2022{\natexlab{b}})\citenamefont {Alvey}, \citenamefont {Escudero},
  \citenamefont {Sabti},\ and\ \citenamefont {Schwetz}}]{Alvey:2021xmq}%
  \BibitemOpen
  \bibfield  {author} {\bibinfo {author} {\bibfnamefont {J.}~\bibnamefont
  {Alvey}}, \bibinfo {author} {\bibfnamefont {M.}~\bibnamefont {Escudero}},
  \bibinfo {author} {\bibfnamefont {N.}~\bibnamefont {Sabti}}, \ and\ \bibinfo
  {author} {\bibfnamefont {T.}~\bibnamefont {Schwetz}},\ }\href {\doibase
  10.1103/PhysRevD.105.063501} {\bibfield  {journal} {\bibinfo  {journal}
  {Phys. Rev. D}\ }\textbf {\bibinfo {volume} {105}},\ \bibinfo {pages}
  {063501} (\bibinfo {year} {2022}{\natexlab{b}})},\ \Eprint
  {http://arxiv.org/abs/2111.14870} {arXiv:2111.14870 [hep-ph]} \BibitemShut
  {NoStop}%
\bibitem [{\citenamefont {Beacom}\ \emph {et~al.}(2004)\citenamefont {Beacom},
  \citenamefont {Bell},\ and\ \citenamefont {Dodelson}}]{Beacom:2004yd}%
  \BibitemOpen
  \bibfield  {author} {\bibinfo {author} {\bibfnamefont {J.~F.}\ \bibnamefont
  {Beacom}}, \bibinfo {author} {\bibfnamefont {N.~F.}\ \bibnamefont {Bell}}, \
  and\ \bibinfo {author} {\bibfnamefont {S.}~\bibnamefont {Dodelson}},\ }\href
  {\doibase 10.1103/PhysRevLett.93.121302} {\bibfield  {journal} {\bibinfo
  {journal} {Phys. Rev. Lett.}\ }\textbf {\bibinfo {volume} {93}},\ \bibinfo
  {pages} {121302} (\bibinfo {year} {2004})},\ \Eprint
  {http://arxiv.org/abs/astro-ph/0404585} {arXiv:astro-ph/0404585} \BibitemShut
  {NoStop}%
\bibitem [{\citenamefont {Chun}\ \emph {et~al.}(1995)\citenamefont {Chun},
  \citenamefont {Joshipura},\ and\ \citenamefont {Smirnov}}]{Chun:1995js}%
  \BibitemOpen
  \bibfield  {author} {\bibinfo {author} {\bibfnamefont {E.~J.}\ \bibnamefont
  {Chun}}, \bibinfo {author} {\bibfnamefont {A.~S.}\ \bibnamefont {Joshipura}},
  \ and\ \bibinfo {author} {\bibfnamefont {A.~Y.}\ \bibnamefont {Smirnov}},\
  }\href {\doibase 10.1016/0370-2693(95)00967-P} {\bibfield  {journal}
  {\bibinfo  {journal} {Phys. Lett. B}\ }\textbf {\bibinfo {volume} {357}},\
  \bibinfo {pages} {608} (\bibinfo {year} {1995})},\ \Eprint
  {http://arxiv.org/abs/hep-ph/9505275} {arXiv:hep-ph/9505275} \BibitemShut
  {NoStop}%
\bibitem [{\citenamefont {Barry}\ \emph {et~al.}(2011)\citenamefont {Barry},
  \citenamefont {Rodejohann},\ and\ \citenamefont {Zhang}}]{Barry:2011wb}%
  \BibitemOpen
  \bibfield  {author} {\bibinfo {author} {\bibfnamefont {J.}~\bibnamefont
  {Barry}}, \bibinfo {author} {\bibfnamefont {W.}~\bibnamefont {Rodejohann}}, \
  and\ \bibinfo {author} {\bibfnamefont {H.}~\bibnamefont {Zhang}},\ }\href
  {\doibase 10.1007/JHEP07(2011)091} {\bibfield  {journal} {\bibinfo  {journal}
  {JHEP}\ }\textbf {\bibinfo {volume} {07}},\ \bibinfo {pages} {091} (\bibinfo
  {year} {2011})},\ \Eprint {http://arxiv.org/abs/1105.3911} {arXiv:1105.3911
  [hep-ph]} \BibitemShut {NoStop}%
\bibitem [{\citenamefont {Zhang}(2012)}]{Zhang:2011vh}%
  \BibitemOpen
  \bibfield  {author} {\bibinfo {author} {\bibfnamefont {H.}~\bibnamefont
  {Zhang}},\ }\href {\doibase 10.1016/j.physletb.2012.06.074} {\bibfield
  {journal} {\bibinfo  {journal} {Phys. Lett. B}\ }\textbf {\bibinfo {volume}
  {714}},\ \bibinfo {pages} {262} (\bibinfo {year} {2012})},\ \Eprint
  {http://arxiv.org/abs/1110.6838} {arXiv:1110.6838 [hep-ph]} \BibitemShut
  {NoStop}%
\bibitem [{\citenamefont {Heeck}\ and\ \citenamefont
  {Zhang}(2013)}]{Heeck:2012bz}%
  \BibitemOpen
  \bibfield  {author} {\bibinfo {author} {\bibfnamefont {J.}~\bibnamefont
  {Heeck}}\ and\ \bibinfo {author} {\bibfnamefont {H.}~\bibnamefont {Zhang}},\
  }\href {\doibase 10.1007/JHEP05(2013)164} {\bibfield  {journal} {\bibinfo
  {journal} {JHEP}\ }\textbf {\bibinfo {volume} {05}},\ \bibinfo {pages} {164}
  (\bibinfo {year} {2013})},\ \Eprint {http://arxiv.org/abs/1211.0538}
  {arXiv:1211.0538 [hep-ph]} \BibitemShut {NoStop}%
\bibitem [{\citenamefont {Ballett}\ \emph {et~al.}(2019)\citenamefont
  {Ballett}, \citenamefont {Hostert},\ and\ \citenamefont
  {Pascoli}}]{Ballett:2019cqp}%
  \BibitemOpen
  \bibfield  {author} {\bibinfo {author} {\bibfnamefont {P.}~\bibnamefont
  {Ballett}}, \bibinfo {author} {\bibfnamefont {M.}~\bibnamefont {Hostert}}, \
  and\ \bibinfo {author} {\bibfnamefont {S.}~\bibnamefont {Pascoli}},\ }\href
  {\doibase 10.1103/PhysRevD.99.091701} {\bibfield  {journal} {\bibinfo
  {journal} {Phys. Rev. D}\ }\textbf {\bibinfo {volume} {99}},\ \bibinfo
  {pages} {091701} (\bibinfo {year} {2019})},\ \Eprint
  {http://arxiv.org/abs/1903.07590} {arXiv:1903.07590 [hep-ph]} \BibitemShut
  {NoStop}%
\bibitem [{\citenamefont {Bringmann}\ \emph {et~al.}(2023)\citenamefont
  {Bringmann}, \citenamefont {Depta}, \citenamefont {Hufnagel}, \citenamefont
  {Kersten}, \citenamefont {Ruderman},\ and\ \citenamefont
  {Schmidt-Hoberg}}]{Bringmann:2022aim}%
  \BibitemOpen
  \bibfield  {author} {\bibinfo {author} {\bibfnamefont {T.}~\bibnamefont
  {Bringmann}}, \bibinfo {author} {\bibfnamefont {P.~F.}\ \bibnamefont
  {Depta}}, \bibinfo {author} {\bibfnamefont {M.}~\bibnamefont {Hufnagel}},
  \bibinfo {author} {\bibfnamefont {J.}~\bibnamefont {Kersten}}, \bibinfo
  {author} {\bibfnamefont {J.~T.}\ \bibnamefont {Ruderman}}, \ and\ \bibinfo
  {author} {\bibfnamefont {K.}~\bibnamefont {Schmidt-Hoberg}},\ }\href
  {\doibase 10.1103/PhysRevD.107.L071702} {\bibfield  {journal} {\bibinfo
  {journal} {Phys. Rev. D}\ }\textbf {\bibinfo {volume} {107}},\ \bibinfo
  {pages} {L071702} (\bibinfo {year} {2023})},\ \Eprint
  {http://arxiv.org/abs/2206.10630} {arXiv:2206.10630 [hep-ph]} \BibitemShut
  {NoStop}%
\bibitem [{\citenamefont {Ko}\ and\ \citenamefont {Tang}(2014)}]{Ko:2014bka}%
  \BibitemOpen
  \bibfield  {author} {\bibinfo {author} {\bibfnamefont {P.}~\bibnamefont
  {Ko}}\ and\ \bibinfo {author} {\bibfnamefont {Y.}~\bibnamefont {Tang}},\
  }\href {\doibase 10.1016/j.physletb.2014.10.035} {\bibfield  {journal}
  {\bibinfo  {journal} {Phys. Lett. B}\ }\textbf {\bibinfo {volume} {739}},\
  \bibinfo {pages} {62} (\bibinfo {year} {2014})},\ \Eprint
  {http://arxiv.org/abs/1404.0236} {arXiv:1404.0236 [hep-ph]} \BibitemShut
  {NoStop}%
\bibitem [{\citenamefont {Escudero~Abenza}(2020)}]{EscuderoAbenza:2020cmq}%
  \BibitemOpen
  \bibfield  {author} {\bibinfo {author} {\bibfnamefont {M.}~\bibnamefont
  {Escudero~Abenza}},\ }\href {\doibase 10.1088/1475-7516/2020/05/048}
  {\bibfield  {journal} {\bibinfo  {journal} {JCAP}\ }\textbf {\bibinfo
  {volume} {05}},\ \bibinfo {pages} {048} (\bibinfo {year} {2020})},\ \Eprint
  {http://arxiv.org/abs/2001.04466} {arXiv:2001.04466 [hep-ph]} \BibitemShut
  {NoStop}%
\bibitem [{\citenamefont {Akita}\ and\ \citenamefont
  {Yamaguchi}(2020)}]{Akita:2020szl}%
  \BibitemOpen
  \bibfield  {author} {\bibinfo {author} {\bibfnamefont {K.}~\bibnamefont
  {Akita}}\ and\ \bibinfo {author} {\bibfnamefont {M.}~\bibnamefont
  {Yamaguchi}},\ }\href {\doibase 10.1088/1475-7516/2020/08/012} {\bibfield
  {journal} {\bibinfo  {journal} {JCAP}\ }\textbf {\bibinfo {volume} {08}},\
  \bibinfo {pages} {012} (\bibinfo {year} {2020})},\ \Eprint
  {http://arxiv.org/abs/2005.07047} {arXiv:2005.07047 [hep-ph]} \BibitemShut
  {NoStop}%
\bibitem [{\citenamefont {Froustey}\ \emph {et~al.}(2020)\citenamefont
  {Froustey}, \citenamefont {Pitrou},\ and\ \citenamefont
  {Volpe}}]{Froustey:2020mcq}%
  \BibitemOpen
  \bibfield  {author} {\bibinfo {author} {\bibfnamefont {J.}~\bibnamefont
  {Froustey}}, \bibinfo {author} {\bibfnamefont {C.}~\bibnamefont {Pitrou}}, \
  and\ \bibinfo {author} {\bibfnamefont {M.~C.}\ \bibnamefont {Volpe}},\ }\href
  {\doibase 10.1088/1475-7516/2020/12/015} {\bibfield  {journal} {\bibinfo
  {journal} {JCAP}\ }\textbf {\bibinfo {volume} {12}},\ \bibinfo {pages} {015}
  (\bibinfo {year} {2020})},\ \Eprint {http://arxiv.org/abs/2008.01074}
  {arXiv:2008.01074 [hep-ph]} \BibitemShut {NoStop}%
\bibitem [{\citenamefont {Bennett}\ \emph {et~al.}(2021)\citenamefont
  {Bennett}, \citenamefont {Buldgen}, \citenamefont {De~Salas}, \citenamefont
  {Drewes}, \citenamefont {Gariazzo}, \citenamefont {Pastor},\ and\
  \citenamefont {Wong}}]{Bennett:2020zkv}%
  \BibitemOpen
  \bibfield  {author} {\bibinfo {author} {\bibfnamefont {J.~J.}\ \bibnamefont
  {Bennett}}, \bibinfo {author} {\bibfnamefont {G.}~\bibnamefont {Buldgen}},
  \bibinfo {author} {\bibfnamefont {P.~F.}\ \bibnamefont {De~Salas}}, \bibinfo
  {author} {\bibfnamefont {M.}~\bibnamefont {Drewes}}, \bibinfo {author}
  {\bibfnamefont {S.}~\bibnamefont {Gariazzo}}, \bibinfo {author}
  {\bibfnamefont {S.}~\bibnamefont {Pastor}}, \ and\ \bibinfo {author}
  {\bibfnamefont {Y.~Y.~Y.}\ \bibnamefont {Wong}},\ }\href {\doibase
  10.1088/1475-7516/2021/04/073} {\bibfield  {journal} {\bibinfo  {journal}
  {JCAP}\ }\textbf {\bibinfo {volume} {04}},\ \bibinfo {pages} {073} (\bibinfo
  {year} {2021})},\ \Eprint {http://arxiv.org/abs/2012.02726} {arXiv:2012.02726
  [hep-ph]} \BibitemShut {NoStop}%
\bibitem [{\citenamefont {Allahverdi}\ \emph {et~al.}(2020)\citenamefont
  {Allahverdi} \emph {et~al.}}]{Allahverdi:2020bys}%
  \BibitemOpen
  \bibfield  {author} {\bibinfo {author} {\bibfnamefont {R.}~\bibnamefont
  {Allahverdi}} \emph {et~al.},\ }\href {\doibase 10.21105/astro.2006.16182} {\
   (\bibinfo {year} {2020}),\ 10.21105/astro.2006.16182},\ \Eprint
  {http://arxiv.org/abs/2006.16182} {arXiv:2006.16182 [astro-ph.CO]}
  \BibitemShut {NoStop}%
\bibitem [{\citenamefont {Pospelov}\ and\ \citenamefont
  {Pradler}(2010)}]{Pospelov:2010hj}%
  \BibitemOpen
  \bibfield  {author} {\bibinfo {author} {\bibfnamefont {M.}~\bibnamefont
  {Pospelov}}\ and\ \bibinfo {author} {\bibfnamefont {J.}~\bibnamefont
  {Pradler}},\ }\href {\doibase 10.1146/annurev.nucl.012809.104521} {\bibfield
  {journal} {\bibinfo  {journal} {Ann. Rev. Nucl. Part. Sci.}\ }\textbf
  {\bibinfo {volume} {60}},\ \bibinfo {pages} {539} (\bibinfo {year} {2010})},\
  \Eprint {http://arxiv.org/abs/1011.1054} {arXiv:1011.1054 [hep-ph]}
  \BibitemShut {NoStop}%
\bibitem [{\citenamefont {Taule}\ \emph {et~al.}(2022)\citenamefont {Taule},
  \citenamefont {Escudero},\ and\ \citenamefont {Garny}}]{Taule:2022jrz}%
  \BibitemOpen
  \bibfield  {author} {\bibinfo {author} {\bibfnamefont {P.}~\bibnamefont
  {Taule}}, \bibinfo {author} {\bibfnamefont {M.}~\bibnamefont {Escudero}}, \
  and\ \bibinfo {author} {\bibfnamefont {M.}~\bibnamefont {Garny}},\ }\href
  {\doibase 10.1103/PhysRevD.106.063539} {\bibfield  {journal} {\bibinfo
  {journal} {Phys. Rev. D}\ }\textbf {\bibinfo {volume} {106}},\ \bibinfo
  {pages} {063539} (\bibinfo {year} {2022})},\ \Eprint
  {http://arxiv.org/abs/2207.04062} {arXiv:2207.04062 [astro-ph.CO]}
  \BibitemShut {NoStop}%
\bibitem [{\citenamefont {Georgi}\ and\ \citenamefont
  {Glashow}(2000)}]{Georgi:1998bf}%
  \BibitemOpen
  \bibfield  {author} {\bibinfo {author} {\bibfnamefont {H.}~\bibnamefont
  {Georgi}}\ and\ \bibinfo {author} {\bibfnamefont {S.~L.}\ \bibnamefont
  {Glashow}},\ }\href {\doibase 10.1103/PhysRevD.61.097301} {\bibfield
  {journal} {\bibinfo  {journal} {Phys. Rev. D}\ }\textbf {\bibinfo {volume}
  {61}},\ \bibinfo {pages} {097301} (\bibinfo {year} {2000})},\ \Eprint
  {http://arxiv.org/abs/hep-ph/9808293} {arXiv:hep-ph/9808293} \BibitemShut
  {NoStop}%
\bibitem [{\citenamefont {Ellis}\ and\ \citenamefont
  {Lola}(1999)}]{Ellis:1999my}%
  \BibitemOpen
  \bibfield  {author} {\bibinfo {author} {\bibfnamefont {J.~R.}\ \bibnamefont
  {Ellis}}\ and\ \bibinfo {author} {\bibfnamefont {S.}~\bibnamefont {Lola}},\
  }\href {\doibase 10.1016/S0370-2693(99)00545-6} {\bibfield  {journal}
  {\bibinfo  {journal} {Phys. Lett. B}\ }\textbf {\bibinfo {volume} {458}},\
  \bibinfo {pages} {310} (\bibinfo {year} {1999})},\ \Eprint
  {http://arxiv.org/abs/hep-ph/9904279} {arXiv:hep-ph/9904279} \BibitemShut
  {NoStop}%
\bibitem [{\citenamefont {Casas}\ \emph {et~al.}(1999)\citenamefont {Casas},
  \citenamefont {Espinosa}, \citenamefont {Ibarra},\ and\ \citenamefont
  {Navarro}}]{Casas:1999tp}%
  \BibitemOpen
  \bibfield  {author} {\bibinfo {author} {\bibfnamefont {J.~A.}\ \bibnamefont
  {Casas}}, \bibinfo {author} {\bibfnamefont {J.~R.}\ \bibnamefont {Espinosa}},
  \bibinfo {author} {\bibfnamefont {A.}~\bibnamefont {Ibarra}}, \ and\ \bibinfo
  {author} {\bibfnamefont {I.}~\bibnamefont {Navarro}},\ }\href {\doibase
  10.1016/S0550-3213(99)00383-1} {\bibfield  {journal} {\bibinfo  {journal}
  {Nucl. Phys. B}\ }\textbf {\bibinfo {volume} {556}},\ \bibinfo {pages} {3}
  (\bibinfo {year} {1999})},\ \Eprint {http://arxiv.org/abs/hep-ph/9904395}
  {arXiv:hep-ph/9904395} \BibitemShut {NoStop}%
\bibitem [{\citenamefont {de~Medeiros~Varzielas}\ \emph
  {et~al.}(2009)\citenamefont {de~Medeiros~Varzielas}, \citenamefont {Ross},\
  and\ \citenamefont {Serna}}]{deMedeirosVarzielas:2008uzc}%
  \BibitemOpen
  \bibfield  {author} {\bibinfo {author} {\bibfnamefont {I.}~\bibnamefont
  {de~Medeiros~Varzielas}}, \bibinfo {author} {\bibfnamefont {G.~G.}\
  \bibnamefont {Ross}}, \ and\ \bibinfo {author} {\bibfnamefont
  {M.}~\bibnamefont {Serna}},\ }\href {\doibase 10.1103/PhysRevD.80.073002}
  {\bibfield  {journal} {\bibinfo  {journal} {Phys. Rev. D}\ }\textbf {\bibinfo
  {volume} {80}},\ \bibinfo {pages} {073002} (\bibinfo {year} {2009})},\
  \Eprint {http://arxiv.org/abs/0811.2226} {arXiv:0811.2226 [hep-ph]}
  \BibitemShut {NoStop}%
\bibitem [{\citenamefont {Boudjemaa}\ and\ \citenamefont
  {King}(2009)}]{Boudjemaa:2008jf}%
  \BibitemOpen
  \bibfield  {author} {\bibinfo {author} {\bibfnamefont {S.}~\bibnamefont
  {Boudjemaa}}\ and\ \bibinfo {author} {\bibfnamefont {S.~F.}\ \bibnamefont
  {King}},\ }\href {\doibase 10.1103/PhysRevD.79.033001} {\bibfield  {journal}
  {\bibinfo  {journal} {Phys. Rev. D}\ }\textbf {\bibinfo {volume} {79}},\
  \bibinfo {pages} {033001} (\bibinfo {year} {2009})},\ \Eprint
  {http://arxiv.org/abs/0808.2782} {arXiv:0808.2782 [hep-ph]} \BibitemShut
  {NoStop}%
\bibitem [{\citenamefont {Grimus}\ and\ \citenamefont
  {Lavoura}(2000)}]{Grimus:2000vj}%
  \BibitemOpen
  \bibfield  {author} {\bibinfo {author} {\bibfnamefont {W.}~\bibnamefont
  {Grimus}}\ and\ \bibinfo {author} {\bibfnamefont {L.}~\bibnamefont
  {Lavoura}},\ }\href {\doibase 10.1088/1126-6708/2000/11/042} {\bibfield
  {journal} {\bibinfo  {journal} {JHEP}\ }\textbf {\bibinfo {volume} {11}},\
  \bibinfo {pages} {042} (\bibinfo {year} {2000})},\ \Eprint
  {http://arxiv.org/abs/hep-ph/0008179} {arXiv:hep-ph/0008179} \BibitemShut
  {NoStop}%
\bibitem [{\citenamefont {Gherghetta}\ \emph {et~al.}(2019)\citenamefont
  {Gherghetta}, \citenamefont {Kersten}, \citenamefont {Olive},\ and\
  \citenamefont {Pospelov}}]{Gherghetta:2019coi}%
  \BibitemOpen
  \bibfield  {author} {\bibinfo {author} {\bibfnamefont {T.}~\bibnamefont
  {Gherghetta}}, \bibinfo {author} {\bibfnamefont {J.}~\bibnamefont {Kersten}},
  \bibinfo {author} {\bibfnamefont {K.}~\bibnamefont {Olive}}, \ and\ \bibinfo
  {author} {\bibfnamefont {M.}~\bibnamefont {Pospelov}},\ }\href {\doibase
  10.1103/PhysRevD.100.095001} {\bibfield  {journal} {\bibinfo  {journal}
  {Phys. Rev. D}\ }\textbf {\bibinfo {volume} {100}},\ \bibinfo {pages}
  {095001} (\bibinfo {year} {2019})},\ \Eprint
  {http://arxiv.org/abs/1909.00696} {arXiv:1909.00696 [hep-ph]} \BibitemShut
  {NoStop}%
\bibitem [{\citenamefont {Bauer}\ and\ \citenamefont
  {Foldenauer}(2022)}]{Bauer:2022nwt}%
  \BibitemOpen
  \bibfield  {author} {\bibinfo {author} {\bibfnamefont {M.}~\bibnamefont
  {Bauer}}\ and\ \bibinfo {author} {\bibfnamefont {P.}~\bibnamefont
  {Foldenauer}},\ }\href {\doibase 10.1103/PhysRevLett.129.171801} {\bibfield
  {journal} {\bibinfo  {journal} {Phys. Rev. Lett.}\ }\textbf {\bibinfo
  {volume} {129}},\ \bibinfo {pages} {171801} (\bibinfo {year} {2022})},\
  \Eprint {http://arxiv.org/abs/2207.00023} {arXiv:2207.00023 [hep-ph]}
  \BibitemShut {NoStop}%
\bibitem [{\citenamefont {Janka}(2012)}]{Janka:2012wk}%
  \BibitemOpen
  \bibfield  {author} {\bibinfo {author} {\bibfnamefont {H.-T.}\ \bibnamefont
  {Janka}},\ }\href {\doibase 10.1146/annurev-nucl-102711-094901} {\bibfield
  {journal} {\bibinfo  {journal} {Ann. Rev. Nucl. Part. Sci.}\ }\textbf
  {\bibinfo {volume} {62}},\ \bibinfo {pages} {407} (\bibinfo {year} {2012})},\
  \Eprint {http://arxiv.org/abs/1206.2503} {arXiv:1206.2503 [astro-ph.SR]}
  \BibitemShut {NoStop}%
\bibitem [{\citenamefont {Raffelt}(1996)}]{Raffelt:1996wa}%
  \BibitemOpen
  \bibfield  {author} {\bibinfo {author} {\bibfnamefont {G.~G.}\ \bibnamefont
  {Raffelt}},\ }\href@noop {} {\emph {\bibinfo {title} {{Stars as Laboratories
  for Fundamental Physics}: {The Astrophysics of Neutrinos, Axions, and Other
  Weakly Interacting Particles}}}}\ (\bibinfo {year} {1996})\BibitemShut
  {NoStop}%
\bibitem [{\citenamefont {Fiorillo}\ \emph {et~al.}(2022)\citenamefont
  {Fiorillo}, \citenamefont {Raffelt},\ and\ \citenamefont
  {Vitagliano}}]{Fiorillo:2022cdq}%
  \BibitemOpen
  \bibfield  {author} {\bibinfo {author} {\bibfnamefont {D.~F.~G.}\
  \bibnamefont {Fiorillo}}, \bibinfo {author} {\bibfnamefont {G.~G.}\
  \bibnamefont {Raffelt}}, \ and\ \bibinfo {author} {\bibfnamefont
  {E.}~\bibnamefont {Vitagliano}},\ }\href@noop {} {\  (\bibinfo {year}
  {2022})},\ \Eprint {http://arxiv.org/abs/2209.11773} {arXiv:2209.11773
  [hep-ph]} \BibitemShut {NoStop}%
\bibitem [{\citenamefont {Akita}\ \emph {et~al.}(2022)\citenamefont {Akita},
  \citenamefont {Im},\ and\ \citenamefont {Masud}}]{Akita:2022etk}%
  \BibitemOpen
  \bibfield  {author} {\bibinfo {author} {\bibfnamefont {K.}~\bibnamefont
  {Akita}}, \bibinfo {author} {\bibfnamefont {S.~H.}\ \bibnamefont {Im}}, \
  and\ \bibinfo {author} {\bibfnamefont {M.}~\bibnamefont {Masud}},\
  }\href@noop {} {\  (\bibinfo {year} {2022})},\ \Eprint
  {http://arxiv.org/abs/2206.06852} {arXiv:2206.06852 [hep-ph]} \BibitemShut
  {NoStop}%
\bibitem [{\citenamefont {Barbieri}\ and\ \citenamefont
  {Dolgov}(1991)}]{Barbieri:1990vx}%
  \BibitemOpen
  \bibfield  {author} {\bibinfo {author} {\bibfnamefont {R.}~\bibnamefont
  {Barbieri}}\ and\ \bibinfo {author} {\bibfnamefont {A.}~\bibnamefont
  {Dolgov}},\ }\href {\doibase 10.1016/0550-3213(91)90396-F} {\bibfield
  {journal} {\bibinfo  {journal} {Nucl. Phys. B}\ }\textbf {\bibinfo {volume}
  {349}},\ \bibinfo {pages} {743} (\bibinfo {year} {1991})}\BibitemShut
  {NoStop}%
\bibitem [{\citenamefont {Abazajian}(2006)}]{Abazajian:2005gj}%
  \BibitemOpen
  \bibfield  {author} {\bibinfo {author} {\bibfnamefont {K.}~\bibnamefont
  {Abazajian}},\ }\href {\doibase 10.1103/PhysRevD.73.063506} {\bibfield
  {journal} {\bibinfo  {journal} {Phys. Rev. D}\ }\textbf {\bibinfo {volume}
  {73}},\ \bibinfo {pages} {063506} (\bibinfo {year} {2006})},\ \Eprint
  {http://arxiv.org/abs/astro-ph/0511630} {arXiv:astro-ph/0511630} \BibitemShut
  {NoStop}%
\bibitem [{\citenamefont {Pisanti}\ \emph {et~al.}(2021)\citenamefont
  {Pisanti}, \citenamefont {Mangano}, \citenamefont {Miele},\ and\
  \citenamefont {Mazzella}}]{Pisanti:2020efz}%
  \BibitemOpen
  \bibfield  {author} {\bibinfo {author} {\bibfnamefont {O.}~\bibnamefont
  {Pisanti}}, \bibinfo {author} {\bibfnamefont {G.}~\bibnamefont {Mangano}},
  \bibinfo {author} {\bibfnamefont {G.}~\bibnamefont {Miele}}, \ and\ \bibinfo
  {author} {\bibfnamefont {P.}~\bibnamefont {Mazzella}},\ }\href {\doibase
  10.1088/1475-7516/2021/04/020} {\bibfield  {journal} {\bibinfo  {journal}
  {JCAP}\ }\textbf {\bibinfo {volume} {04}},\ \bibinfo {pages} {020} (\bibinfo
  {year} {2021})},\ \Eprint {http://arxiv.org/abs/2011.11537} {arXiv:2011.11537
  [astro-ph.CO]} \BibitemShut {NoStop}%
\bibitem [{\citenamefont {B\"oser}\ \emph {et~al.}(2020)\citenamefont
  {B\"oser}, \citenamefont {Buck}, \citenamefont {Giunti}, \citenamefont
  {Lesgourgues}, \citenamefont {Ludhova}, \citenamefont {Mertens},
  \citenamefont {Schukraft},\ and\ \citenamefont {Wurm}}]{Boser:2019rta}%
  \BibitemOpen
  \bibfield  {author} {\bibinfo {author} {\bibfnamefont {S.}~\bibnamefont
  {B\"oser}}, \bibinfo {author} {\bibfnamefont {C.}~\bibnamefont {Buck}},
  \bibinfo {author} {\bibfnamefont {C.}~\bibnamefont {Giunti}}, \bibinfo
  {author} {\bibfnamefont {J.}~\bibnamefont {Lesgourgues}}, \bibinfo {author}
  {\bibfnamefont {L.}~\bibnamefont {Ludhova}}, \bibinfo {author} {\bibfnamefont
  {S.}~\bibnamefont {Mertens}}, \bibinfo {author} {\bibfnamefont
  {A.}~\bibnamefont {Schukraft}}, \ and\ \bibinfo {author} {\bibfnamefont
  {M.}~\bibnamefont {Wurm}},\ }\href {\doibase 10.1016/j.ppnp.2019.103736}
  {\bibfield  {journal} {\bibinfo  {journal} {Prog. Part. Nucl. Phys.}\
  }\textbf {\bibinfo {volume} {111}},\ \bibinfo {pages} {103736} (\bibinfo
  {year} {2020})},\ \Eprint {http://arxiv.org/abs/1906.01739} {arXiv:1906.01739
  [hep-ex]} \BibitemShut {NoStop}%
\bibitem [{\citenamefont {Dentler}\ \emph {et~al.}(2018)\citenamefont
  {Dentler}, \citenamefont {Hern\'andez-Cabezudo}, \citenamefont {Kopp},
  \citenamefont {Machado}, \citenamefont {Maltoni}, \citenamefont
  {Martinez-Soler},\ and\ \citenamefont {Schwetz}}]{Dentler:2018sju}%
  \BibitemOpen
  \bibfield  {author} {\bibinfo {author} {\bibfnamefont {M.}~\bibnamefont
  {Dentler}}, \bibinfo {author} {\bibfnamefont {A.}~\bibnamefont
  {Hern\'andez-Cabezudo}}, \bibinfo {author} {\bibfnamefont {J.}~\bibnamefont
  {Kopp}}, \bibinfo {author} {\bibfnamefont {P.~A.~N.}\ \bibnamefont
  {Machado}}, \bibinfo {author} {\bibfnamefont {M.}~\bibnamefont {Maltoni}},
  \bibinfo {author} {\bibfnamefont {I.}~\bibnamefont {Martinez-Soler}}, \ and\
  \bibinfo {author} {\bibfnamefont {T.}~\bibnamefont {Schwetz}},\ }\href
  {\doibase 10.1007/JHEP08(2018)010} {\bibfield  {journal} {\bibinfo  {journal}
  {JHEP}\ }\textbf {\bibinfo {volume} {08}},\ \bibinfo {pages} {010} (\bibinfo
  {year} {2018})},\ \Eprint {http://arxiv.org/abs/1803.10661} {arXiv:1803.10661
  [hep-ph]} \BibitemShut {NoStop}%
\bibitem [{\citenamefont {Forastieri}\ \emph {et~al.}(2019)\citenamefont
  {Forastieri}, \citenamefont {Lattanzi},\ and\ \citenamefont
  {Natoli}}]{Forastieri:2019cuf}%
  \BibitemOpen
  \bibfield  {author} {\bibinfo {author} {\bibfnamefont {F.}~\bibnamefont
  {Forastieri}}, \bibinfo {author} {\bibfnamefont {M.}~\bibnamefont
  {Lattanzi}}, \ and\ \bibinfo {author} {\bibfnamefont {P.}~\bibnamefont
  {Natoli}},\ }\href {\doibase 10.1103/PhysRevD.100.103526} {\bibfield
  {journal} {\bibinfo  {journal} {Phys. Rev. D}\ }\textbf {\bibinfo {volume}
  {100}},\ \bibinfo {pages} {103526} (\bibinfo {year} {2019})},\ \Eprint
  {http://arxiv.org/abs/1904.07810} {arXiv:1904.07810 [astro-ph.CO]}
  \BibitemShut {NoStop}%
\bibitem [{\citenamefont {Escudero}\ and\ \citenamefont
  {Witte}(2021)}]{Escudero:2021rfi}%
  \BibitemOpen
  \bibfield  {author} {\bibinfo {author} {\bibfnamefont {M.}~\bibnamefont
  {Escudero}}\ and\ \bibinfo {author} {\bibfnamefont {S.~J.}\ \bibnamefont
  {Witte}},\ }\href {\doibase 10.1140/epjc/s10052-021-09276-5} {\bibfield
  {journal} {\bibinfo  {journal} {Eur. Phys. J. C}\ }\textbf {\bibinfo {volume}
  {81}},\ \bibinfo {pages} {515} (\bibinfo {year} {2021})},\ \Eprint
  {http://arxiv.org/abs/2103.03249} {arXiv:2103.03249 [hep-ph]} \BibitemShut
  {NoStop}%
\bibitem [{\citenamefont {Staub}(2008)}]{Staub:2008uz}%
  \BibitemOpen
  \bibfield  {author} {\bibinfo {author} {\bibfnamefont {F.}~\bibnamefont
  {Staub}},\ }\href@noop {} {\  (\bibinfo {year} {2008})},\ \Eprint
  {http://arxiv.org/abs/0806.0538} {arXiv:0806.0538 [hep-ph]} \BibitemShut
  {NoStop}%
\bibitem [{\citenamefont {Staub}(2015)}]{Staub:2015kfa}%
  \BibitemOpen
  \bibfield  {author} {\bibinfo {author} {\bibfnamefont {F.}~\bibnamefont
  {Staub}},\ }\href {\doibase 10.1155/2015/840780} {\bibfield  {journal}
  {\bibinfo  {journal} {Adv. High Energy Phys.}\ }\textbf {\bibinfo {volume}
  {2015}},\ \bibinfo {pages} {840780} (\bibinfo {year} {2015})},\ \Eprint
  {http://arxiv.org/abs/1503.04200} {arXiv:1503.04200 [hep-ph]} \BibitemShut
  {NoStop}%
\bibitem [{\citenamefont {Abada}\ \emph {et~al.}(2022)\citenamefont {Abada},
  \citenamefont {Escribano}, \citenamefont {Marcano},\ and\ \citenamefont
  {Piazza}}]{Abada:2022wvh}%
  \BibitemOpen
  \bibfield  {author} {\bibinfo {author} {\bibfnamefont {A.}~\bibnamefont
  {Abada}}, \bibinfo {author} {\bibfnamefont {P.}~\bibnamefont {Escribano}},
  \bibinfo {author} {\bibfnamefont {X.}~\bibnamefont {Marcano}}, \ and\
  \bibinfo {author} {\bibfnamefont {G.}~\bibnamefont {Piazza}},\ }\href@noop {}
  {\  (\bibinfo {year} {2022})},\ \Eprint {http://arxiv.org/abs/2208.13882}
  {arXiv:2208.13882 [hep-ph]} \BibitemShut {NoStop}%
\bibitem [{\citenamefont {Buchmuller}\ \emph {et~al.}(2005)\citenamefont
  {Buchmuller}, \citenamefont {Di~Bari},\ and\ \citenamefont
  {Plumacher}}]{Buchmuller:2004nz}%
  \BibitemOpen
  \bibfield  {author} {\bibinfo {author} {\bibfnamefont {W.}~\bibnamefont
  {Buchmuller}}, \bibinfo {author} {\bibfnamefont {P.}~\bibnamefont {Di~Bari}},
  \ and\ \bibinfo {author} {\bibfnamefont {M.}~\bibnamefont {Plumacher}},\
  }\href {\doibase 10.1016/j.aop.2004.02.003} {\bibfield  {journal} {\bibinfo
  {journal} {Annals Phys.}\ }\textbf {\bibinfo {volume} {315}},\ \bibinfo
  {pages} {305} (\bibinfo {year} {2005})},\ \Eprint
  {http://arxiv.org/abs/hep-ph/0401240} {arXiv:hep-ph/0401240} \BibitemShut
  {NoStop}%
\bibitem [{\citenamefont {Davidson}\ \emph {et~al.}(2008)\citenamefont
  {Davidson}, \citenamefont {Nardi},\ and\ \citenamefont
  {Nir}}]{Davidson:2008bu}%
  \BibitemOpen
  \bibfield  {author} {\bibinfo {author} {\bibfnamefont {S.}~\bibnamefont
  {Davidson}}, \bibinfo {author} {\bibfnamefont {E.}~\bibnamefont {Nardi}}, \
  and\ \bibinfo {author} {\bibfnamefont {Y.}~\bibnamefont {Nir}},\ }\href
  {\doibase 10.1016/j.physrep.2008.06.002} {\bibfield  {journal} {\bibinfo
  {journal} {Phys. Rept.}\ }\textbf {\bibinfo {volume} {466}},\ \bibinfo
  {pages} {105} (\bibinfo {year} {2008})},\ \Eprint
  {http://arxiv.org/abs/0802.2962} {arXiv:0802.2962 [hep-ph]} \BibitemShut
  {NoStop}%
\bibitem [{\citenamefont {Davidson}\ and\ \citenamefont
  {Ibarra}(2002)}]{Davidson:2002qv}%
  \BibitemOpen
  \bibfield  {author} {\bibinfo {author} {\bibfnamefont {S.}~\bibnamefont
  {Davidson}}\ and\ \bibinfo {author} {\bibfnamefont {A.}~\bibnamefont
  {Ibarra}},\ }\href {\doibase 10.1016/S0370-2693(02)01735-5} {\bibfield
  {journal} {\bibinfo  {journal} {Phys. Lett. B}\ }\textbf {\bibinfo {volume}
  {535}},\ \bibinfo {pages} {25} (\bibinfo {year} {2002})},\ \Eprint
  {http://arxiv.org/abs/hep-ph/0202239} {arXiv:hep-ph/0202239} \BibitemShut
  {NoStop}%
\bibitem [{\citenamefont {Arkani-Hamed}\ \emph {et~al.}(2016)\citenamefont
  {Arkani-Hamed}, \citenamefont {Cohen}, \citenamefont {D'Agnolo},
  \citenamefont {Hook}, \citenamefont {Kim},\ and\ \citenamefont
  {Pinner}}]{Arkani-Hamed:2016rle}%
  \BibitemOpen
  \bibfield  {author} {\bibinfo {author} {\bibfnamefont {N.}~\bibnamefont
  {Arkani-Hamed}}, \bibinfo {author} {\bibfnamefont {T.}~\bibnamefont {Cohen}},
  \bibinfo {author} {\bibfnamefont {R.~T.}\ \bibnamefont {D'Agnolo}}, \bibinfo
  {author} {\bibfnamefont {A.}~\bibnamefont {Hook}}, \bibinfo {author}
  {\bibfnamefont {H.~D.}\ \bibnamefont {Kim}}, \ and\ \bibinfo {author}
  {\bibfnamefont {D.}~\bibnamefont {Pinner}},\ }\href {\doibase
  10.1103/PhysRevLett.117.251801} {\bibfield  {journal} {\bibinfo  {journal}
  {Phys. Rev. Lett.}\ }\textbf {\bibinfo {volume} {117}},\ \bibinfo {pages}
  {251801} (\bibinfo {year} {2016})},\ \Eprint
  {http://arxiv.org/abs/1607.06821} {arXiv:1607.06821 [hep-ph]} \BibitemShut
  {NoStop}%
\bibitem [{\citenamefont {Buchmuller}\ \emph {et~al.}(2007)\citenamefont
  {Buchmuller}, \citenamefont {Hamaguchi}, \citenamefont {Lebedev},
  \citenamefont {Ramos-Sanchez},\ and\ \citenamefont
  {Ratz}}]{Buchmuller:2007zd}%
  \BibitemOpen
  \bibfield  {author} {\bibinfo {author} {\bibfnamefont {W.}~\bibnamefont
  {Buchmuller}}, \bibinfo {author} {\bibfnamefont {K.}~\bibnamefont
  {Hamaguchi}}, \bibinfo {author} {\bibfnamefont {O.}~\bibnamefont {Lebedev}},
  \bibinfo {author} {\bibfnamefont {S.}~\bibnamefont {Ramos-Sanchez}}, \ and\
  \bibinfo {author} {\bibfnamefont {M.}~\bibnamefont {Ratz}},\ }\href {\doibase
  10.1103/PhysRevLett.99.021601} {\bibfield  {journal} {\bibinfo  {journal}
  {Phys. Rev. Lett.}\ }\textbf {\bibinfo {volume} {99}},\ \bibinfo {pages}
  {021601} (\bibinfo {year} {2007})},\ \Eprint
  {http://arxiv.org/abs/hep-ph/0703078} {arXiv:hep-ph/0703078} \BibitemShut
  {NoStop}%
\bibitem [{\citenamefont {Ellis}\ and\ \citenamefont
  {Lebedev}(2007)}]{Ellis:2007wz}%
  \BibitemOpen
  \bibfield  {author} {\bibinfo {author} {\bibfnamefont {J.~R.}\ \bibnamefont
  {Ellis}}\ and\ \bibinfo {author} {\bibfnamefont {O.}~\bibnamefont
  {Lebedev}},\ }\href {\doibase 10.1016/j.physletb.2007.08.031} {\bibfield
  {journal} {\bibinfo  {journal} {Phys. Lett. B}\ }\textbf {\bibinfo {volume}
  {653}},\ \bibinfo {pages} {411} (\bibinfo {year} {2007})},\ \Eprint
  {http://arxiv.org/abs/0707.3419} {arXiv:0707.3419 [hep-ph]} \BibitemShut
  {NoStop}%
\bibitem [{\citenamefont {Hinshaw}\ \emph {et~al.}(2013)\citenamefont {Hinshaw}
  \emph {et~al.}}]{WMAP:2012nax}%
  \BibitemOpen
  \bibfield  {author} {\bibinfo {author} {\bibfnamefont {G.}~\bibnamefont
  {Hinshaw}} \emph {et~al.} (\bibinfo {collaboration} {WMAP}),\ }\href
  {\doibase 10.1088/0067-0049/208/2/19} {\bibfield  {journal} {\bibinfo
  {journal} {Astrophys. J. Suppl.}\ }\textbf {\bibinfo {volume} {208}},\
  \bibinfo {pages} {19} (\bibinfo {year} {2013})},\ \Eprint
  {http://arxiv.org/abs/1212.5226} {arXiv:1212.5226 [astro-ph.CO]} \BibitemShut
  {NoStop}%
\bibitem [{\citenamefont {Lattanzi}\ and\ \citenamefont
  {Gerbino}(2018)}]{Lattanzi:2017ubx}%
  \BibitemOpen
  \bibfield  {author} {\bibinfo {author} {\bibfnamefont {M.}~\bibnamefont
  {Lattanzi}}\ and\ \bibinfo {author} {\bibfnamefont {M.}~\bibnamefont
  {Gerbino}},\ }\href {\doibase 10.3389/fphy.2017.00070} {\bibfield  {journal}
  {\bibinfo  {journal} {Front. in Phys.}\ }\textbf {\bibinfo {volume} {5}},\
  \bibinfo {pages} {70} (\bibinfo {year} {2018})},\ \Eprint
  {http://arxiv.org/abs/1712.07109} {arXiv:1712.07109 [astro-ph.CO]}
  \BibitemShut {NoStop}%
\bibitem [{\citenamefont {Hannestad}(2005)}]{Hannestad:2004qu}%
  \BibitemOpen
  \bibfield  {author} {\bibinfo {author} {\bibfnamefont {S.}~\bibnamefont
  {Hannestad}},\ }\href {\doibase 10.1088/1475-7516/2005/02/011} {\bibfield
  {journal} {\bibinfo  {journal} {JCAP}\ }\textbf {\bibinfo {volume} {02}},\
  \bibinfo {pages} {011} (\bibinfo {year} {2005})},\ \Eprint
  {http://arxiv.org/abs/astro-ph/0411475} {arXiv:astro-ph/0411475} \BibitemShut
  {NoStop}%
\bibitem [{\citenamefont {Escudero}\ \emph {et~al.}(2024)\citenamefont
  {Escudero}, \citenamefont {Schwetz},\ and\ \citenamefont
  {{Terol-Calvo}}}]{Addendum}%
  \BibitemOpen
  \bibfield  {author} {\bibinfo {author} {\bibfnamefont {M.}~\bibnamefont
  {Escudero}}, \bibinfo {author} {\bibfnamefont {T.}~\bibnamefont {Schwetz}}, \
  and\ \bibinfo {author} {\bibfnamefont {J.}~\bibnamefont {{Terol-Calvo}}},\
  }\href {\doibase 10.1007/JHEP06(2024)119} {\bibfield  {journal} {\bibinfo
  {journal} {Journal of High Energy Physics}\ }\textbf {\bibinfo {volume}
  {2024}},\ \bibinfo {pages} {119} (\bibinfo {year} {2024})}\BibitemShut
  {NoStop}%
\bibitem [{\citenamefont {Weldon}(1982)}]{Weldon:1982bn}%
  \BibitemOpen
  \bibfield  {author} {\bibinfo {author} {\bibfnamefont {H.~A.}\ \bibnamefont
  {Weldon}},\ }\href {\doibase 10.1103/PhysRevD.26.2789} {\bibfield  {journal}
  {\bibinfo  {journal} {Phys. Rev. D}\ }\textbf {\bibinfo {volume} {26}},\
  \bibinfo {pages} {2789} (\bibinfo {year} {1982})}\BibitemShut {NoStop}%
\bibitem [{\citenamefont {Dasgupta}\ and\ \citenamefont
  {Kopp}(2014)}]{Dasgupta:2013zpn}%
  \BibitemOpen
  \bibfield  {author} {\bibinfo {author} {\bibfnamefont {B.}~\bibnamefont
  {Dasgupta}}\ and\ \bibinfo {author} {\bibfnamefont {J.}~\bibnamefont
  {Kopp}},\ }\href {\doibase 10.1103/PhysRevLett.112.031803} {\bibfield
  {journal} {\bibinfo  {journal} {Phys. Rev. Lett.}\ }\textbf {\bibinfo
  {volume} {112}},\ \bibinfo {pages} {031803} (\bibinfo {year} {2014})},\
  \Eprint {http://arxiv.org/abs/1310.6337} {arXiv:1310.6337 [hep-ph]}
  \BibitemShut {NoStop}%
\bibitem [{\citenamefont {Chu}\ \emph {et~al.}(2015)\citenamefont {Chu},
  \citenamefont {Dasgupta},\ and\ \citenamefont {Kopp}}]{Chu:2015ipa}%
  \BibitemOpen
  \bibfield  {author} {\bibinfo {author} {\bibfnamefont {X.}~\bibnamefont
  {Chu}}, \bibinfo {author} {\bibfnamefont {B.}~\bibnamefont {Dasgupta}}, \
  and\ \bibinfo {author} {\bibfnamefont {J.}~\bibnamefont {Kopp}},\ }\href
  {\doibase 10.1088/1475-7516/2015/10/011} {\bibfield  {journal} {\bibinfo
  {journal} {JCAP}\ }\textbf {\bibinfo {volume} {10}},\ \bibinfo {pages} {011}
  (\bibinfo {year} {2015})},\ \Eprint {http://arxiv.org/abs/1505.02795}
  {arXiv:1505.02795 [hep-ph]} \BibitemShut {NoStop}%
\bibitem [{\citenamefont {Notzold}\ and\ \citenamefont
  {Raffelt}(1988)}]{Notzold:1987ik}%
  \BibitemOpen
  \bibfield  {author} {\bibinfo {author} {\bibfnamefont {D.}~\bibnamefont
  {Notzold}}\ and\ \bibinfo {author} {\bibfnamefont {G.}~\bibnamefont
  {Raffelt}},\ }\href {\doibase 10.1016/0550-3213(88)90113-7} {\bibfield
  {journal} {\bibinfo  {journal} {Nucl. Phys. B}\ }\textbf {\bibinfo {volume}
  {307}},\ \bibinfo {pages} {924} (\bibinfo {year} {1988})}\BibitemShut
  {NoStop}%
\bibitem [{\citenamefont {Escudero}(2019)}]{Escudero:2018mvt}%
  \BibitemOpen
  \bibfield  {author} {\bibinfo {author} {\bibfnamefont {M.}~\bibnamefont
  {Escudero}},\ }\href {\doibase 10.1088/1475-7516/2019/02/007} {\bibfield
  {journal} {\bibinfo  {journal} {JCAP}\ }\textbf {\bibinfo {volume} {02}},\
  \bibinfo {pages} {007} (\bibinfo {year} {2019})},\ \Eprint
  {http://arxiv.org/abs/1812.05605} {arXiv:1812.05605 [hep-ph]} \BibitemShut
  {NoStop}%
\bibitem [{\citenamefont {Bringmann}\ \emph {et~al.}(2021)\citenamefont
  {Bringmann}, \citenamefont {Depta}, \citenamefont {Hufnagel}, \citenamefont
  {Ruderman},\ and\ \citenamefont {Schmidt-Hoberg}}]{Bringmann:2021tjr}%
  \BibitemOpen
  \bibfield  {author} {\bibinfo {author} {\bibfnamefont {T.}~\bibnamefont
  {Bringmann}}, \bibinfo {author} {\bibfnamefont {P.~F.}\ \bibnamefont
  {Depta}}, \bibinfo {author} {\bibfnamefont {M.}~\bibnamefont {Hufnagel}},
  \bibinfo {author} {\bibfnamefont {J.~T.}\ \bibnamefont {Ruderman}}, \ and\
  \bibinfo {author} {\bibfnamefont {K.}~\bibnamefont {Schmidt-Hoberg}},\ }\href
  {\doibase 10.1103/PhysRevLett.127.191802} {\bibfield  {journal} {\bibinfo
  {journal} {Phys. Rev. Lett.}\ }\textbf {\bibinfo {volume} {127}},\ \bibinfo
  {pages} {191802} (\bibinfo {year} {2021})},\ \Eprint
  {http://arxiv.org/abs/2103.16572} {arXiv:2103.16572 [hep-ph]} \BibitemShut
  {NoStop}%
\end{thebibliography}%
\end{document}